\documentclass[twocolumn,aps,showpacs,showkeys,prd,superscriptaddress,byrevtex, amsmath]{revtex4-1}

\usepackage{amssymb}
\usepackage{amsmath}
\usepackage{verbatim}
\usepackage{mathrsfs}
\usepackage{amsfonts}
\usepackage{latexsym}
\usepackage{epsfig}
\usepackage{color}
\usepackage{graphicx,subfigure}
\usepackage{units}
\usepackage{aas_macros}

\begin{document}


\renewcommand{\t}{\times}
\newcommand{\sg}[1]{\textcolor{blue}{SG: #1}}
\newcommand{\tf}[1]{\textcolor{red}{TF: #1}}
\newcommand{\ch}[1]{\textcolor{cyan}{CH: #1}}

\long\def\symbolfootnote[#1]#2{\begingroup%
\def\thefootnote{\fnsymbol{footnote}}\footnote[#1]{#2}\endgroup}


\title{Magnetized accretion disks around Kerr black holes with scalar hair - I. Constant angular momentum disks} 

\author{Sergio Gimeno-Soler}
\affiliation{Departamento de
  Astronom\'{\i}a y Astrof\'{\i}sica, Universitat de Val\`encia,
  C/ Dr. Moliner 50, 46100, Burjassot (Val\`encia), Spain}

\author{Jos\'e A. Font}
\affiliation{Departamento de
  Astronom\'{\i}a y Astrof\'{\i}sica, Universitat de Val\`encia,
  C/ Dr. Moliner 50, 46100, Burjassot (Val\`encia), Spain}
\affiliation{Observatori Astron\`omic, Universitat de Val\`encia, C/ Catedr\'atico 
  Jos\'e Beltr\'an 2, 46980, Paterna (Val\`encia), Spain}

\author{Carlos Herdeiro}
\affiliation{Centro de Astrof\'\i sica e Gravita\c c\~ao - CENTRA, Departamento de F\'\i sica,
Instituto Superior T\'ecnico - IST, Universidade de Lisboa - UL, Avenida
Rovisco Pais 1, 1049-001, Portugal}

\author{Eugen Radu}
\affiliation{School of Theoretical Physics, Dublin Institute for Advanced Studies, 
 10 Burlington Road, Dublin 4, Ireland and 
CIDMA, Universidade de Aveiro,
Campus de Santiago, 3810-183 Aveiro, Portugal}


\date{\today}


\begin{abstract} 
Testing the true nature of black holes -- the no-hair hypothesis -- will become increasingly more precise in the next few years as new observational data is collected in both the gravitational wave channel and the electromagnetic channel. In this paper we consider numerically generated spacetimes of Kerr black holes with synchronised scalar hair and build stationary models of magnetized thick disks (or tori)  around them. Our approach assumes that the disks are not self-gravitating, they obey a polytropic equation of state, the distribution of their specific angular momentum is constant, and they are marginally stable, i.e.~the disks completely fill their Roche lobe. Moreover, contrary to existing approaches in the literature, our models are thermodinamically relativist, as the specific enthalpy of the fluid can adopt values significantly larger than unity. We study the dependence of the morphology and properties of the accretion tori on the type of black hole considered, from purely Kerr black holes with varying degrees of spin parameter, namely from a Schwarzschild black hole to a nearly extremal Kerr case,  to Kerr black holes with scalar hair with different ADM mass and horizon angular velocity. Comparisons between the disk properties for both types of black holes are presented. The sequences of magnetized, equilibrium disks models discussed in this study can be used as initial data for numerical relativity codes to investigate their dynamical (non-linear) stability and used in tandem with ray-tracing codes to obtain synthetic images of black holes (i.e.~shadows) in astrophysically relevant situations where the light source is provided by an emitting accretion disk.
\end{abstract}




\maketitle


\section{Introduction}

In recent years, new families of stationary, asymptotically flat black holes (BHs) avoiding the so-called ``no hair" theorems, have been obtained both in general relativity and in modified gravity (see e.g.~\cite{Herdeiro:2015_review} and references therein). Among those, Kerr BHs with synchronised hair~\cite{Herdeiro:2014a,Herdeiro:2016} are a counterexample to the no hair conjecture resulting from minimally coupling Einstein's gravity to simple (bosonic) matter fields obeying all energy conditions. The physical conditions and stability properties of these classes of {\it hairy} BHs  (HBHs) have been recently investigated to assess their potential viability as  alternatives to astrophysical Kerr BHs.  On the one hand, Kerr BHs with Proca hair have been shown to form dynamically as the end-product of the superradiant instability~\cite{East:2017,Herdeiro:2017} (see also~\cite{Sanchis-Gual:2016,Bosch:2016} for the case of a charged scalar field around a charged BH in spherical  symmetry). On the other hand, even though the hairy BHs themselves are (like Kerr BHs) afflicted by superradiant instabilities~\cite{Herdeiro:2014b,Ganchev:2018}, these instabilities are weaker than for Kerr and, at least in some regions of parameter space, are inefficient for astrophysical time scales, making the hairy BHs {\it effectively} stable against superradiance~\cite{Degollado:2018}. 

In the observational arena, the LIGO/Virgo detection of gravitational waves from binary BHs~\cite{Abbott2016, Abbott:2016nmj, Abbott:2017vtc, Abbott:2017oio, Abbott:2017gyy} and the exciting prospects of observing the first image -- the black hole {\it shadow} -- of a BH by the Event Horizon Telescope (EHT)~\cite{Fish:2016} opens the opportunity to test the true nature of BHs -- the no-hair hypothesis -- and, in particular, the astrophysical relevance of HBHs. It is not yet known whether the LIGO/Virgo binary BH signals are consistent with alternative scenarios, such as the merger of ultracompact boson stars or non-Kerr BHs, because the latter possibilities remain thus far insufficiently modelled. Likewise, Kerr BHs with scalar hair (KBHsSH) can exhibit very distinct shadows from those of (bald) Kerr BHs, as shown by~\cite{Cunha:2015} and~\cite{Vincent:2016} for two different setups for the light source, either a celestial sphere far from the compact object or an emitting torus of matter surrounding the BH, respectively. It is therefore an intriguing open possibility if the very long baseline interferometric observations of BH candidates in Sgr A* and M87 envisaged by the EHT may constrain the astrophysical significance of HBHs.

The setup considered by~\cite{Vincent:2016} in which the light source producing the BH shadow is an accretion disk, is arguably more realistic than the distant celestial sphere of~\cite{Cunha:2015}. Thick accretion disks (or tori) are common systems in astrophysics, either surrounding the supermassive central BHs of quasars and active galactic nuclei or, at  stellar scale, surrounding the compact objects in X-ray binaries, microquasars, and gamma-ray bursts (see~\cite{Abramowicz:2013} and references therein). In this paper we present new families of stationary solutions of magnetized thick accretion disks around KBHsSH that differ from those considered by~\cite{Vincent:2016}. Our procedure, which combines earlier approaches put forward by~\cite{Komissarov:2006,Qian:2009} was presented in~\cite{Gimeno-Soler:2017} for the Kerr BH case. In Ref.~\cite{Gimeno-Soler:2017} we built equilibrium sequences of accretion disks  in the test-fluid approximation endowed with a purely toroidal magnetic field, assuming a form of the angular momentum distribution that departs from the constant case considered by~\cite{Komissarov:2006} and from which the location and morphology of the equipotential surfaces can be numerically computed. Our goal in the present work is to extend this approach to KBHsSH and to assess the dependence of the morphology and properties of accretion disks on the type of BH considered, either Kerr BHs of varying spins or KBHsSH. In this first investigation we focus on disks with a constant distribution of specific angular momentum. In the  purely hydrodynamical case, such a model is commonly refereed to as a `Polish doughnut', after the seminal work by~\cite{Abramowicz:1978} (but see also~\cite{Fishbone:1976}). In a companion paper we will present the non-constant (power-law) case, whose sequences have already been computed. The dynamical (non-linear) stability of these solutions as well as the analysis of the corresponding shadows will be discussed elsewhere.

The organization of this paper is as follows: Section~\ref{framework} presents  the mathematical framework we employ to build magnetized disks in the numerically generated spacetimes of KBHsSH. Section~\ref{procedure} discusses the corresponding numerical methodology to build the disks. Sequences of equilibrium models are presented in Section~\ref{results} along with the discussion of their morphological features and properties and the comparison with models around Kerr BHs. Finally, our conclusions are summarized in Section~\ref{conclusions}. Geometrized units ($G=c=1$) are used throughout. 
 
\section{Framework}
\label{framework}

\subsection{Spacetime metric and KBHsSH models}

The models of KBHsSH we use in this study are built following the procedure described in~\cite{Herdeiro:2015b}. The underlying theoretical framework is the Einstein-Klein-Gordon (EKG) field theory, describing a massive complex scalar field $\Psi$ minimally coupled to Einstein gravity. KBHsSH solutions are obtained by using the following ansatz for the metric and the scalar field~\cite{Herdeiro:2014a}
\begin{eqnarray}
\mathrm{d}s^2 &=& e^{2F_1}\left(\frac{\mathrm{d}r^2}{N} + r^2\mathrm{d}\theta^2\right) +  e^{2F_2}r^2\sin^2 \theta(\mathrm{d}\phi-W\mathrm{d}t)^2 
\nonumber \\ 
&-&  e^{2F_0}N\mathrm{d}t^2\,,
\label{metric}
\end{eqnarray}
\begin{eqnarray}
\Psi = \phi(r, \theta) e^{\mathrm{i}(m\varphi - \omega t)} \,,
\end{eqnarray}
with $N = 1 - r_{\rm H}/r$, where $r_{\rm H}$ is the radius of the event horizon of the BH, and $W$, $F_1$, $F_2$, $F_0$ are functions of $r$ and $\theta$. Moreover, $\omega$ is the scalar field frequency and $m$ is the azimuthal harmonic index.
We note that the radial coordinate $r$ is related to the Boyer-Lindquist radial coordinate $r_{\rm BL}$ by $r=r_{\rm BL}-a^2/r_{\rm H,BL}$, in the Kerr limit, where $a = J/M$ stands for the spin of the BH and $r_{\rm H,BL}$ is the location of the horizon in Boyer-Lindquist coordinates.

The stationary and axisymmetric metric ansatz is a solution to the EKG field equations $R_{ab} - \frac{1}{2}R g _{ab} = 8 \pi (T_{\mathrm{SF}})_{ab}$ with 
\begin{eqnarray}\label{eq:e-m_scalaf_field}
(T_{\mathrm{SF}})_{ab} &=& \partial_a \Psi^* \partial_b \Psi + \partial_b \Psi^* \partial_a \Psi 
 \\ 
&-& g_{ab} \left(\frac{1}{2} g^{cd}(\partial_c \Psi^* \partial_d \Psi + \partial_d \Psi^* \partial_c \Psi) + \mu^2 \Psi^* \Psi \right), \nonumber
\end{eqnarray}
where $\mu$ is the mass of the scalar field and superscript $(^*)$ denotes complex conjugation. The interested reader is addressed to~\cite{Herdeiro:2015b} for details on the equations of motion for the scalar field $\Psi$ and the four metric functions ${W, F_0,F_1,F_2}$, along with their solution.

\begin{figure}
\centering
\includegraphics[scale=0.43]{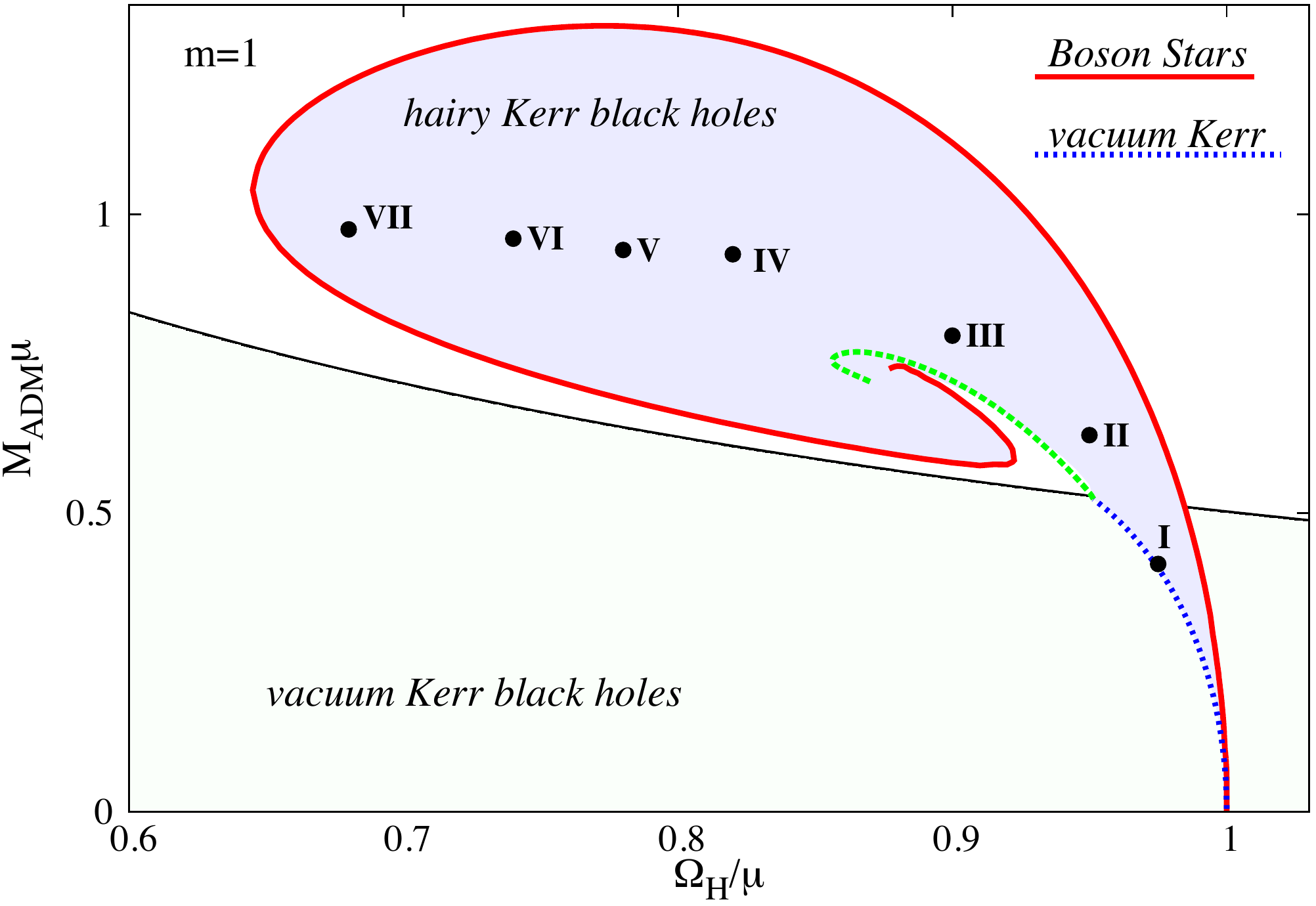}
\caption{Domain of existence for KBHsSH (shaded blue area) in an ADM mass versus scalar
field frequency diagram. The seven solutions to be studied herein are highlighted in this diagram.}
\label{existence}
\end{figure}

Table~\ref{models_list} lists the seven KBHsSH models we use in this work. The models have been selected to span all regions of interest in the parameter space. Model I corresponds to a Kerr-like model, with almost all the mass and angular momentum stored in the BH (namely, $94.7\%$ of the total mass and $87.2\%$ of the total angular momentum of the spacetime are stored in the BH), while model VII corresponds to a hairy Kerr BH with almost all the mass (98.15\%) and angular momentum (99.76\%) stored in the scalar field. It is worth mentioning that some of the models violate the Kerr bound (i.e.~the normalized spin parameter is larger than unity) in terms of both ADM or horizon quantities. This is not a source of concern because, as shown in~\cite{Herdeiro:2015c}, the linear velocity of the horizon, $v_{\mathrm{H}}$, never exceeds the speed of light. For comparison, we also show in Table~\ref{models_list} the spin parameter $a_{\mathrm{H_{eq}}}$ corresponding to a Kerr BH with a horizon linear velocity $v_{\mathrm{H}}$. In the last column of Table~\ref{models_list} we indicate the horizon sphericity of the KBHsSH, defined in~\cite{Delgado:2018} as the quotient of the equatorial and polar proper lengths of the event horizon
\begin{equation}
\mathfrak{s} = \frac{L_{\mathrm{e}}}{L_{\mathrm{p}}} = \frac{\int^{2\pi}_{0} \mathrm{d}\phi \, e^{F_2(r_{\mathrm{H}}, \pi/2)} r_{\mathrm{H}}}{2\int^{\pi}_{0} \mathrm{d} \theta \, e^{F_1(r_{\mathrm{H}}, \theta)} r_{\mathrm{H}}}\,.
\end{equation}
In addition to the information provided in Table~\ref{models_list}, Figure~\ref{existence} plots the location of our models in the domain of existence of KBHsSH in an ADM mass versus scalar field frequency diagram. 

\begin{table*}[t]
\caption{List of models of KBHsSH used in this work. From left to right the columns report the name of the model, the ADM mass, $M_{\mathrm{ADM}}$, the ADM angular momentum, $J_{\mathrm{ADM}}$, the horizon mass, $M_{\mathrm{H}}$, the horizon angular momentum, $J_{\mathrm{H}}$, the mass of the scalar field, $M_{\mathrm{SF}}$, the angular momentum of the scalar field, $J_{\mathrm{SF}}$, the radius of the event horizon, $r_{\mathrm{H}}$, the values of the normalized spin parameter for the ADM quantities, $a_{\mathrm{ADM}}$, and for the BH horizon quantities, $a_{\mathrm{H}}$, the horizon linear velocity, $v_{\mathrm{H}}$, the spin parameter corresponding to a Kerr BH with a linear velocity equal to $v_{\mathrm{H}}$, $a_{\mathrm{H_{eq}}}$, and the sphericity of the horizon, $\mathfrak{s}$. Here $\mu=1$.}        
\label{models_list}      
\centering          
\begin{tabular}{c c c c  c c c c   c c c c c}
\hline\hline       
 Model & $M_{\mathrm{ADM}}$ & $J_{\mathrm{ADM}}$ & $M_{\mathrm{H}}$ &  $J_{\mathrm{H}}$ & $M_{\mathrm{SF}}$ & $J_{\mathrm{SF}}$ & $r_{\mathrm{H}}$ & $a_{\mathrm{ADM}}$ & $a_{\mathrm{H}}$ & $v_{\mathrm{H}}$ & $a_{\mathrm{H_{eq}}}$ & $\mathfrak{s}$\\ 
\hline           
I & $0.415$ & $0.172$ & $0.393$ &  $0.150$  & $0.022$ & $0.022$ & $0.200$ & $0.9987$ & $0.971$ & $0.7685$ & $0.9663$ & $1.404$\\ 
II & $0.630$ & $0.403$ & $0.340$ &  $0.121$  & $0.290$ & $0.282$ & $0.221$ & $1.0140$ & $0.376$ & $0.6802$ & $0.9301$ & $1.352$ \\
III & $0.797$ & $0.573$ & $0.365$ &  $0.172$  & $0.432$ & $0.401$ & $0.111$ & $0.9032$ & $1.295$ & $0.7524$ & $0.9608$ & $1.489$ \\ 
IV & $0.933$ & $0.739$ & $0.234$ &  $0.114$  & $0.699$ & $0.625$ & $0.100$ & $0.8489$ & $2.082$ & $0.5635$ & $0.8554$ & $1.425$ \\ 
V & $0.940$ & $0.757$ & $0.159$ &  $0.076$  & $0.781$ & $0.680$ & $0.091$ & $0.8560$ & $3.017$ & $0.4438$ & $0.7415$ & $1.357$ \\ 
VI & $0.959$ & $0.795$ & $0.087$ &  $0.034$  & $0.872$ & $0.747$ & $0.088$ & $0.8644$ & $3.947$ & $0.2988$ & $0.5487$ & $1.222$ \\ 
VII & $0.975$ & $0.850$ & $0.018$ &  $0.002$  & $0.957$ & $0.848$ & $0.040$ & $0.8941$ & $6.173$ & $0.0973$ & $0.1928$ & $1.039$ \\ 
\hline      
\end{tabular}
\end{table*}

\subsection{Distribution of angular momentum in the disk}

Equilibrium models of thick disks around Kerr BHs are built assuming that the spacetime metric and the fluid fields are stationary and  axisymmetric (see, e.g.~\cite{Font:2002,Daigne:2004,Gimeno-Soler:2017} and references therein). For disks around KBHsSH we can follow the same approach as the metric ansatz given by Eq.~(\ref{metric}) is stationary and axisymmetric.

We start by introducing the specific angular momentum $l$ and the angular velocity $\Omega$ employing the standard definitions,
\begin{equation}
l = - \frac{u_{\phi}}{u_t}, \;\;\; \Omega = \frac{u^{\phi}}{u^t},
\end{equation}
where $u^{\mu}$ is the fluid four-velocity.
The relationship between $l$ and $\Omega$ is given by the equations
\begin{equation}
l = - \frac{\Omega g_{\phi\phi} + g_{t\phi}}{\Omega g_{t\phi} + g_{tt}}, \;\;\; \Omega = - \frac{l g_{tt} + g_{t\phi}}{l g_{t\phi} + g_{\phi\phi}},
\end{equation}
where we are assuming circular motion, i.e.~the four-velocity can be written as
\begin{equation}
u^{\mu} = (u^t, 0, 0, u^{\phi})\,.
\end{equation}

The approach we followed in~\cite{Gimeno-Soler:2017} for the angular momentum distribution of the disks was introduced by~\cite{Qian:2009}, and it is characterized by three free parameters, $\beta$, $\gamma$, and $\eta$ (see Eq.~(7) in~\cite{Gimeno-Soler:2017}). In this work, for simplicity and to reduce the ample space of parameters of the system, we consider a constant angular momentum distribution, $l(r,\theta) = \mathrm{const}$, which corresponds to setting $\beta=\gamma=0$ in~\cite{Gimeno-Soler:2017}. This choice also allows for the presence of a cusp (and hence matter accretion onto the BH) and a centre. Following~\citep{Daigne:2004}, the specific value of the angular momentum corresponding to bound fluid elements ($-u_t<1$) is computed as the minimum of the following equation
\begin{equation}\label{eq:mb_ang_mom}
l^{\pm}_{\mathrm{b}}(r, \theta) = \frac{g_{t\phi} \pm \sqrt{ (g_{t\phi}^2-g_{tt}g_{\phi\phi})  (1+g_{tt}) } }{-g_{tt}}\,,
\end{equation}
where the plus sign solution corresponds to prograde orbits and the minus sign solution to retrograde orbits. Our convention is that the angular momentum of the BH is positive and the matter of the disk rotates in the positive (negative) direction of $\phi$ for a prograde (retrograde) disk. Equation~(\ref{eq:mb_ang_mom}) is given by~\citep{Daigne:2004} for Kerr BHs, but it is valid for any stationary and axisymmetric spacetime. For prograde motion, the function has a minimum outside the event horizon. The location of this minimum corresponds with the marginally bound orbit $r_{\mathrm{mb}}$ (also known as ICO, innermost circular orbit, in the literature), and the angular momentum corresponds to the Keplerian angular momentum $l_{\mathrm{mb}}$ at that point. We show the proof of this statement in Appendix~\ref{ang_mom_appendix}.

\subsection{Magnetized disks}

To account for the magnetic field in the disks we use the procedure described by~\cite{Komissarov:2006,Montero:2007}. First, we write the equations of ideal general relativistic MHD as the following conservation laws, $\nabla_{\mu} T^{\mu\nu} = 0$, $\nabla_{\mu} \,^\ast F^{\mu\nu} = 0$, and 
$\nabla_{\mu} (\rho u^{\mu}) = 0$, 
where $\nabla_{\mu}$ is the covariant derivative and
\begin{equation}\label{eq:e-m_tensor}
T^{\mu\nu} = (\rho h + b^2)u^{\mu}u^{\nu} + (p + p_{\mathrm{m}})g^{\mu\nu} - b^{\mu}b^{\nu},
\end{equation}
is the energy-momentum tensor of a magnetized perfect fluid, with $h$, $\rho$, $p$, and $p_{\mathrm{m}}$ being the fluid specific enthalpy, density, fluid pressure, and magnetic pressure, respectively, the latter defined as $p_{\mathrm{m}} = b^2/2$. The ratio of fluid pressure to magnetic pressure defines the magnetization parameter $\beta_{\mathrm{m}} = p/p_{\mathrm{m}}$.
Moreover, $^\ast F^{\mu\nu} = b^{\mu}u^{\nu} - b^{\nu}u^{\mu}$ is the (dual of the) Faraday tensor relative to an observer with 
four-velocity $u^{\mu}$, and $b^{\mu}$ is the magnetic field in that frame, with
$b^2=b^{\mu}b_{\mu}$ (see~\cite{Anton:2006} for further details). Assuming the magnetic field is purely azimuthal, i.e.~$b^r = b^{\theta} = 0$,
and taking into account that the flow is stationary and axisymmetric, the conservation of the current density and of the Faraday tensor follow. Contracting the divergence of Eq.~\eqref{eq:e-m_tensor} with the projection tensor $h^{\alpha}_{\,\,\beta} = \delta^{\alpha}_{\,\,\beta} + u^{\alpha}u_{\beta}$, we arrive at
\begin{equation}
(\rho h + b^2)u_{\nu}\partial_i u^{\nu} + \partial_i\left(p + \frac{b^2}{2}\right) - b_{\nu}\partial_i b^{\nu}=0\,,
\end{equation}
where $i = r, \theta$. This equation can be rewritten in terms of the specific angular momentum $l$ and of the angular velocity $\Omega$, 
\begin{equation}\label{eq:diff_ver}
\partial_i(\ln |u_t|) - \frac{\Omega \partial_i l}{1-l\Omega} + \frac{\partial_i p}{\rho h} + \frac{\partial_i(\mathcal{L}b^2)}{2\mathcal{L}\rho h} = 0\,,
\end{equation}
where $\mathcal{L} = g_{t\phi}^2 - g_{tt}g_{\phi\phi}$.

To integrate Eq.~\eqref{eq:diff_ver} we need to assume an equation of state (EOS). We assume a polytropic EOS of the form
\begin{equation}\label{eq:eos_fluid}
p = K \rho^{\Gamma},
\end{equation}
with $K$ and $\Gamma$ constants.
 By introducing the definitions $\tilde{p}_{\mathrm{m}} = \mathcal{L} p_{\mathrm{m}}$, $w = \rho h$ and $\tilde{w} = \mathcal{L} (w)$, we can write equations equivalent to Eq.~\eqref{eq:eos_fluid} for both $\tilde{p}_{\mathrm{m}}$ and $p_{\mathrm{m}}$
\begin{eqnarray}
\label{eq:eos_mag_tilde}
\tilde{p}_{\mathrm{m}} &=& K_{\rm m} \tilde{w}^q,
\\
p_{\mathrm{m}} &=& K_{\rm m} \mathcal{L}^{q-1} (\rho h)^q,
\end{eqnarray}
where $K_{\rm m}$ and $q$ are constants. Then we can integrate Eq.~\eqref{eq:diff_ver} as
\begin{equation}\label{eq:final}
W - W_{\mathrm{in}} + \ln \left(1 + \frac{\Gamma K}{\Gamma -1}\rho^{\Gamma -1}\right) + \frac{q}{q-1}K_{\rm m}(\mathcal{L}\rho h)^{q-1}=0,
\end{equation}
where $W \equiv \ln |u_t|$ stands for the (gravitational plus centrifugal) potential and $W_{\mathrm{in}}$ is the potential at the inner edge of the disk.

\begin{figure*}
\centering
\includegraphics[scale=0.14]{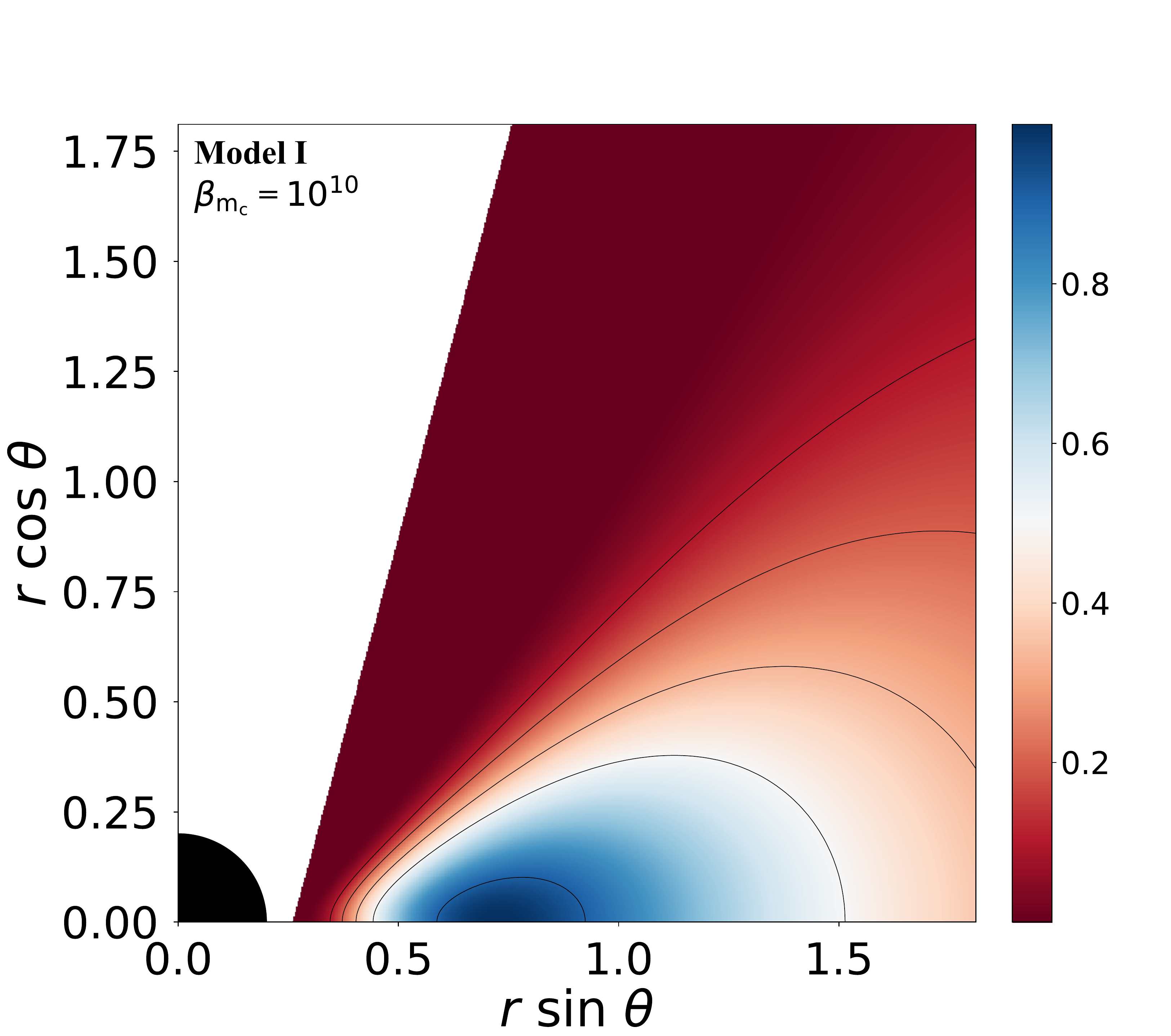}
\hspace{-0.3cm}
\includegraphics[scale=0.14]{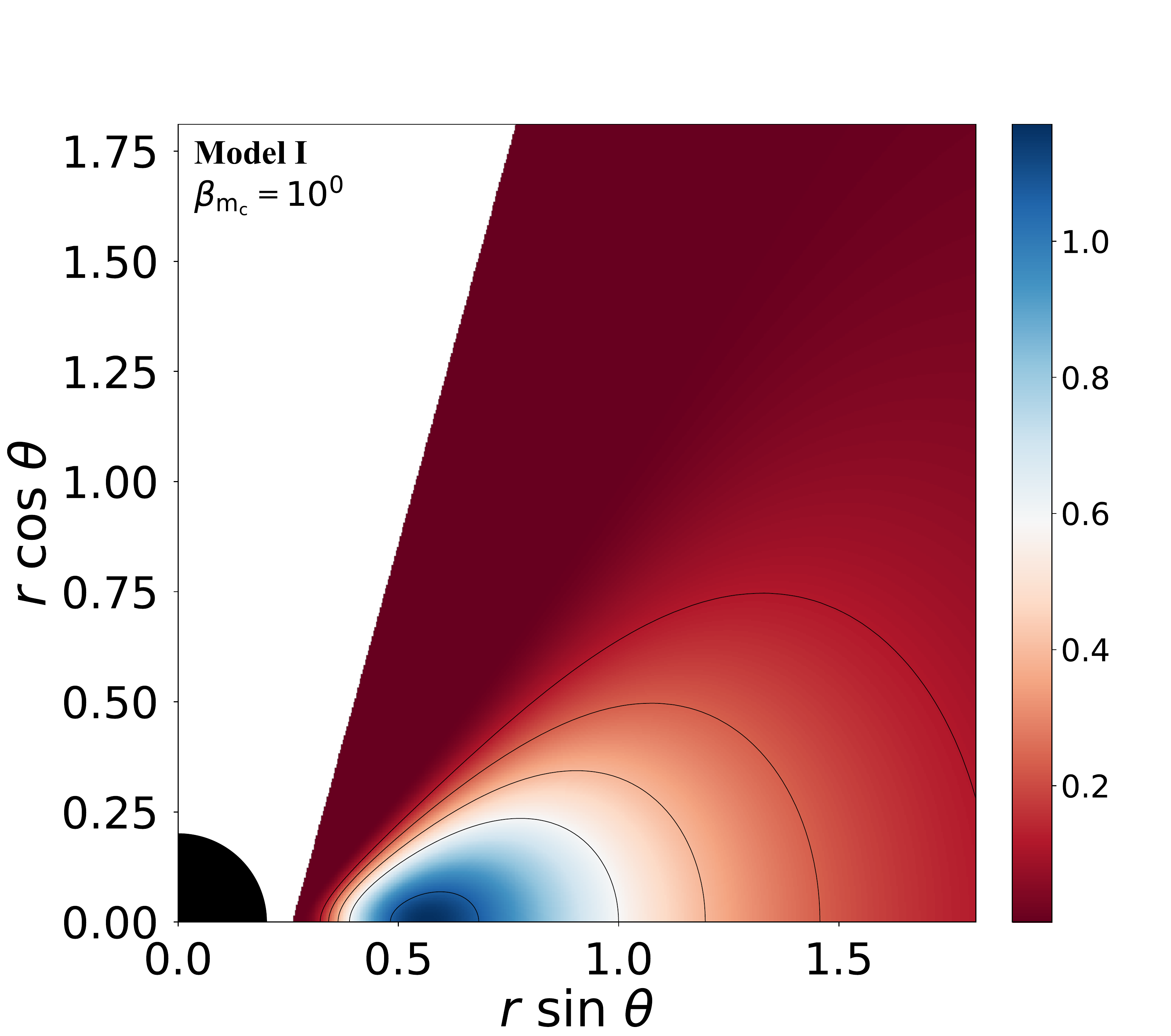}
\hspace{-0.2cm}
\includegraphics[scale=0.14]{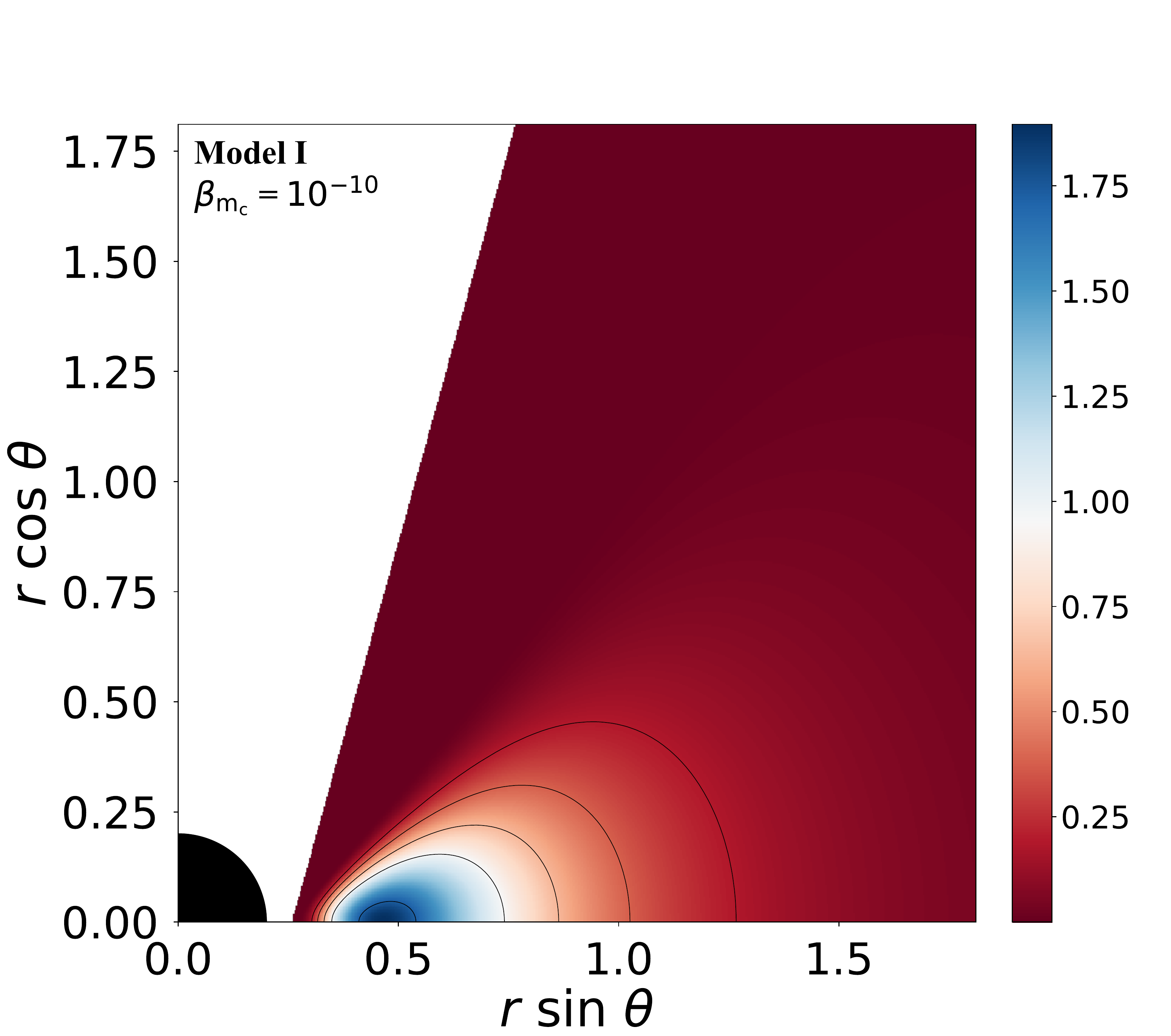}
\\
\includegraphics[scale=0.14]{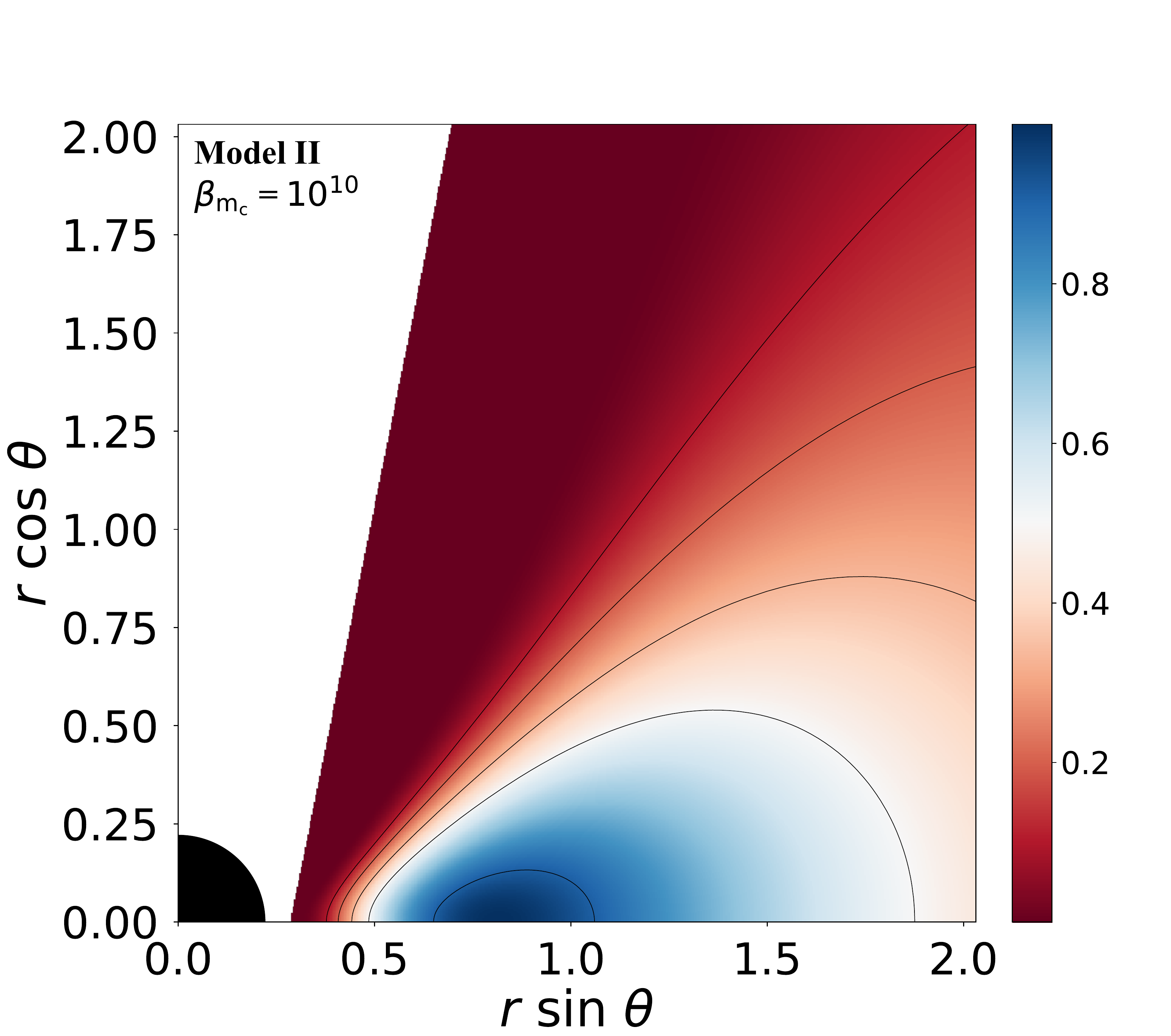}
\hspace{-0.3cm}
\includegraphics[scale=0.14]{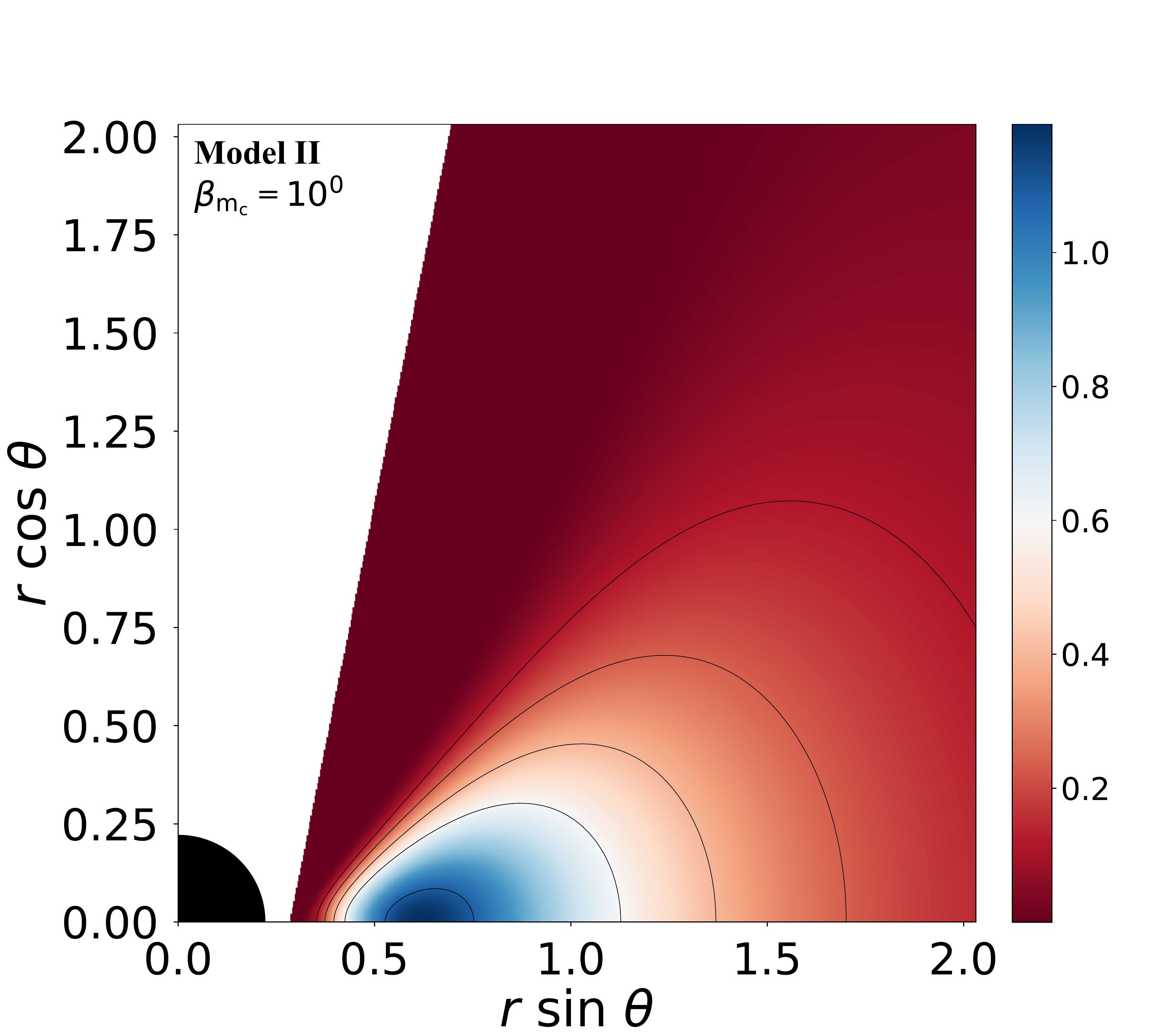}
\hspace{-0.2cm}
\includegraphics[scale=0.14]{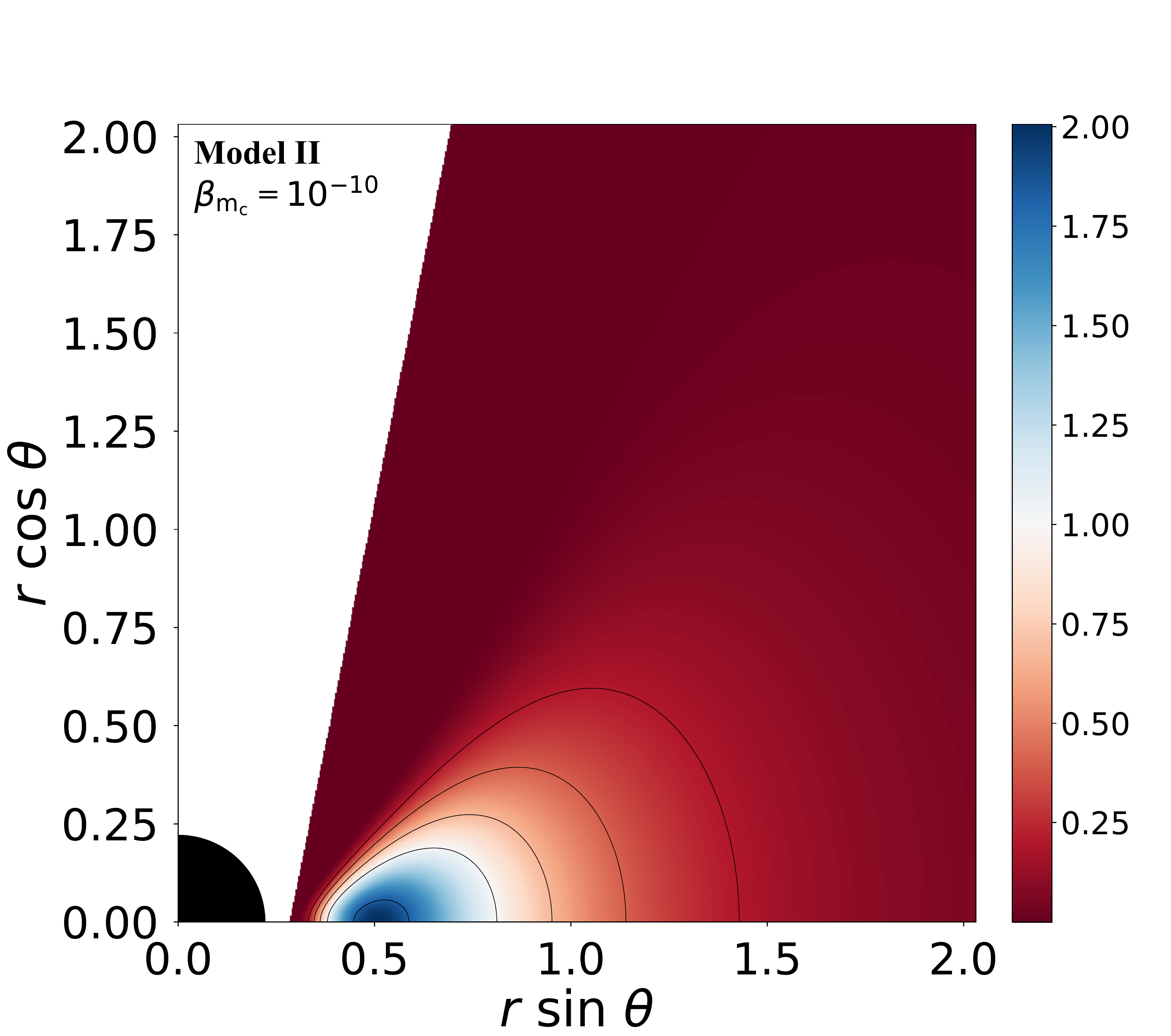}
\\
\includegraphics[scale=0.14]{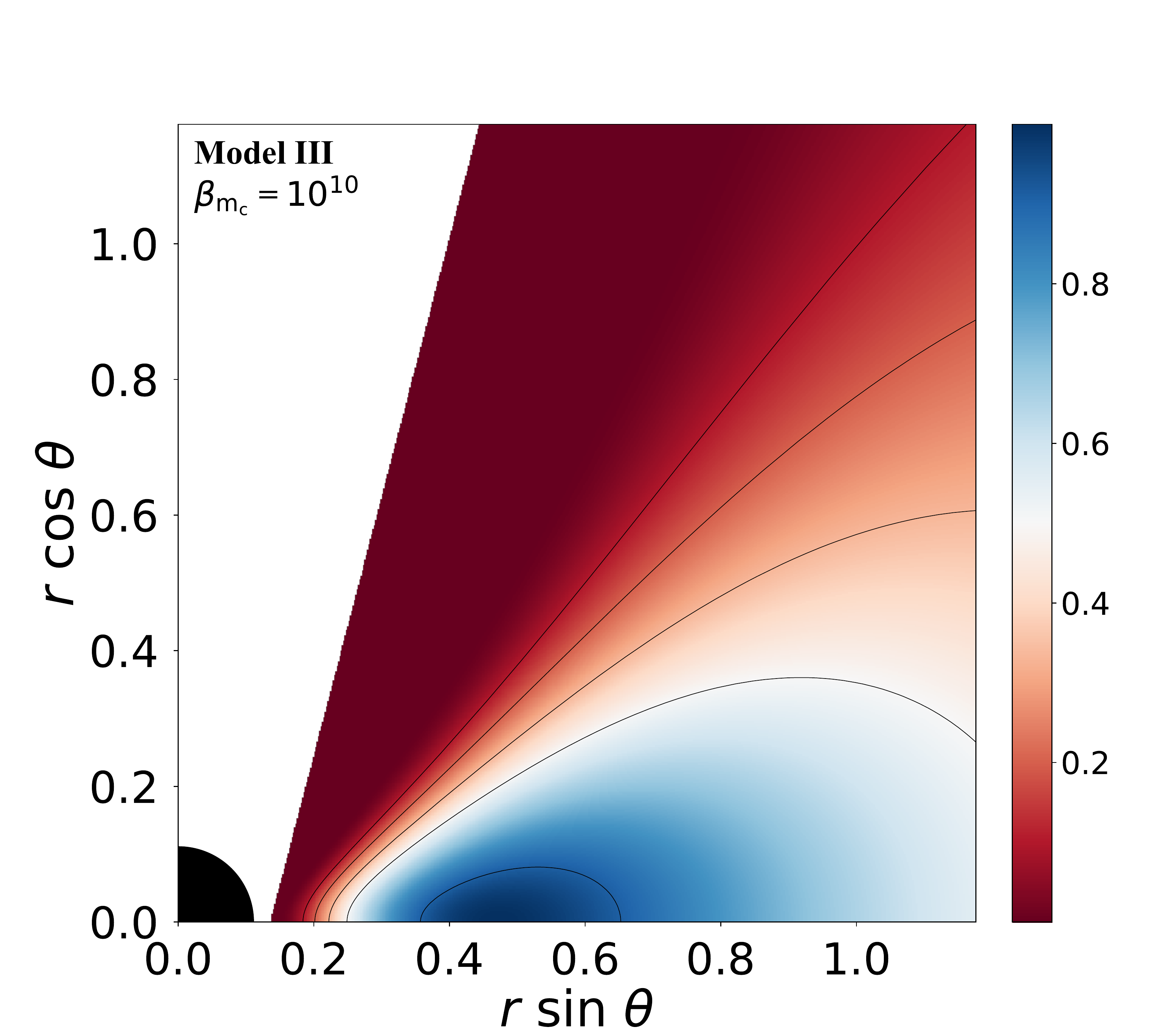}
\hspace{-0.3cm}
\includegraphics[scale=0.14]{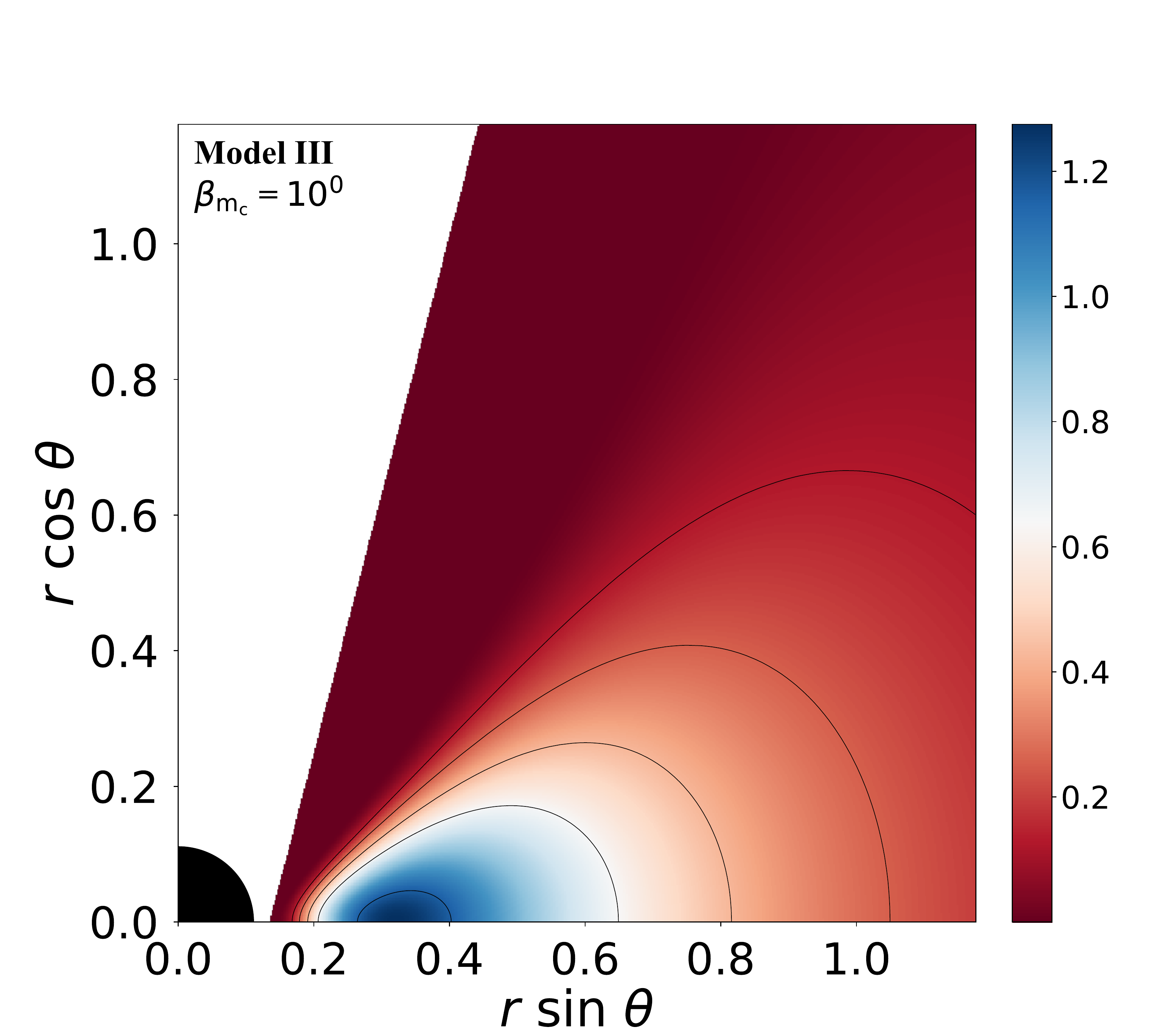}
\hspace{-0.2cm}
\includegraphics[scale=0.14]{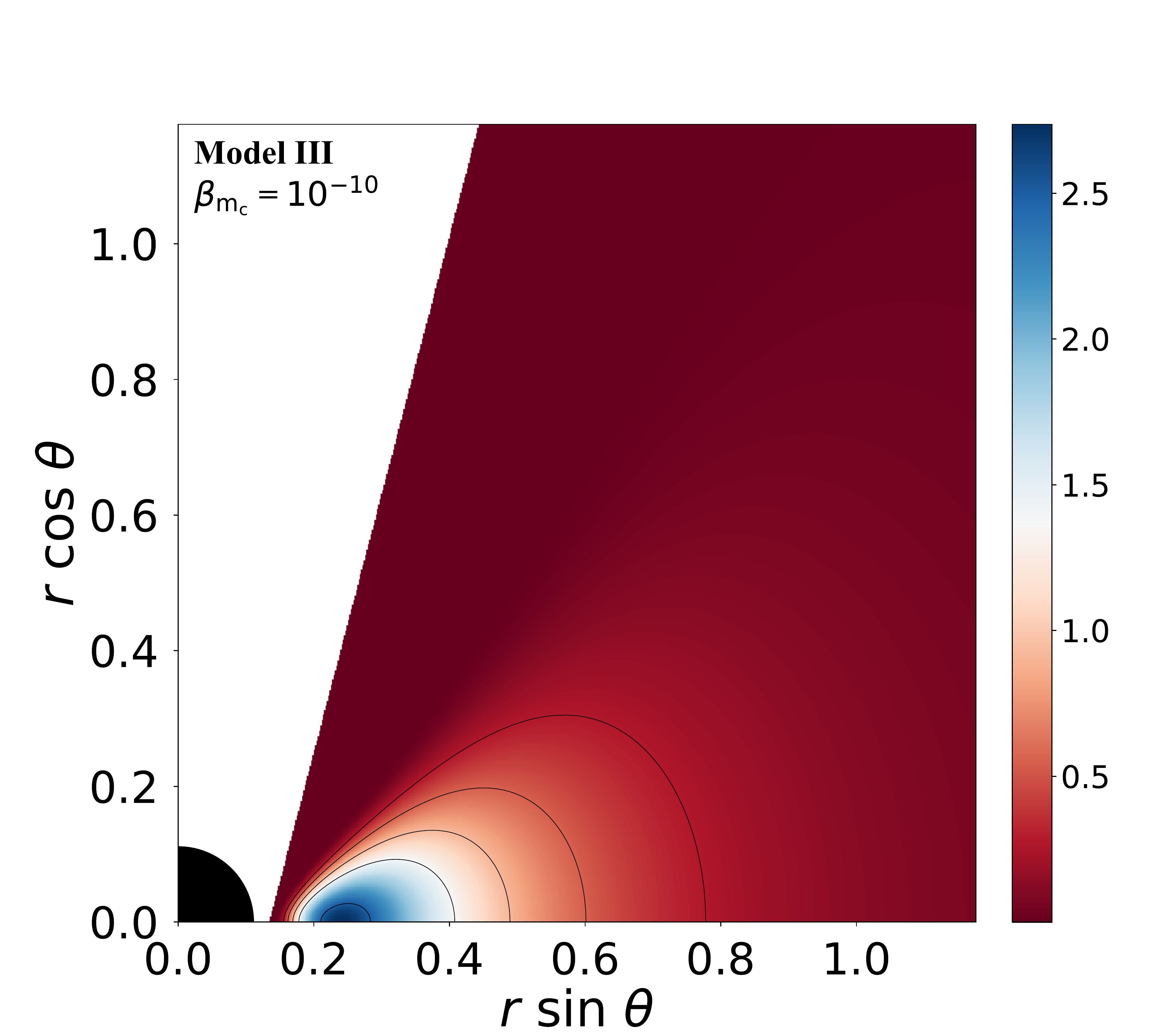}
\\
\includegraphics[scale=0.14]{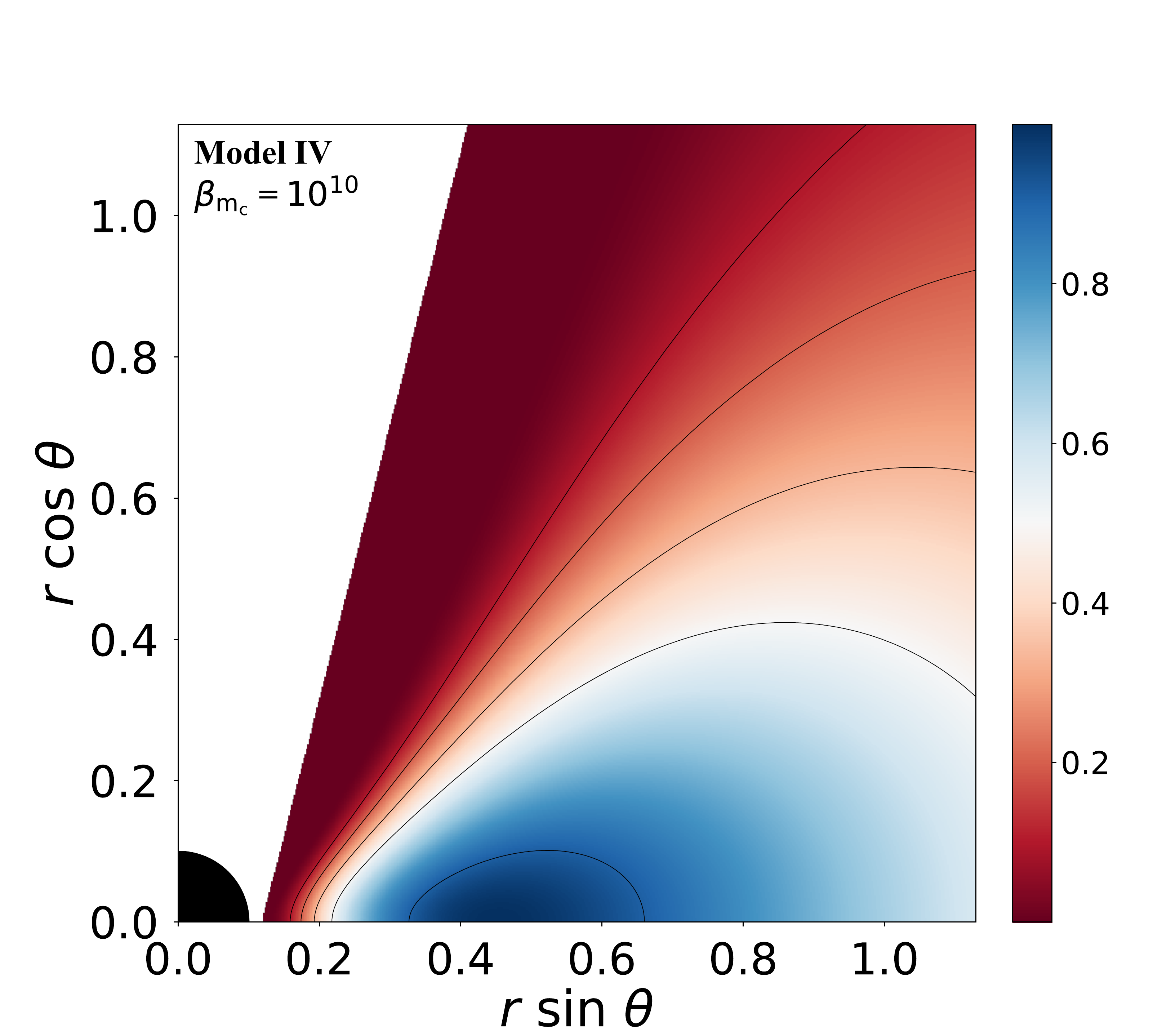}
\hspace{-0.3cm}
\includegraphics[scale=0.14]{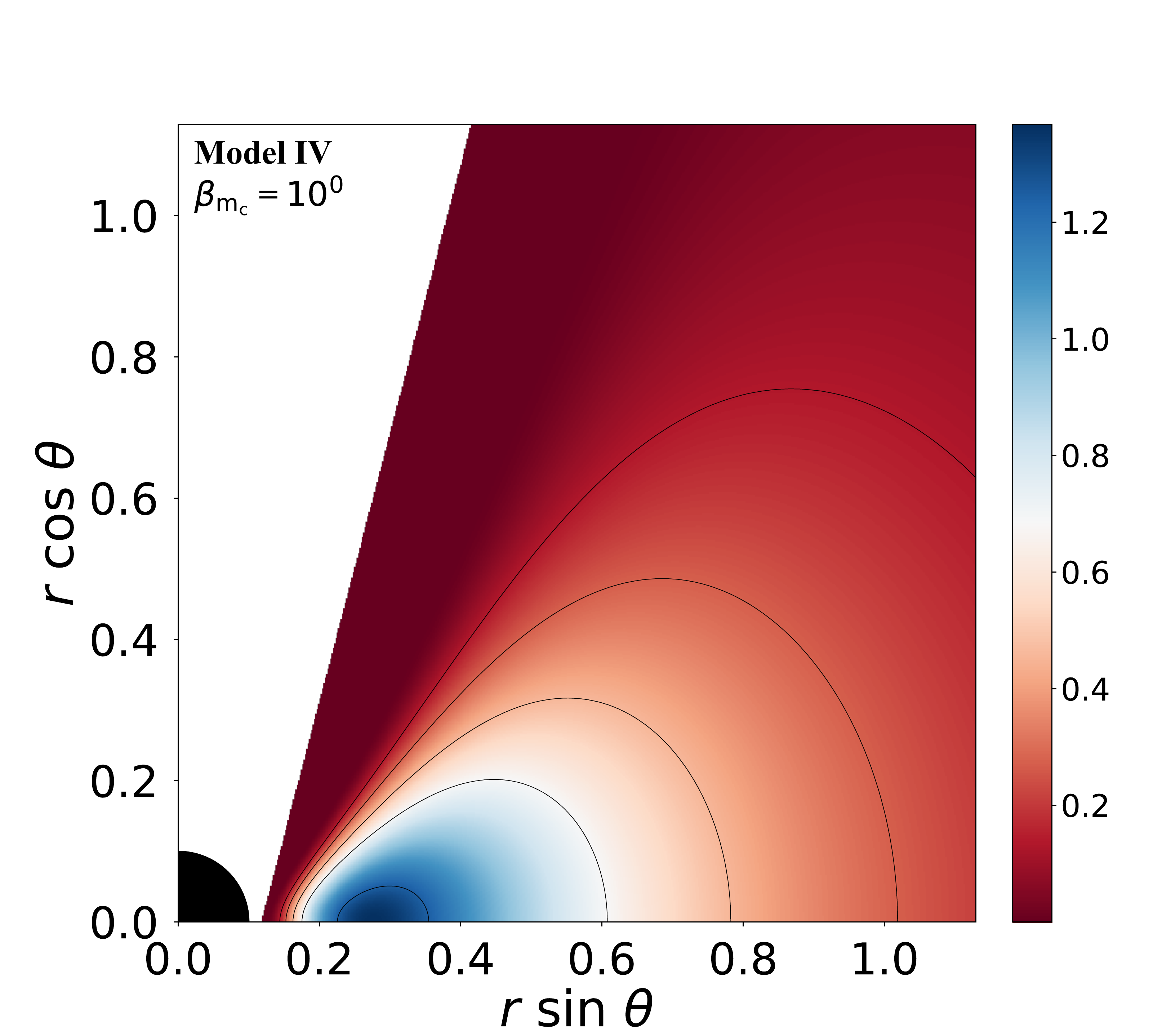}
\hspace{-0.2cm}
\includegraphics[scale=0.14]{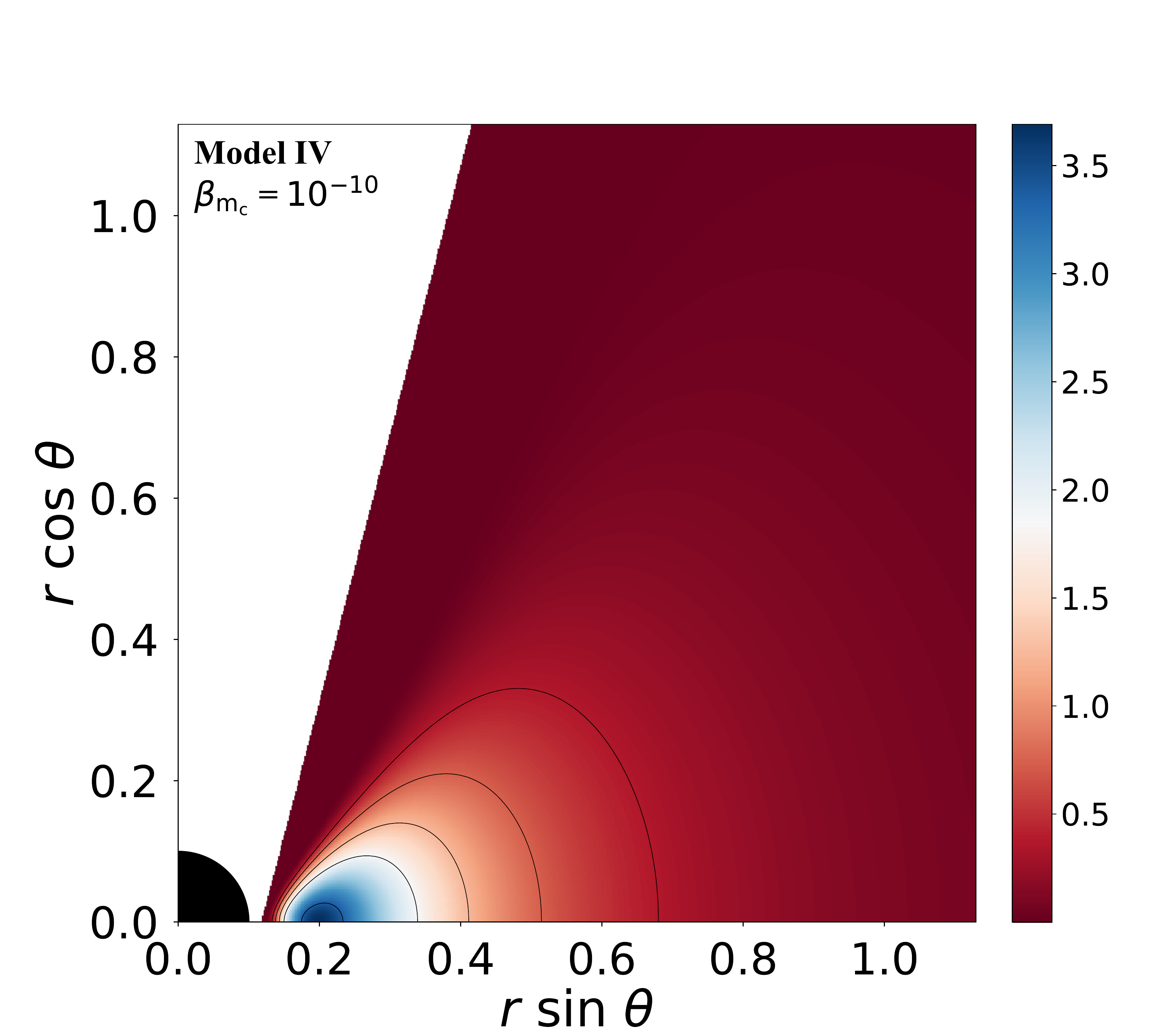}
\hspace{-0.2cm}
\caption{Distribution of the rest-mass density. From top to bottom the rows correspond to the first four models of KBHsSH (I, II, III and IV). From left to right the columns correspond to different values of the magnetization parameter, namely non-magnetized ($\beta_{\mathrm{m}_{\mathrm{c}}} = 10^{10}$), mildly magnetized ($\beta_{\mathrm{m}_{\mathrm{c}}} = 1$) and strongly magnetized ($\beta_{\mathrm{m}_{\mathrm{c}}} = 10^{-10}$). Note that the range of the colour scale is not the same for all plots.}
\label{models_I}
\end{figure*}

\begin{figure*}
\centering
\includegraphics[scale=0.14]{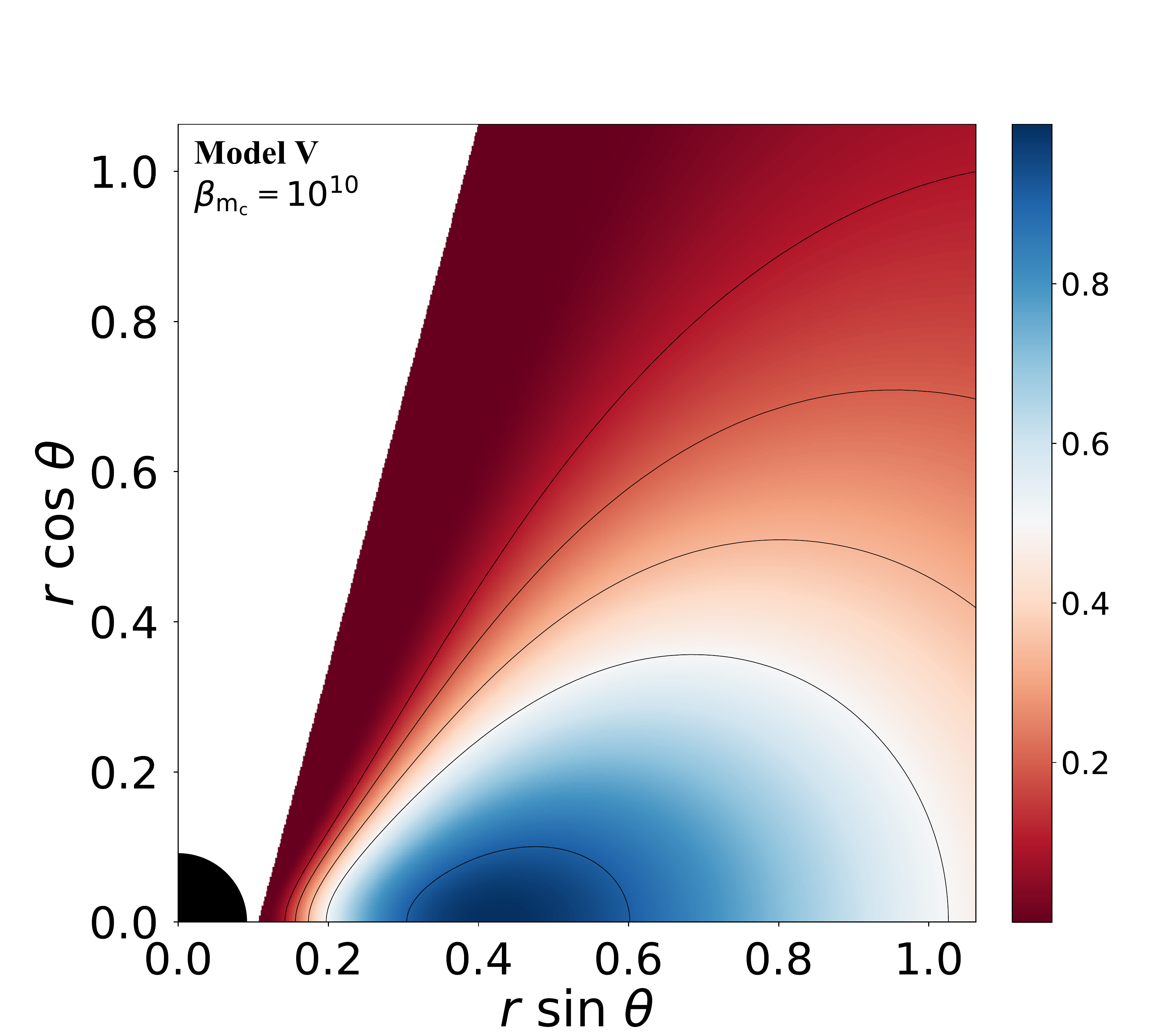}
\hspace{-0.3cm}
\includegraphics[scale=0.14]{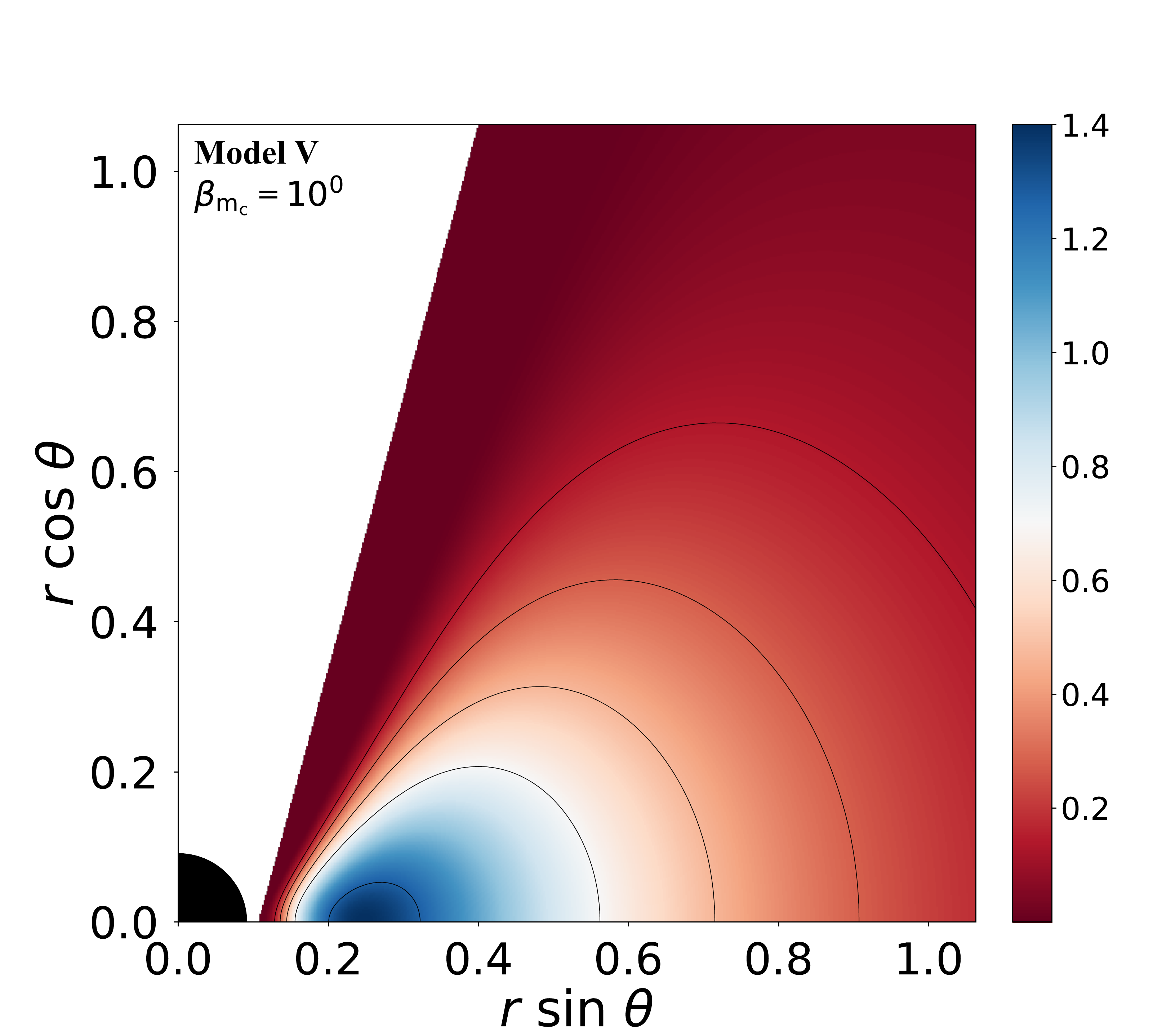}
\hspace{-0.2cm}
\includegraphics[scale=0.14]{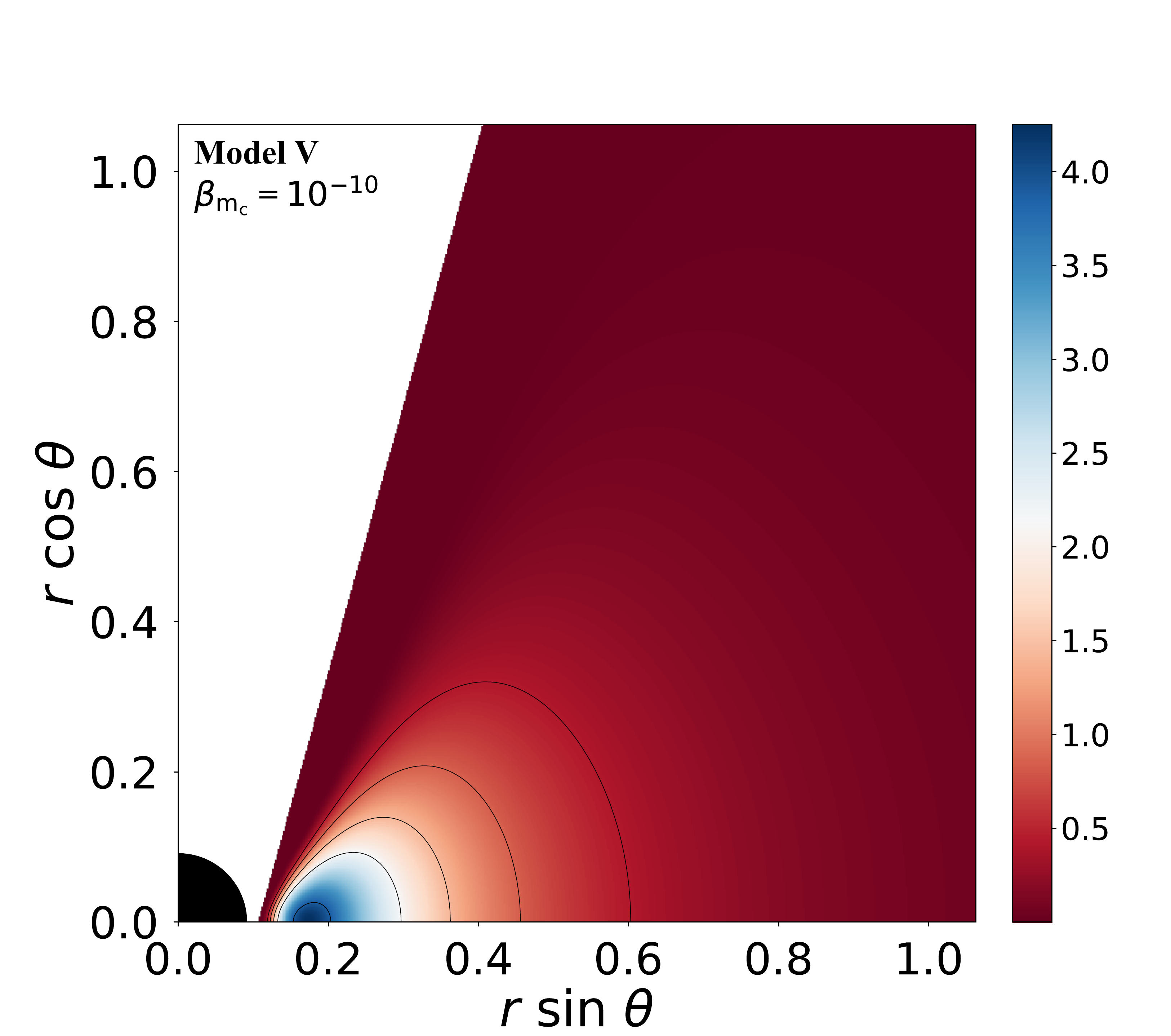}
\\
\includegraphics[scale=0.14]{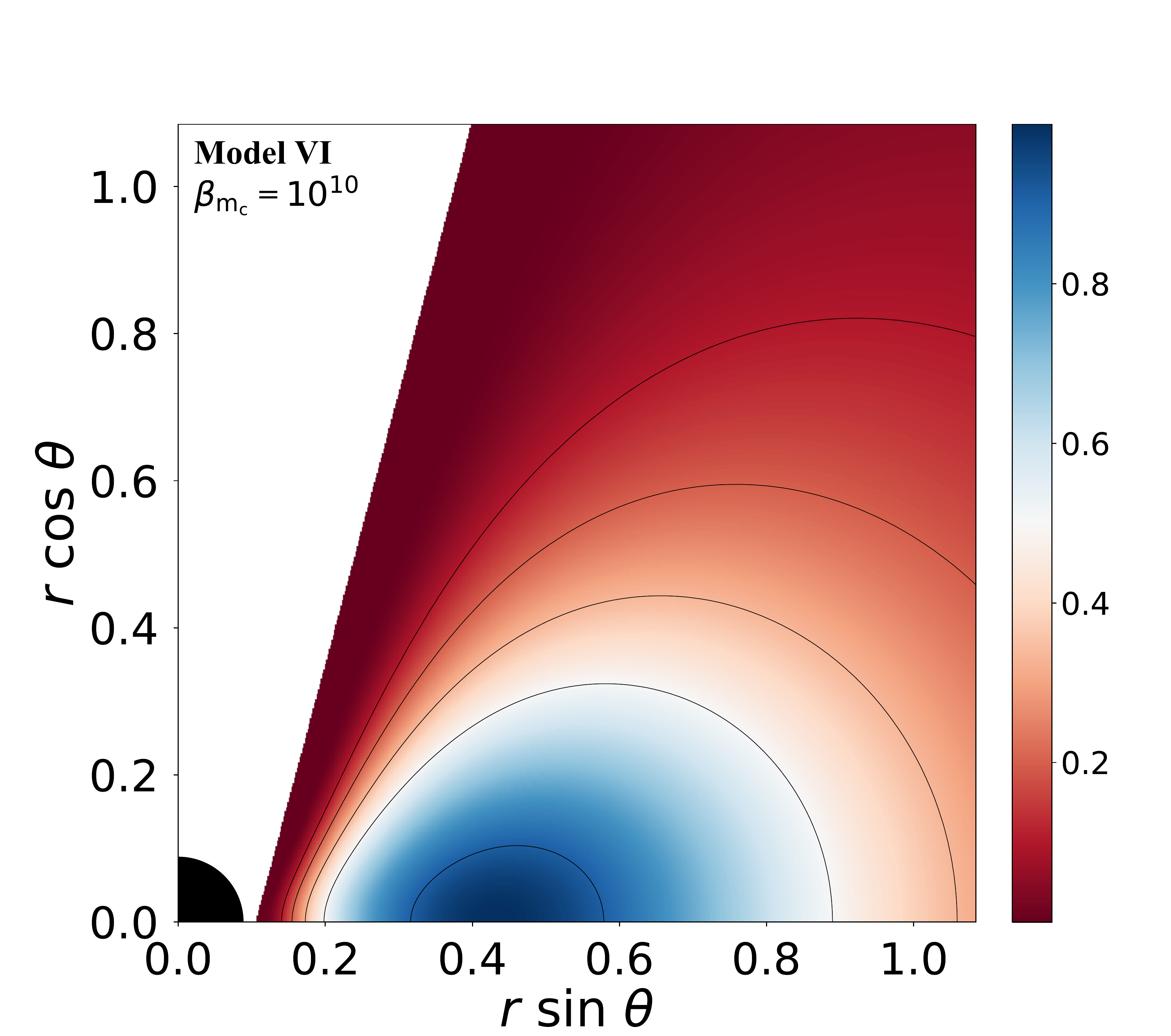}
\hspace{-0.3cm}
\includegraphics[scale=0.14]{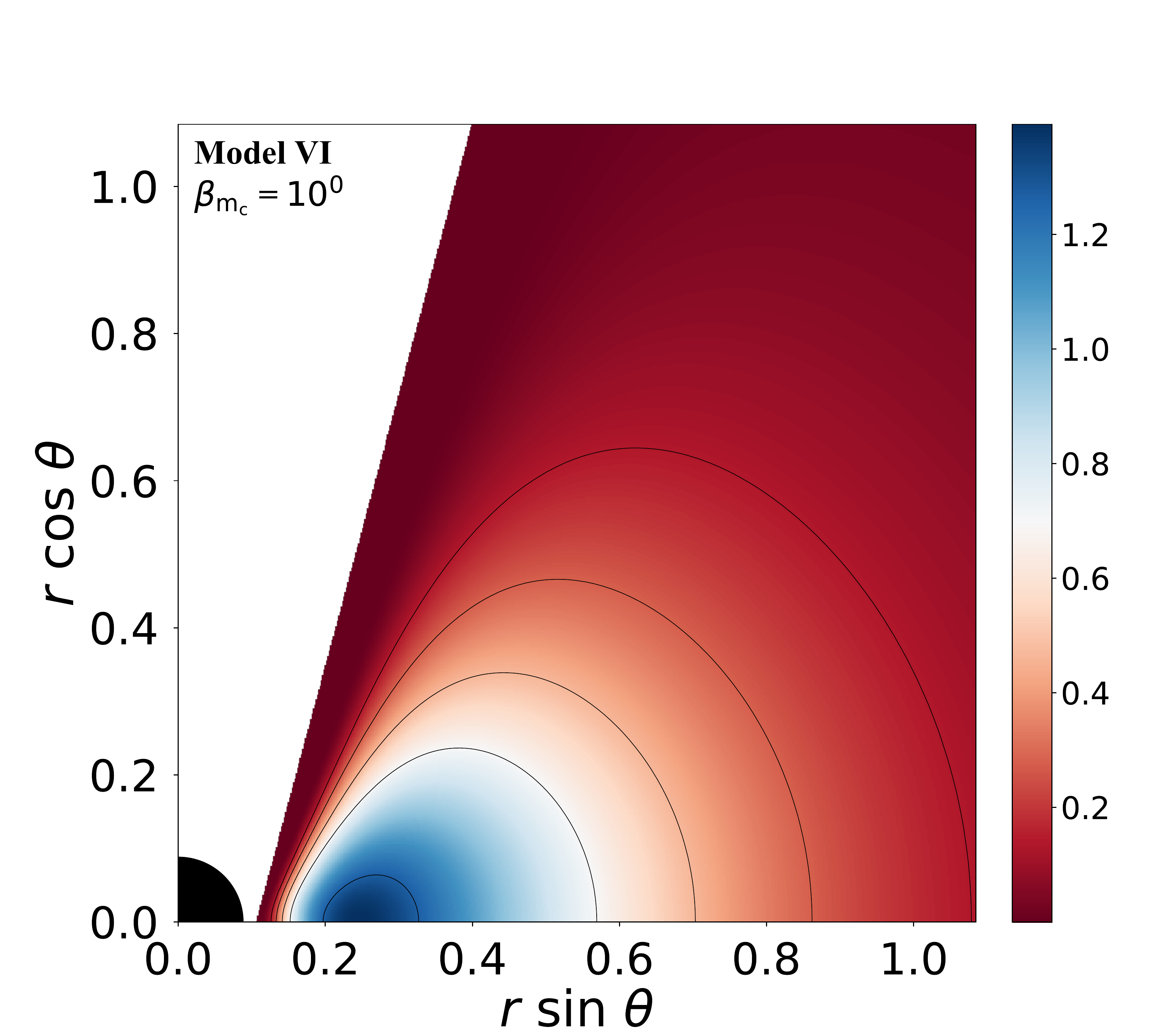}
\hspace{-0.2cm}
\includegraphics[scale=0.14]{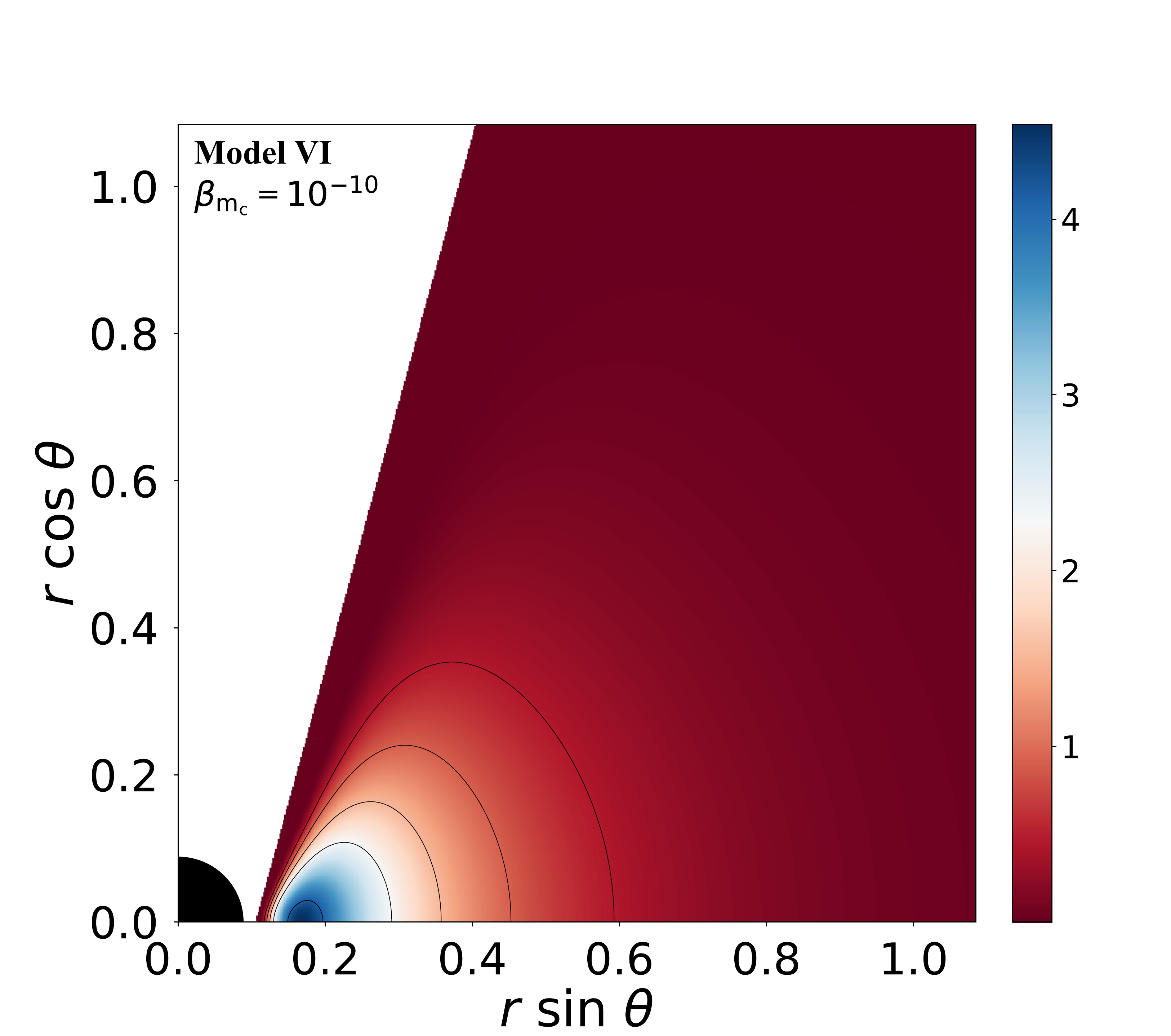}
\\
\includegraphics[scale=0.14]{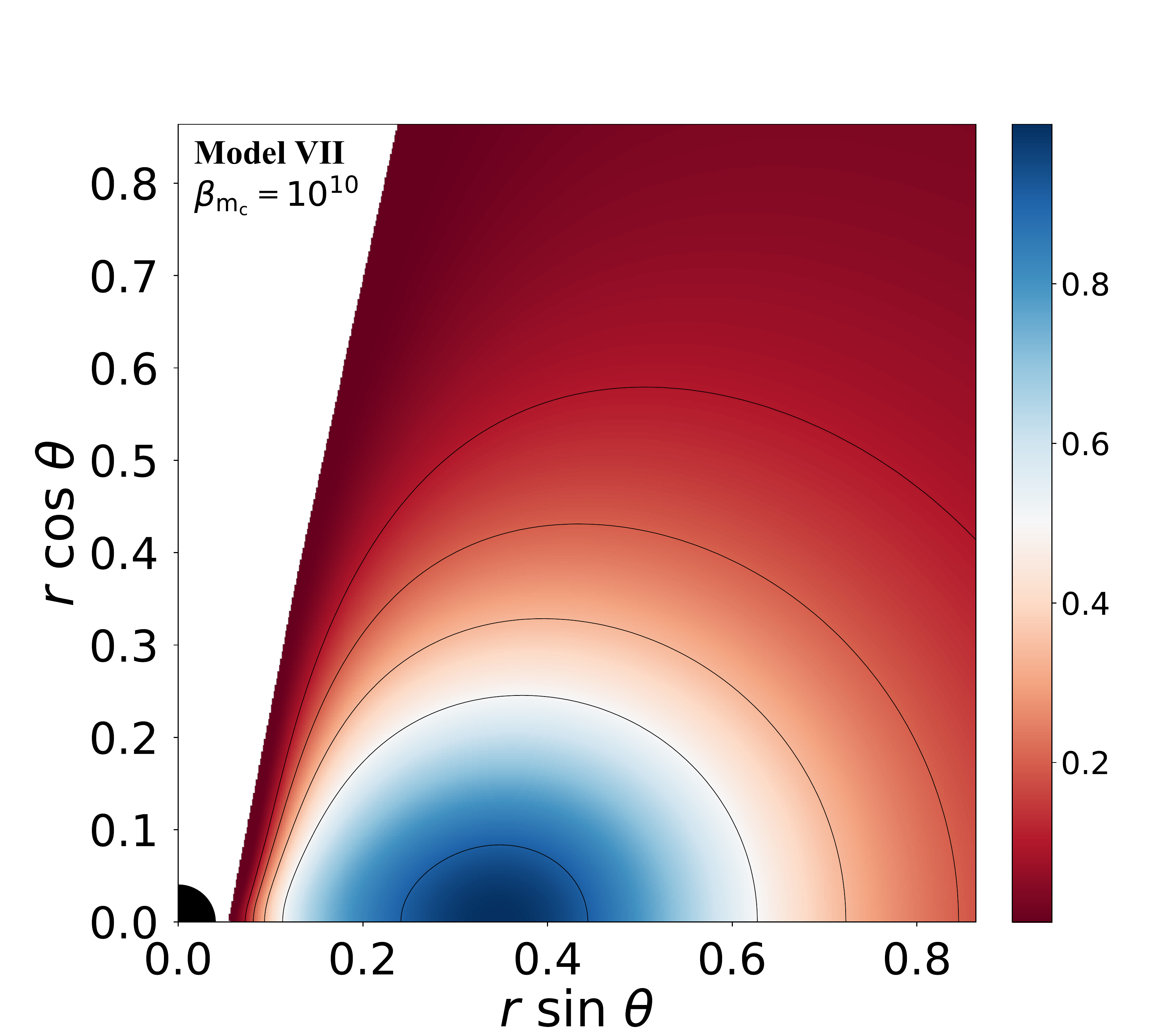}
\hspace{-0.3cm}
\includegraphics[scale=0.14]{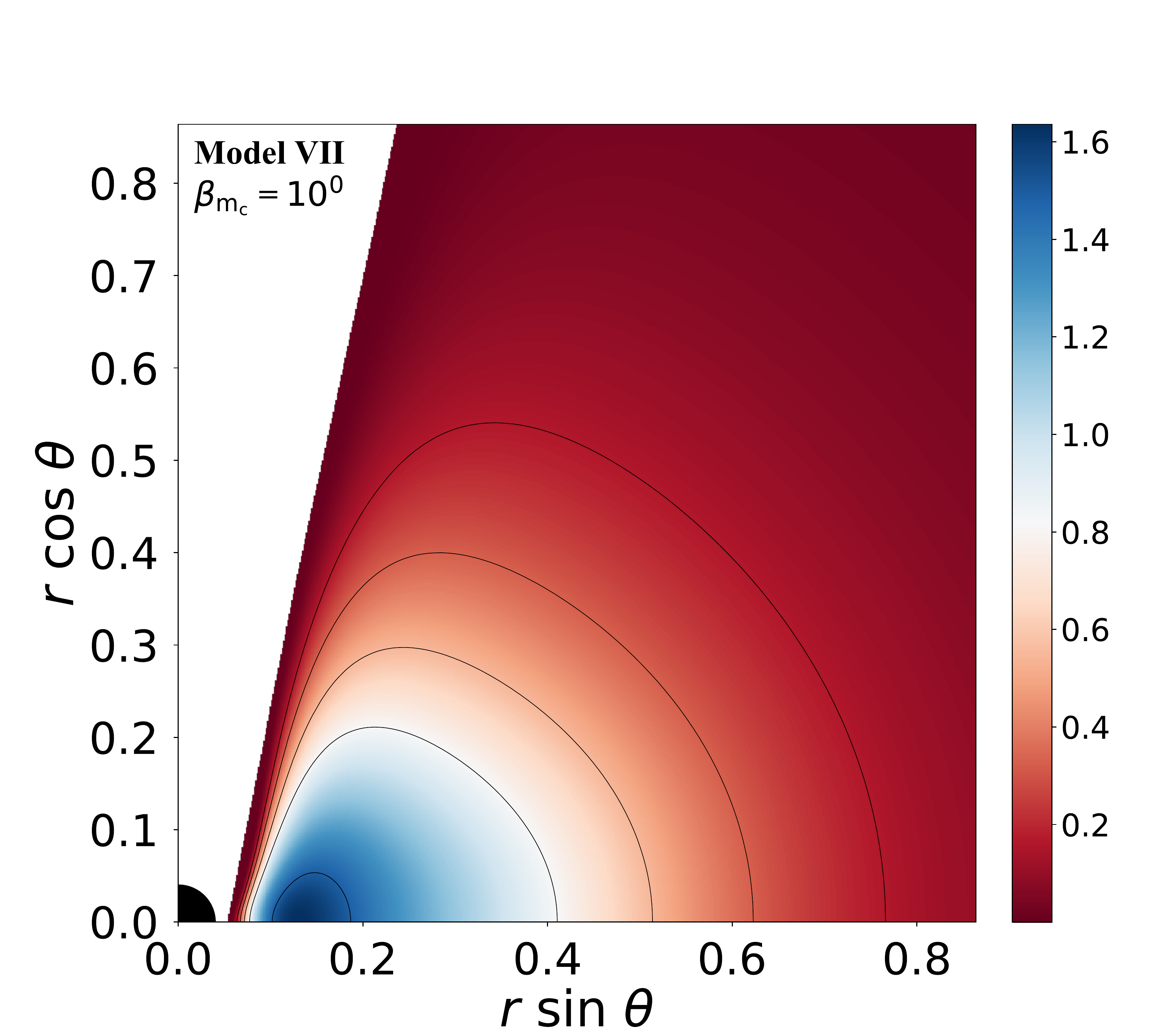}
\hspace{-0.2cm}
\includegraphics[scale=0.14]{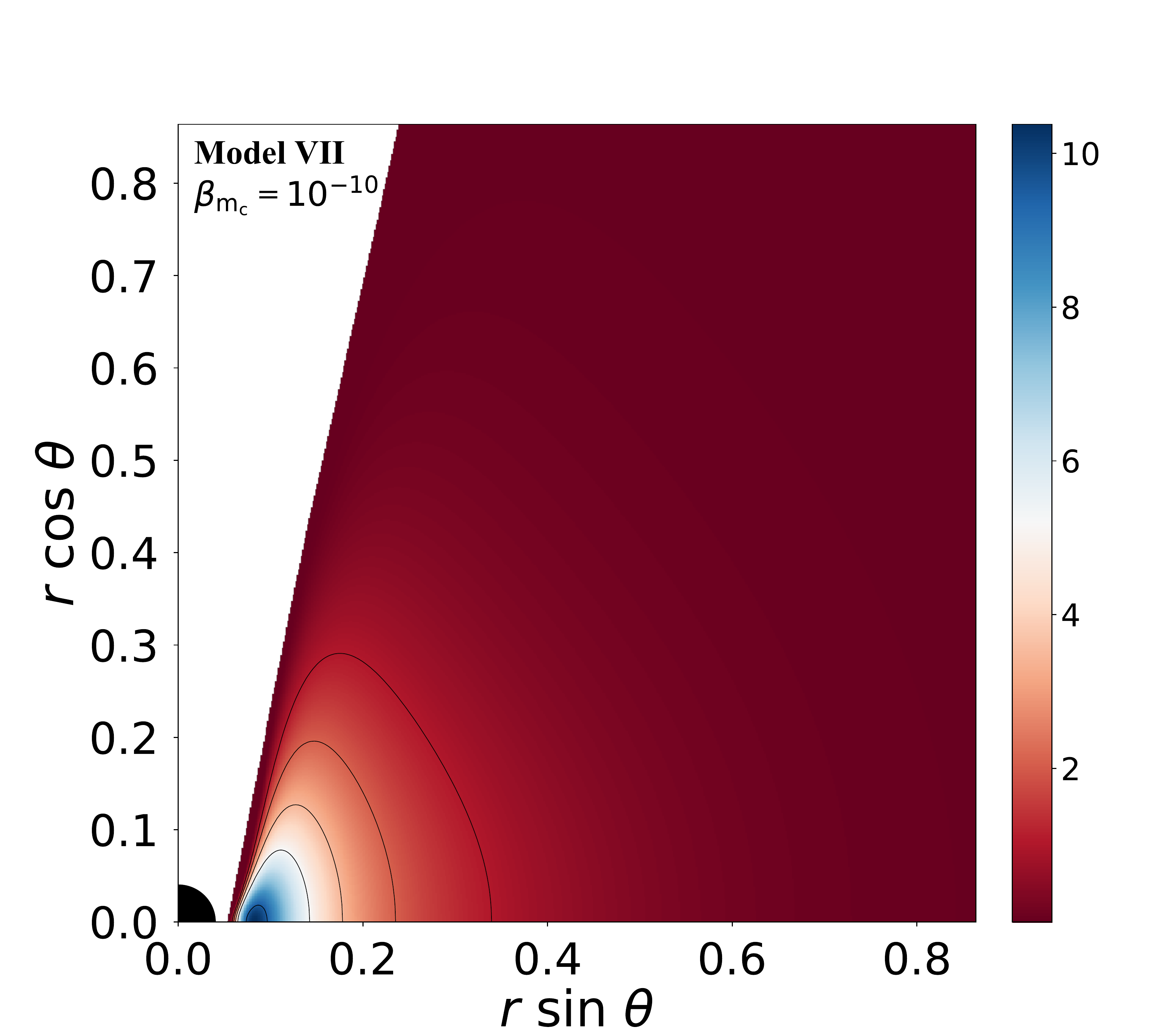}
\hspace{-0.2cm}
\caption{Same as Fig.~\ref{models_I} but for the last three models of KBHsSH (V, VI, and VII).}
\label{models_II}
\end{figure*}

We can also define the total energy density for the torus, $\rho_{\mathrm{T}}=-T^t_t + T^i_i$, and for the scalar field, $\rho_{\mathrm{SF}}=-(T_{\mathrm{SF}})^t_t + (T_{\mathrm{SF}})^i_i$. These are given by
\begin{eqnarray}\label{eq:torus_energy_density}
\rho_{\mathrm{T}} &=&  \frac{\rho h (g_{\phi\phi} - g_{tt} l^2)}{g_{\phi\phi} + 2 g_{t\phi} l + g_{tt} l^2} + 2 (p + p_{\mathrm{m}}),
\\
\rho_{\mathrm{SF}} &=&  2 \left(\frac{2 e^{-2 F_0} \omega (\omega-m W)}{N} - \mu^2\right) \phi^2.
\end{eqnarray}

Using these expressions, we can compute the total gravitational mass of the torus and the scalar field as the following expression
\begin{equation}\label{eq:mass_integral}
\mathcal{M} = \int  \rho \sqrt{-g}\,\mathrm{d}^3x\,,
\end{equation}
where $g$ is the determinant of the metric tensor and $\rho\equiv \rho_{\mathrm{T}}, \rho_{\mathrm{SF}}$.

In this work we take an approach to construct the magnetized disks different to the one proposed by~\citep{Komissarov:2006} and used by~\cite{Vincent:2016} for building disks around KBHsSH. As noted by~\cite{Gimeno-Soler:2017}, the approach of~\citep{Komissarov:2006} implicitly assumes that the specific enthalpy of the fluid is close to unity ($w = \rho h \simeq \rho$). This means that the polytropic EOS Eq.~\eqref{eq:eos_fluid} can be written as $p = K w^{\Gamma}$ (see Eq.~(27) of~\cite{Komissarov:2006}). We do not make this assumption here. To better understand the differences between these two approaches, we consider their behaviour in two limiting cases, namely the non-magnetized case and the extremely magnetized case.

For the former, we can rewrite Eq.~\eqref{eq:final} in the limiting case of $\beta_{\mathrm{m_c}} \rightarrow \infty$ ($K_{\rm m} \rightarrow 0$) as
\begin{equation}
W - W_{\mathrm{in}} + \ln \left(1 + \frac{\Gamma K}{\Gamma -1}\rho^{\Gamma -1}\right) = 0.
\end{equation}
Then, we can solve this equation for the specific enthalpy
\begin{equation}\label{eq:rel_non_mag_enth}
h = e^{W_{\mathrm{in}} - W}.
\end{equation}
Now, we want to obtain an analogous equation for the $h \simeq 1$ case. We start by considering Eq.~(20) of~\cite{Gimeno-Soler:2017} and taking the limit $\beta_{\mathrm{m_c}} \rightarrow \infty$ (in this equation, this means $K_{\mathrm{m}} \rightarrow 0$), to obtain
\begin{equation}
W - W_{\mathrm{in}} + \frac{\Gamma K}{\Gamma -1}w^{\Gamma -1}.
\end{equation}
If we consider the $h \simeq 1$ approximmation, we can use the definition of $h$ and solve the equation to arrive at
\begin{equation}\label{eq:non_rel_non_mag_enth}
h = 1 + (W_{\mathrm{in}} - W).
\end{equation}
If we compare both results, we can see that Eq.~\eqref{eq:non_rel_non_mag_enth} is the first-order Taylor series expansion of Eq.~\eqref{eq:rel_non_mag_enth} for a sufficiently small value of $W_{\mathrm{in}} - W$.

\begin{figure*}
\centering
\includegraphics[scale=0.22]{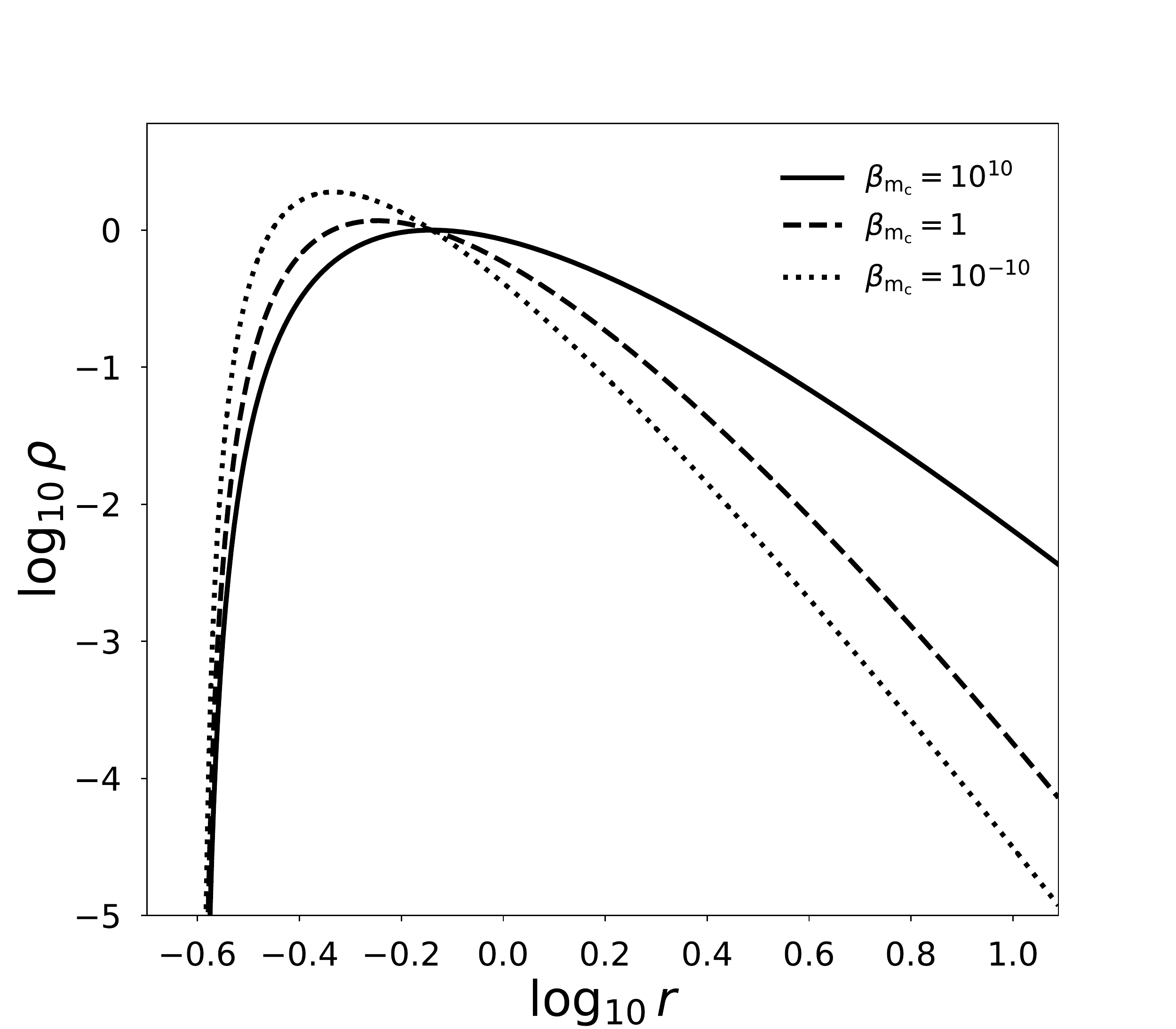}
\hspace{-0.6cm}
\includegraphics[scale=0.22]{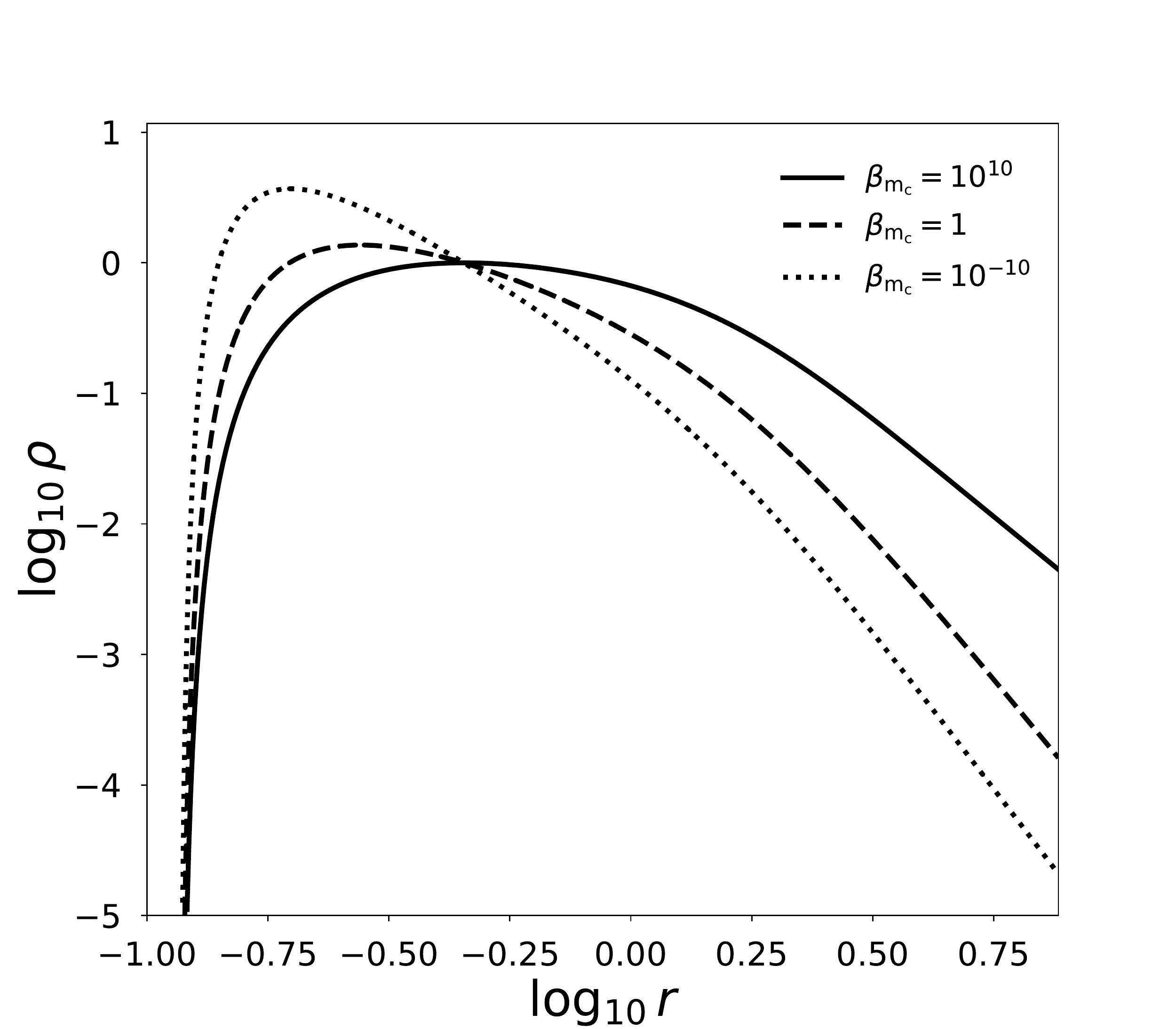}
\hspace{-0.6cm}
\includegraphics[scale=0.22]{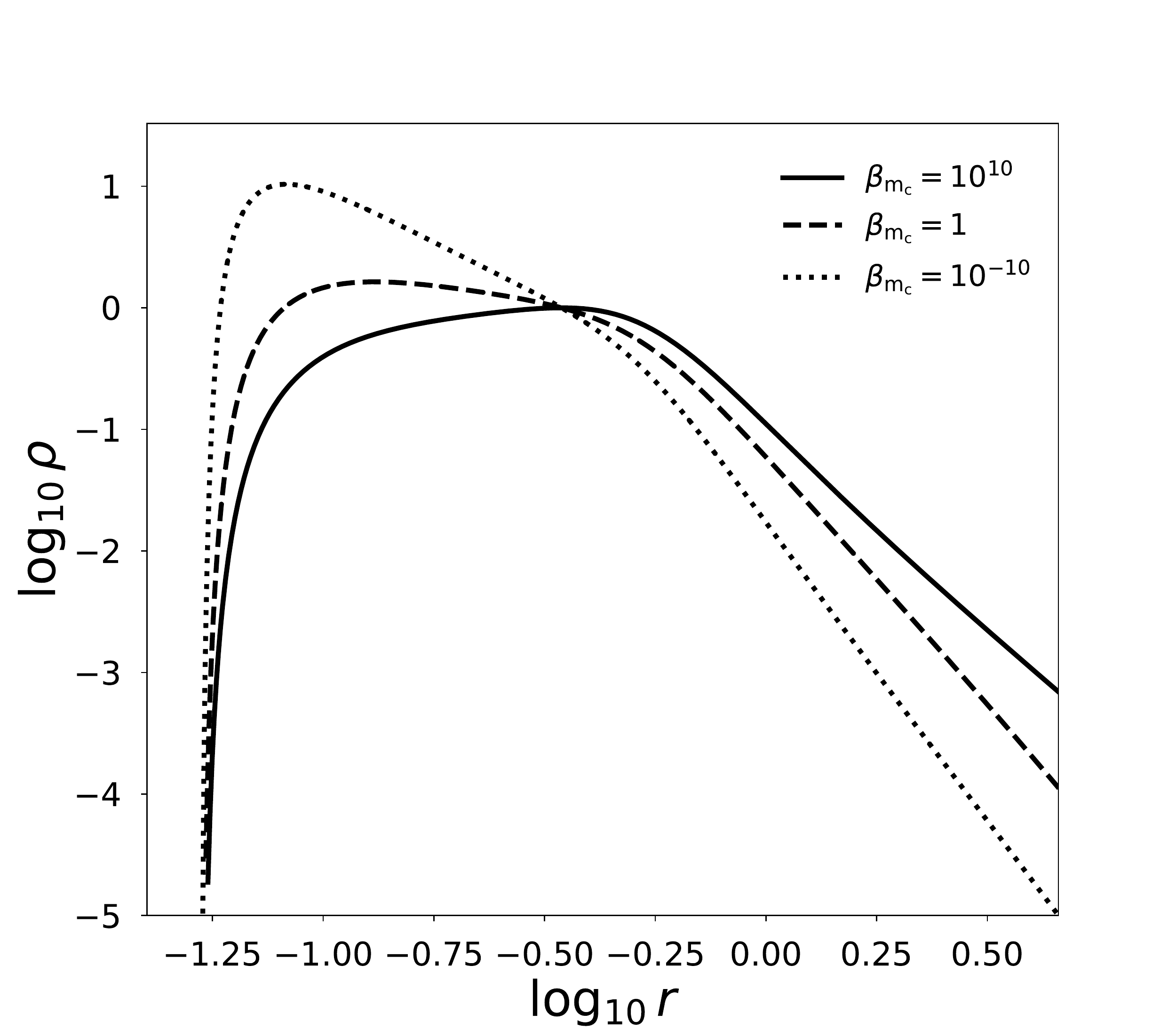}
\hspace{-0.6cm}
\\
\hspace{-0.6cm}
\includegraphics[scale=0.22]{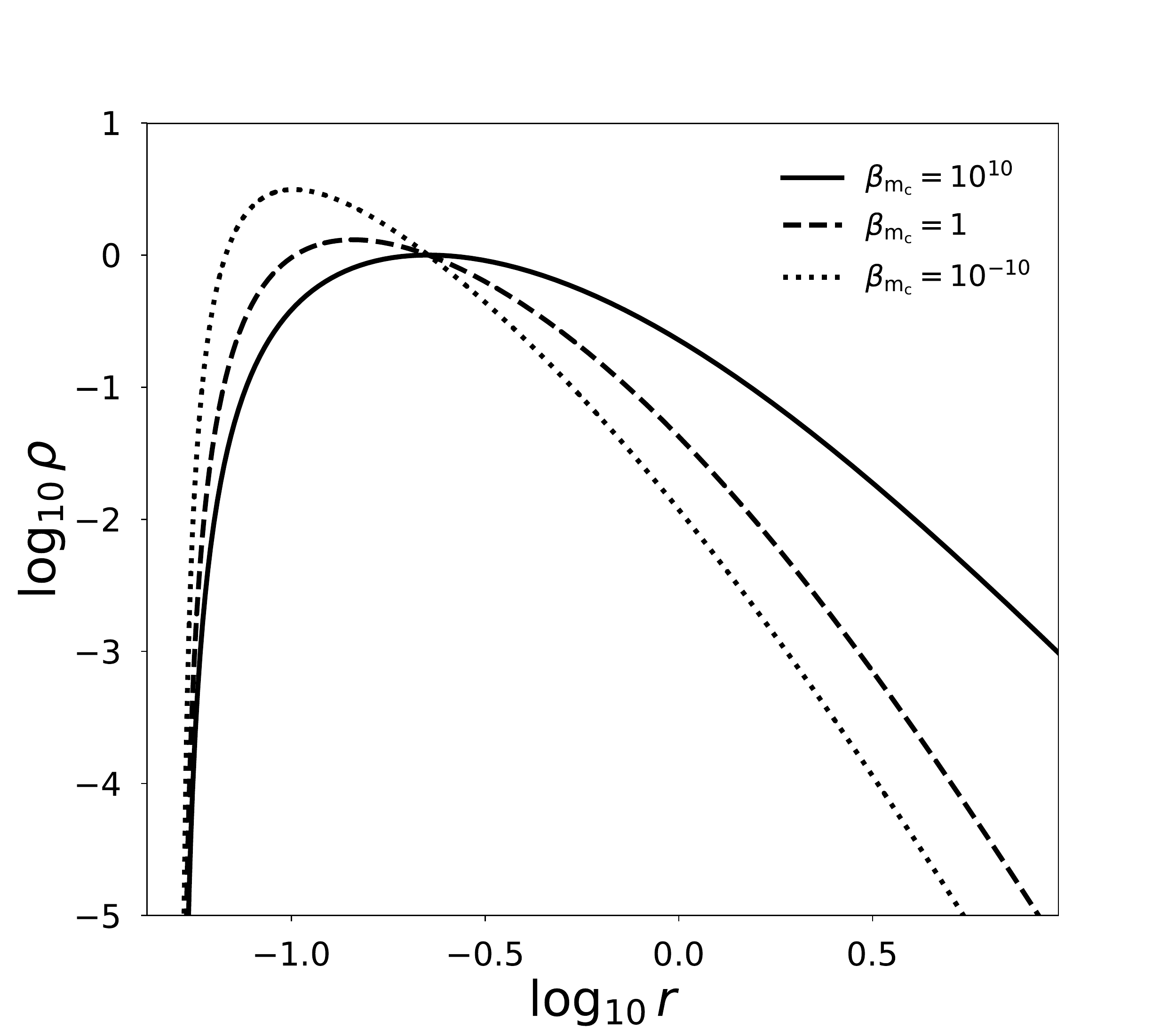}
\hspace{-0.6cm}
\includegraphics[scale=0.22]{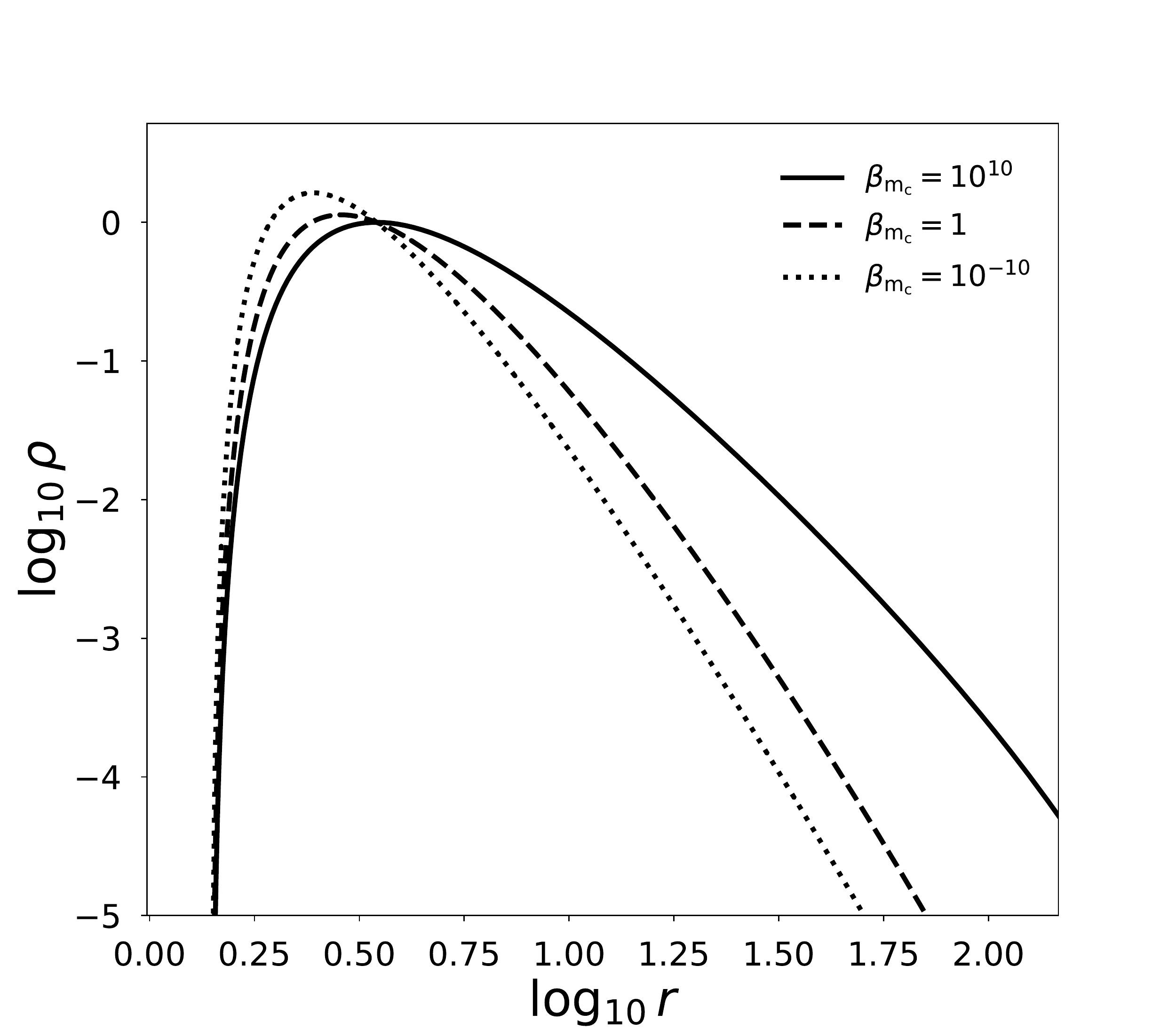}
\hspace{-0.6cm}
\includegraphics[scale=0.22]{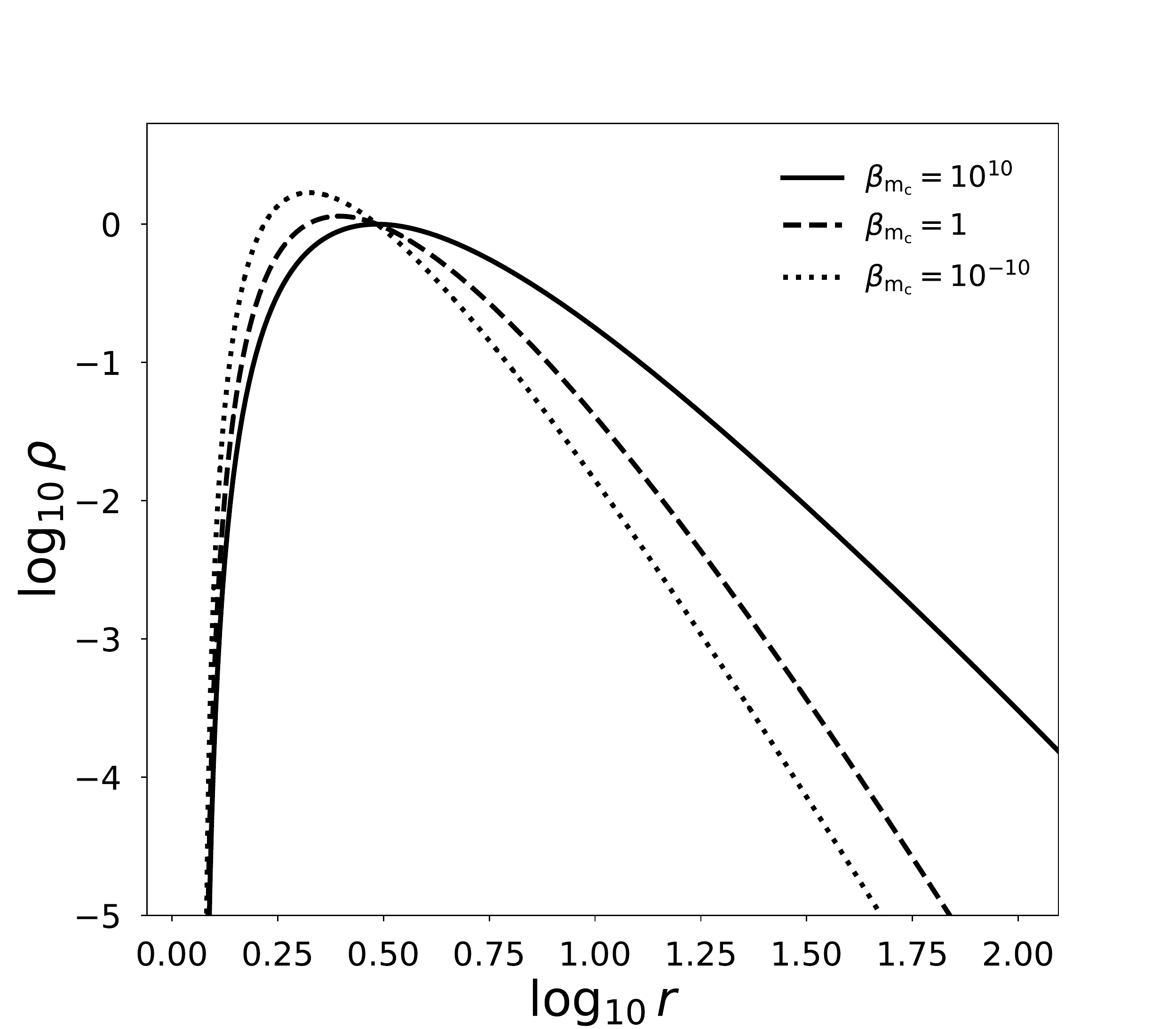}
\hspace{-0.6cm}
\caption{Size of the disks. Top panel: Effects of the magnetization on the radial profiles of the logarithm of the density at the equatorial plane for different KBHsSH models. From left to right we show model I, IV, and VII, respectively. Bottom panel: same as the top panel but for Kerr BHs. From left to right the cases shown have the same ADM quantities as the KBHsSH model I, IV, and VII, respectively, shown in the top panel. Note that the scale shown in the horizontal axes is different in all plots.}
\label{radial_profiles}
\end{figure*}

For the extremely magnetised case, we consider again Eq.~\eqref{eq:final} and Eq.~(20) of~\cite{Gimeno-Soler:2017}, but this time around we take $\beta_{\mathrm{m_c}} \rightarrow 0$ ($K \rightarrow 0$). This yields the same result for both equations

\begin{equation}
W - W_{\mathrm{in}} + \frac{q}{q-1}K_{\rm m}(\mathcal{L}\rho h)^{q-1}=0.
\end{equation}
In addition, we could consider the expression for the specific enthalpy in terms of the density $h = 1 + \frac{K\Gamma\rho^{\Gamma-1}}{\Gamma - 1}$ to 
see that we will have $h \rightarrow 1$. This shows that, for the extremely magnetized limit, the two approaches coincide.

Taking into account these two limits we can obtain the range of validity of the $h \simeq 1$ approximation: As magnetized disks exist between the two considered cases, for disks with a sufficiently small value of the potential well, $\Delta W \equiv W_{\mathrm{in}} - W_{\mathrm{c}}$, the $h \simeq 1$ approximation is valid. On the contrary, if the value of $\Delta W$ is large enough, the approximation does not hold even for disks with a fairly low value of magnetization.

\begin{figure*}
\centering
\includegraphics[scale=0.14]{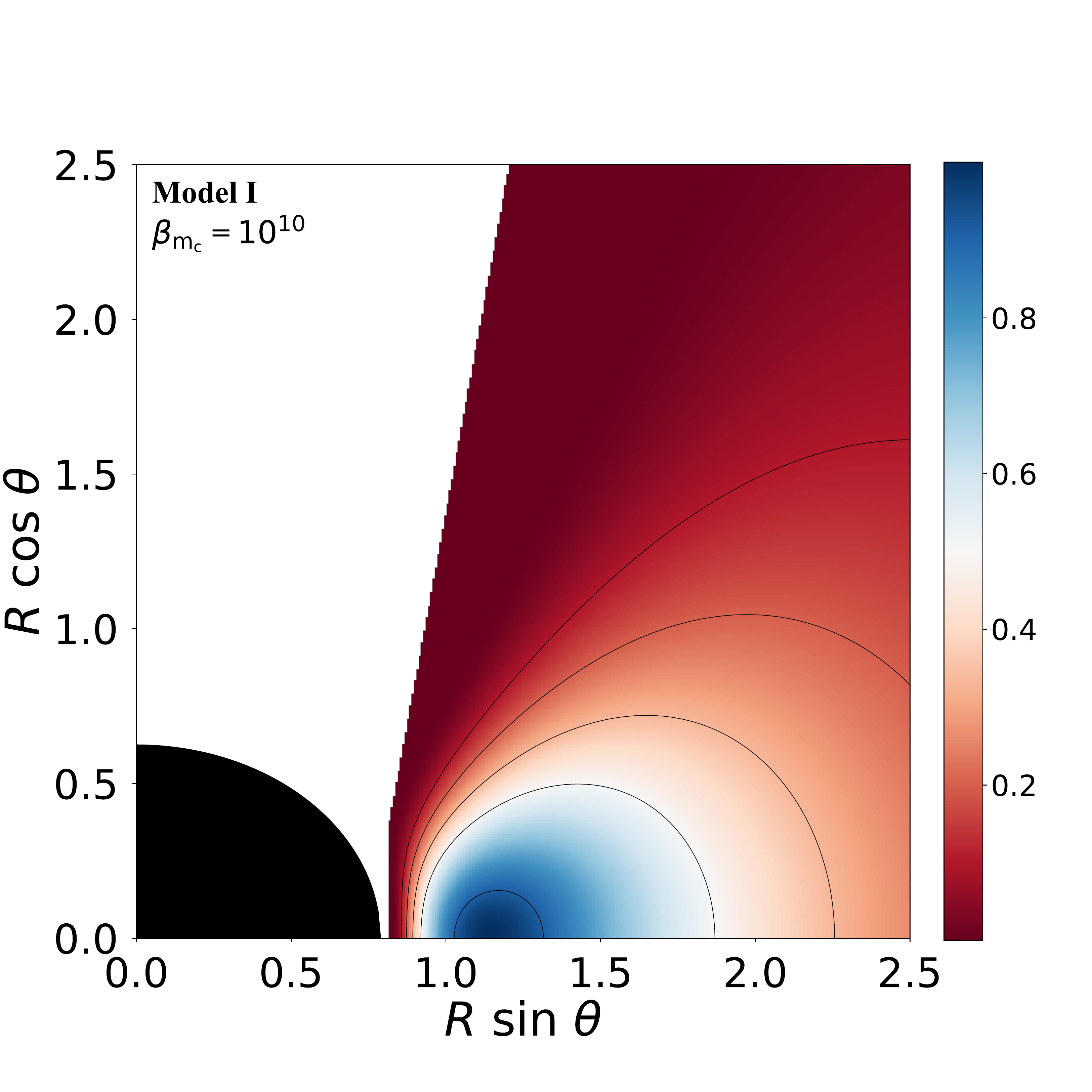}
\hspace{-0.3cm}
\includegraphics[scale=0.14]{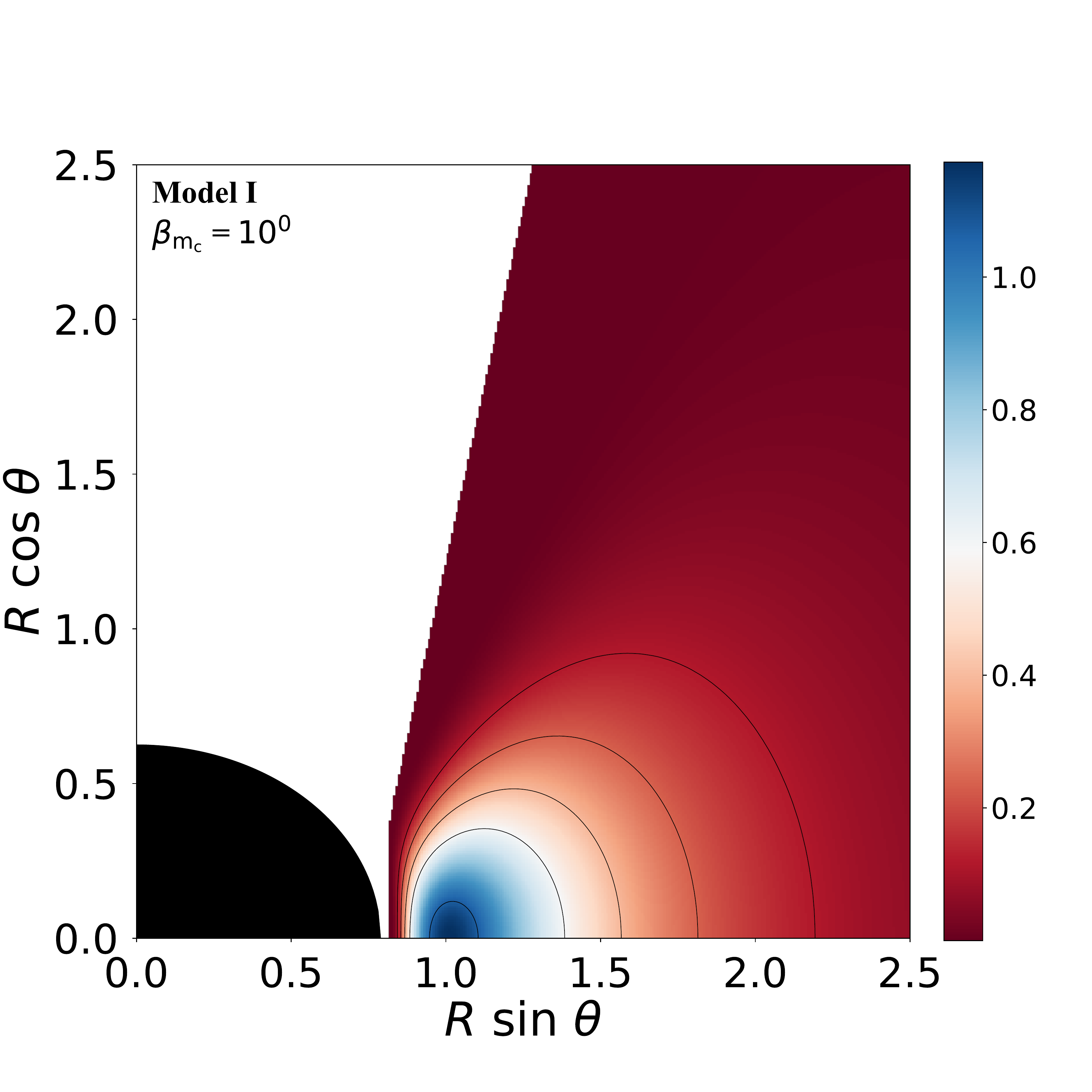}
\hspace{-0.2cm}
\includegraphics[scale=0.14]{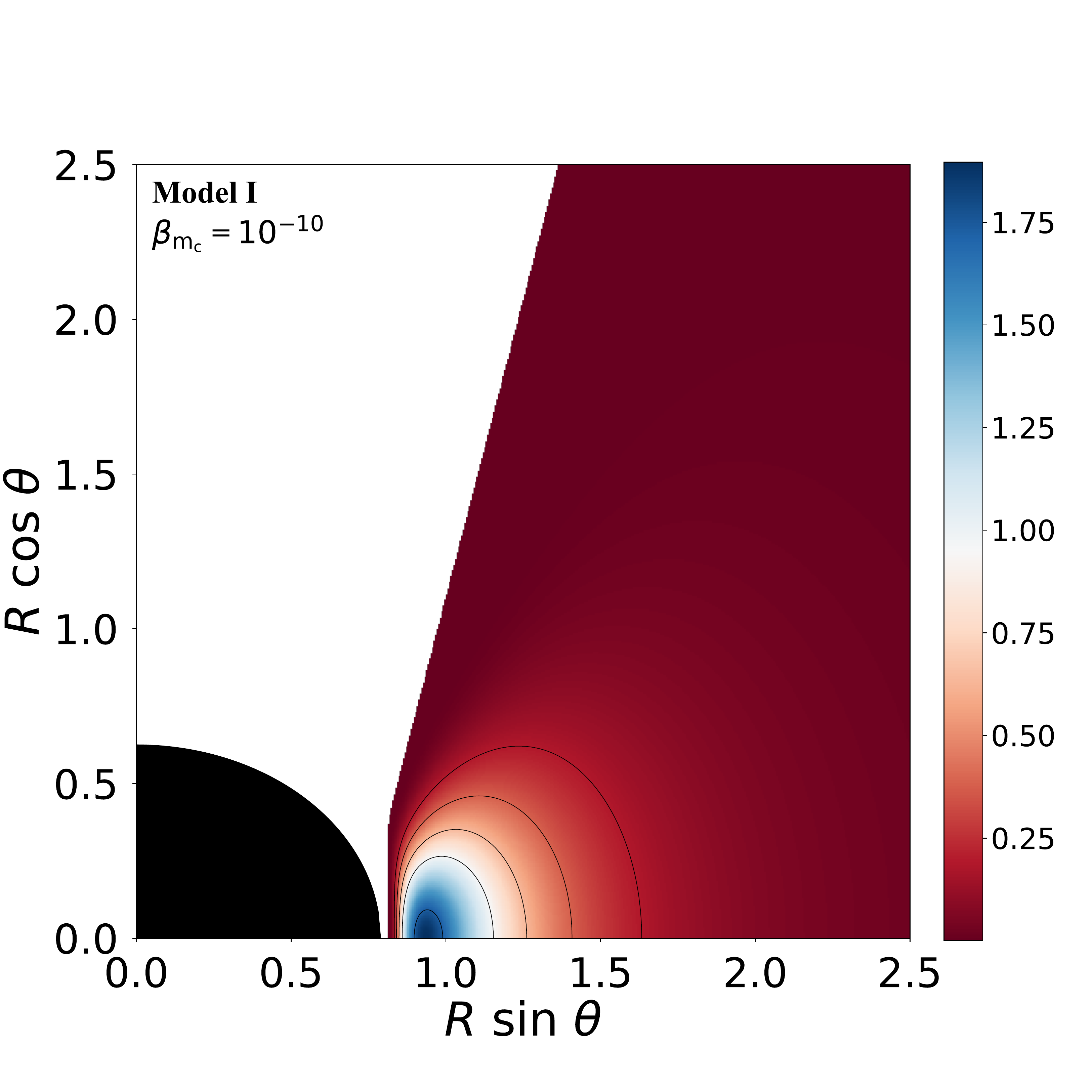}
\\
\includegraphics[scale=0.14]{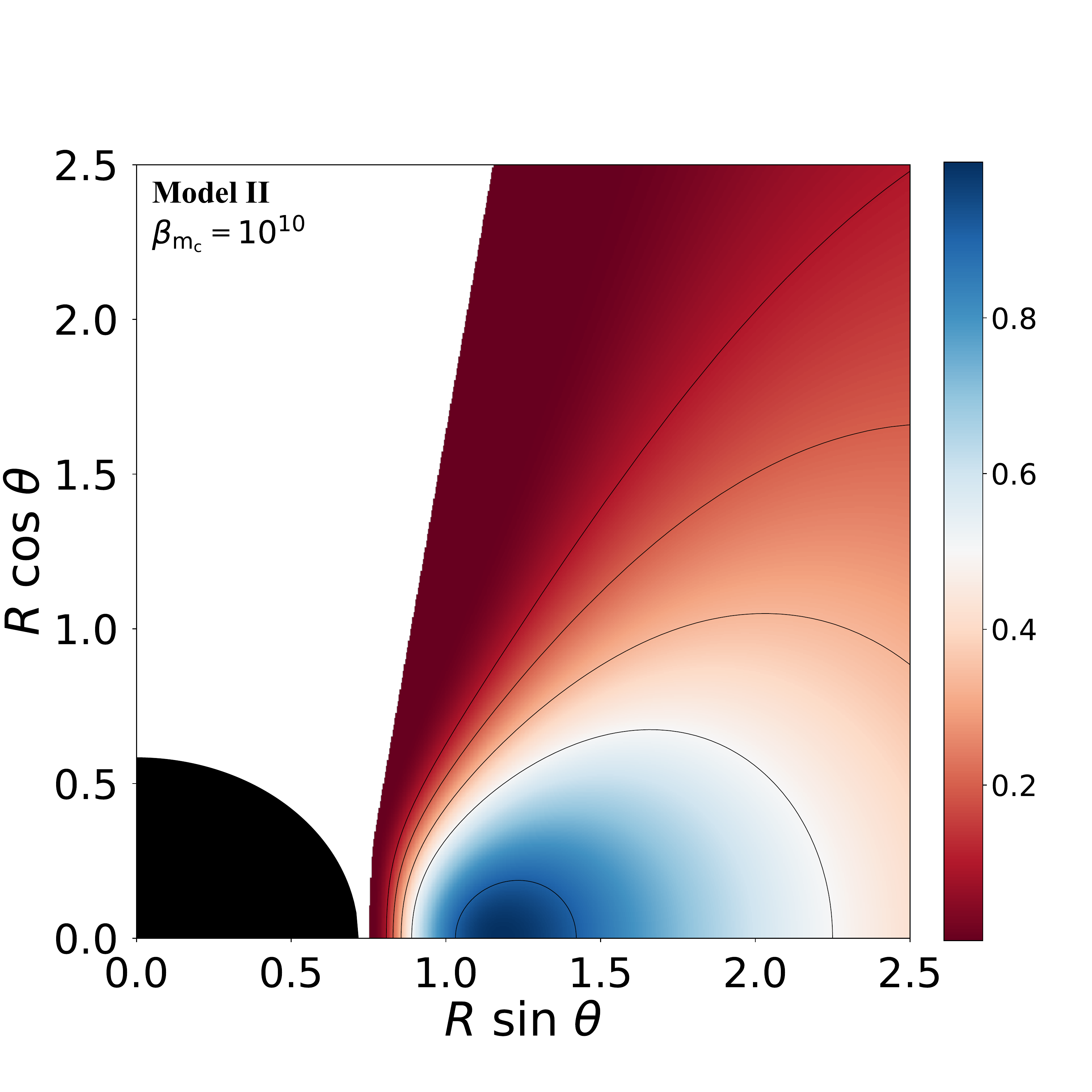}
\hspace{-0.3cm}
\includegraphics[scale=0.14]{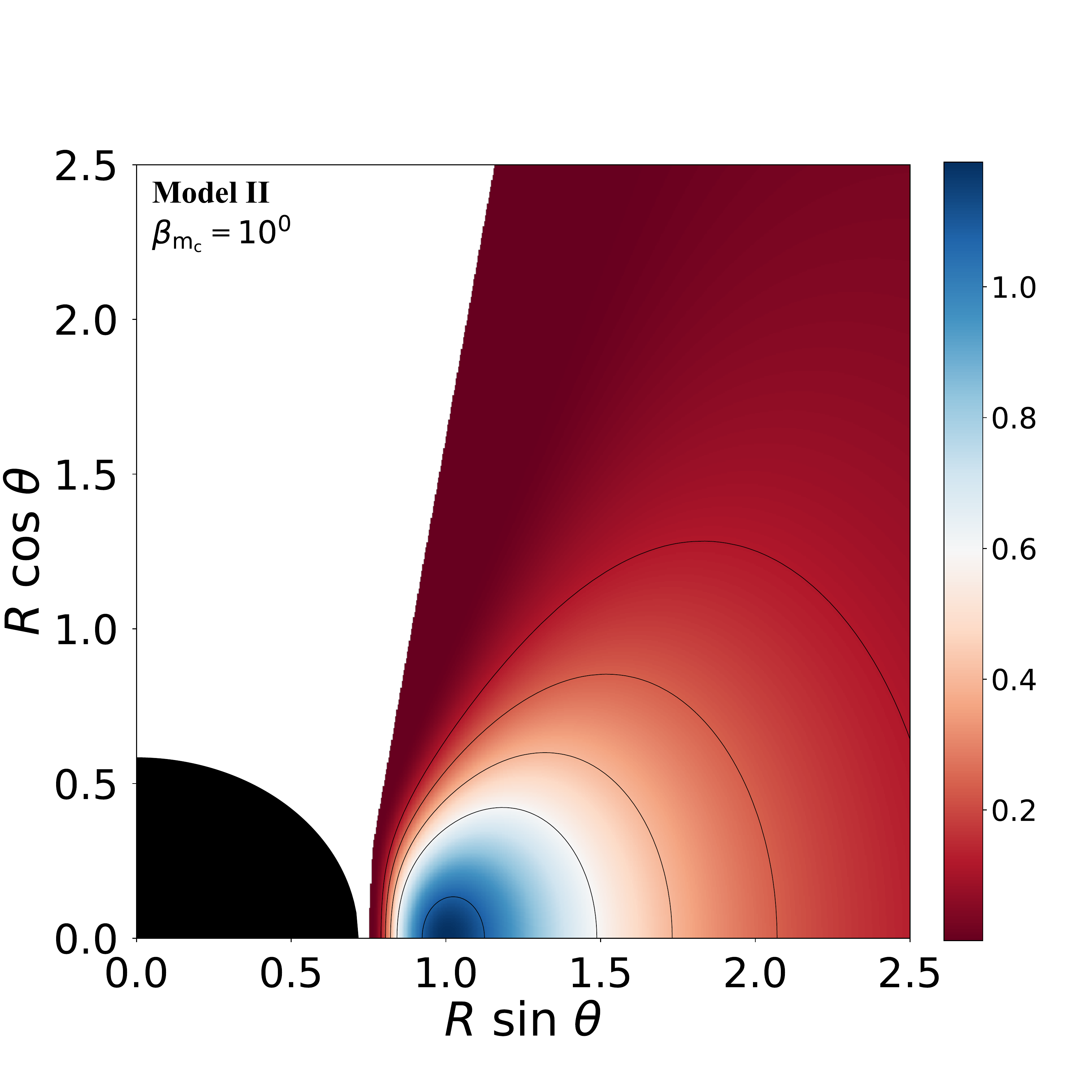}
\hspace{-0.2cm}
\includegraphics[scale=0.14]{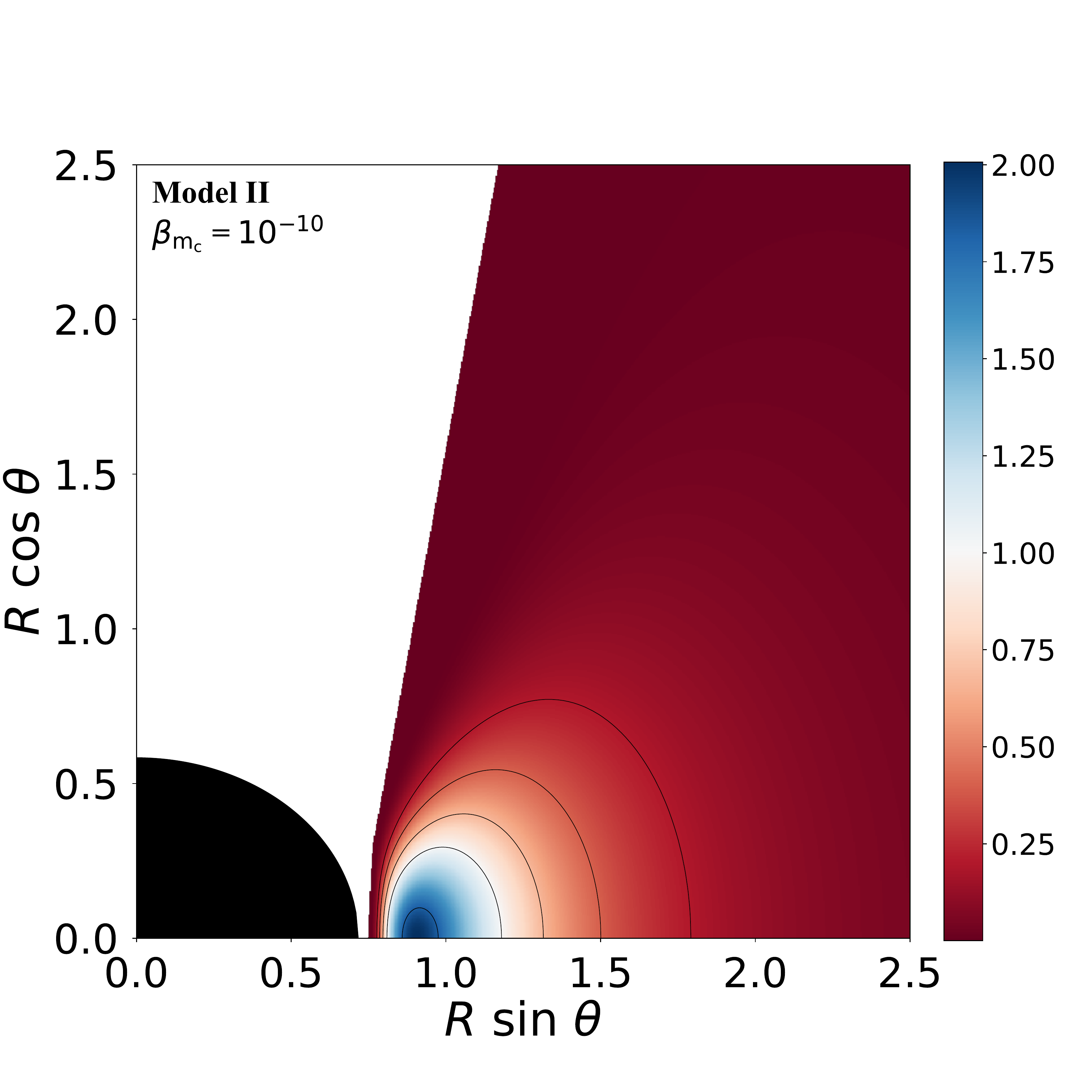}
\\
\includegraphics[scale=0.14]{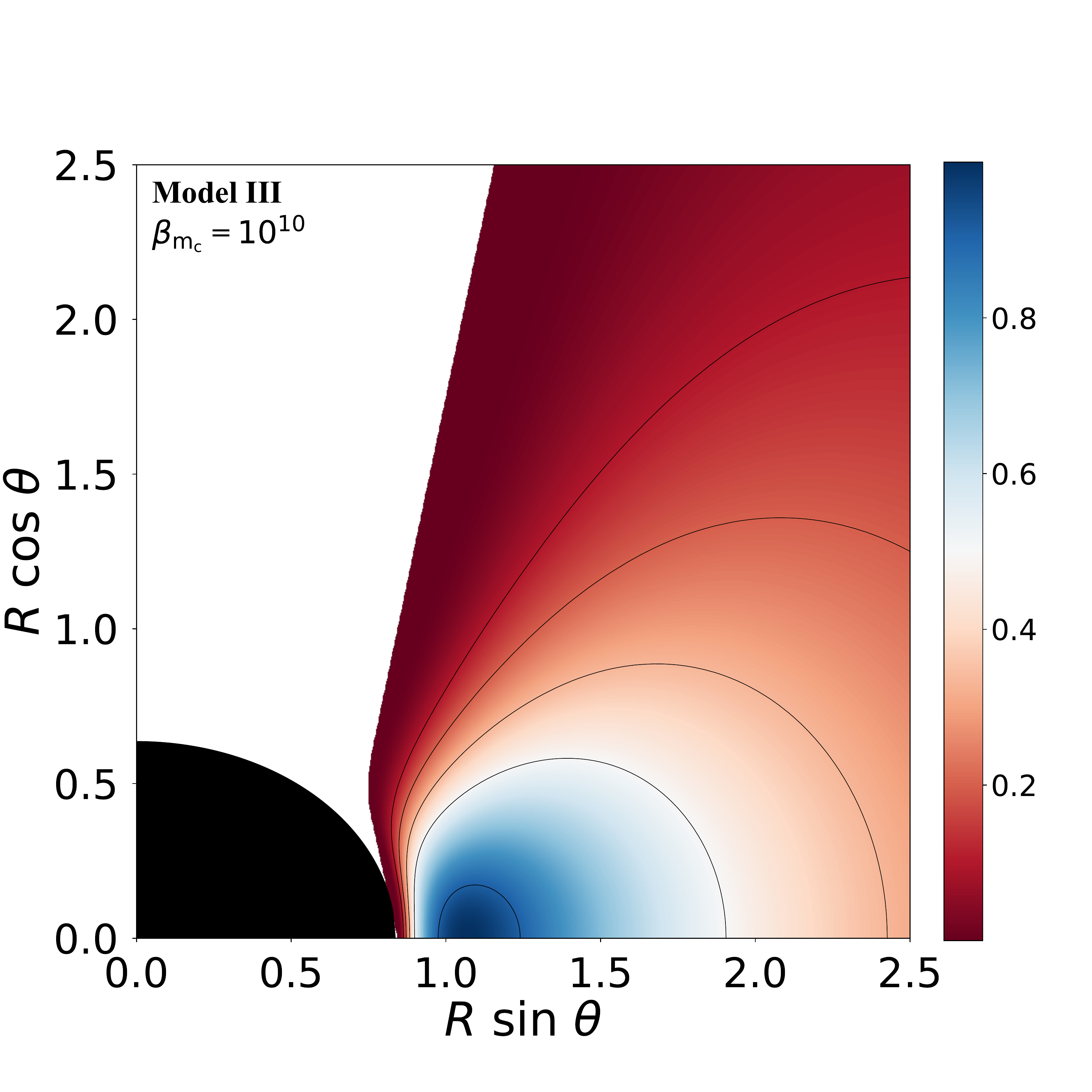}
\hspace{-0.3cm}
\includegraphics[scale=0.14]{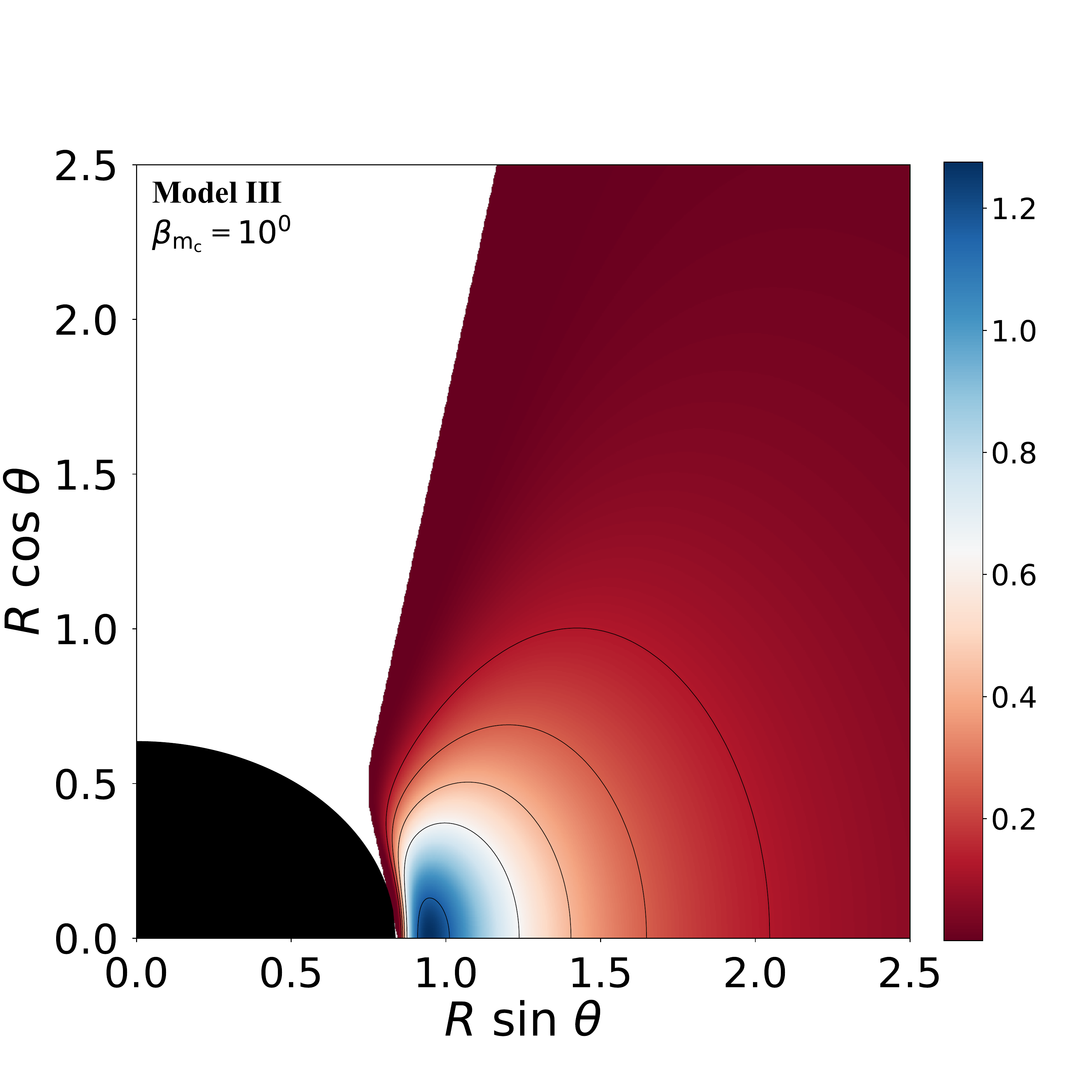}
\hspace{-0.2cm}
\includegraphics[scale=0.14]{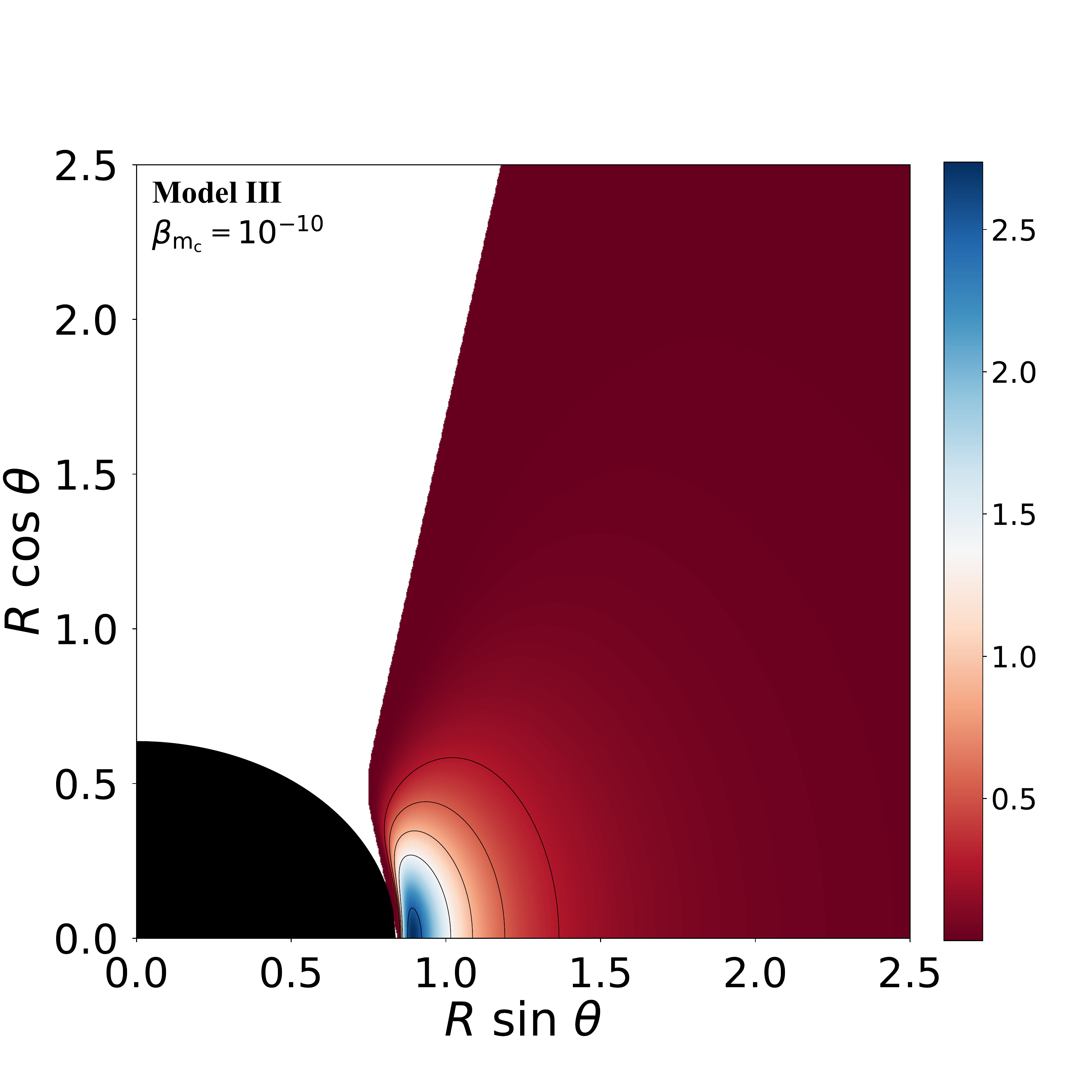}
\\
\includegraphics[scale=0.14]{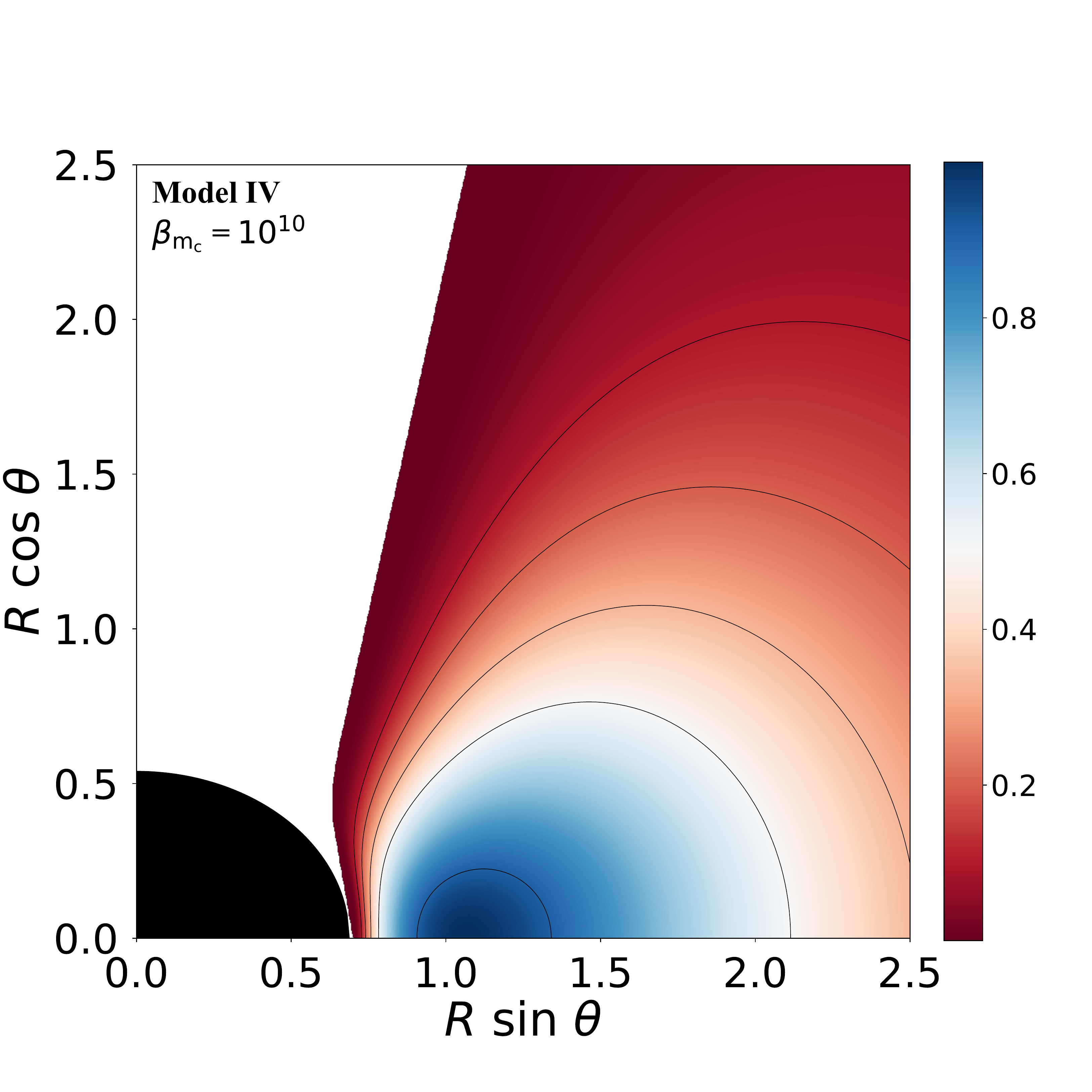}
\hspace{-0.3cm}
\includegraphics[scale=0.14]{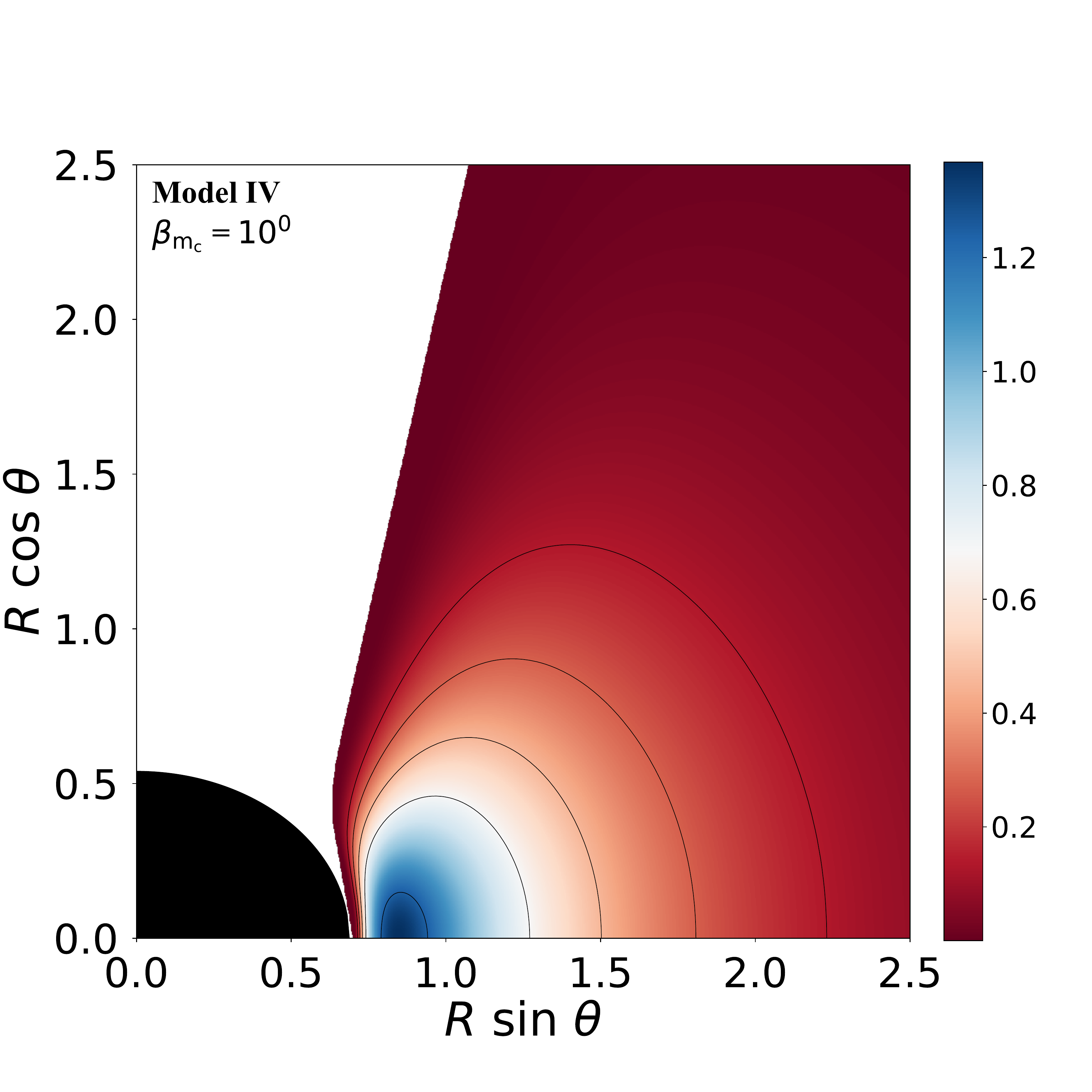}
\hspace{-0.2cm}
\includegraphics[scale=0.14]{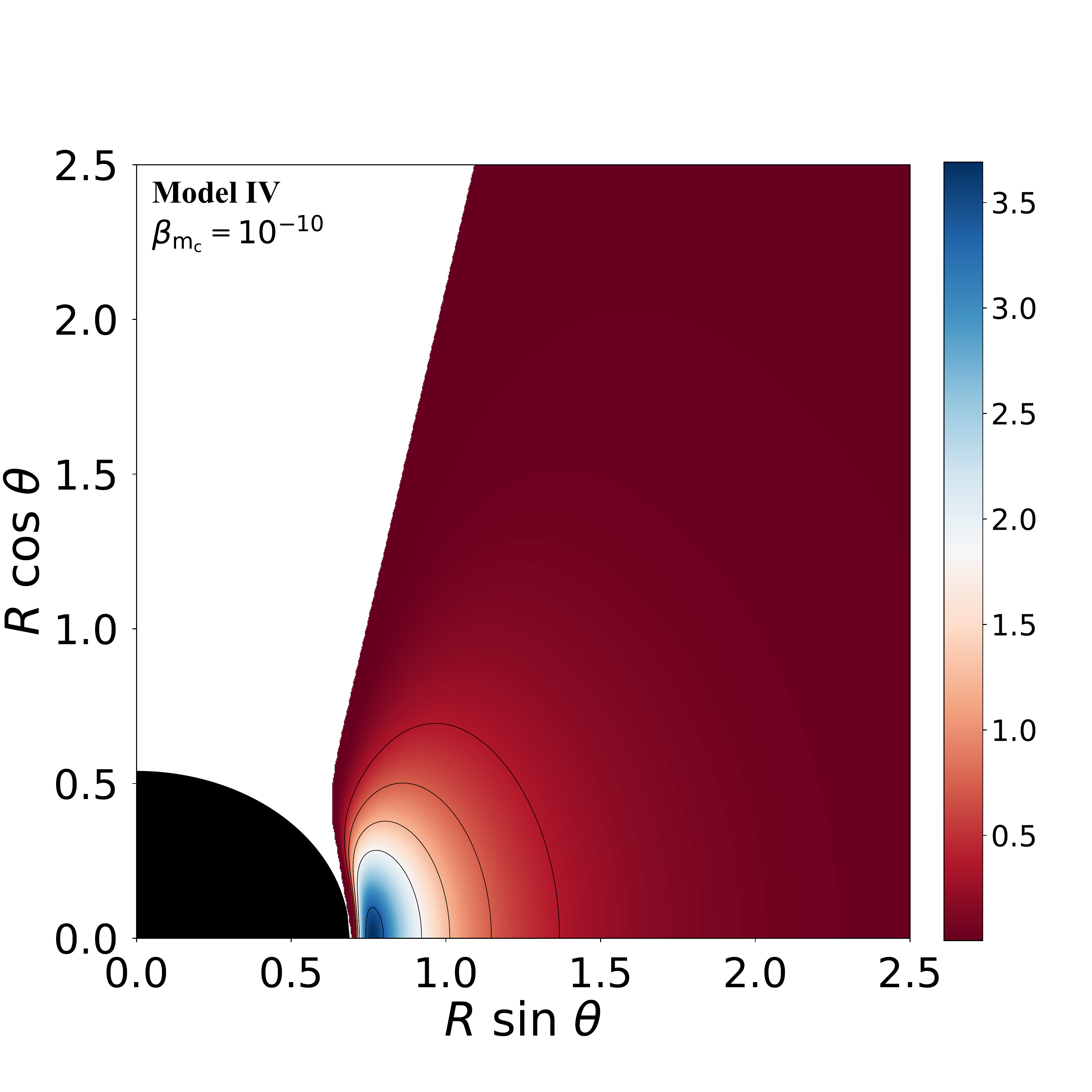}
\hspace{-0.2cm}
\caption{Same as Fig.~\ref{models_I} but using the perimeteral radial coordinate $R$.}
\label{models_peri_I}
\end{figure*}

\begin{figure*}
\centering
\includegraphics[scale=0.14]{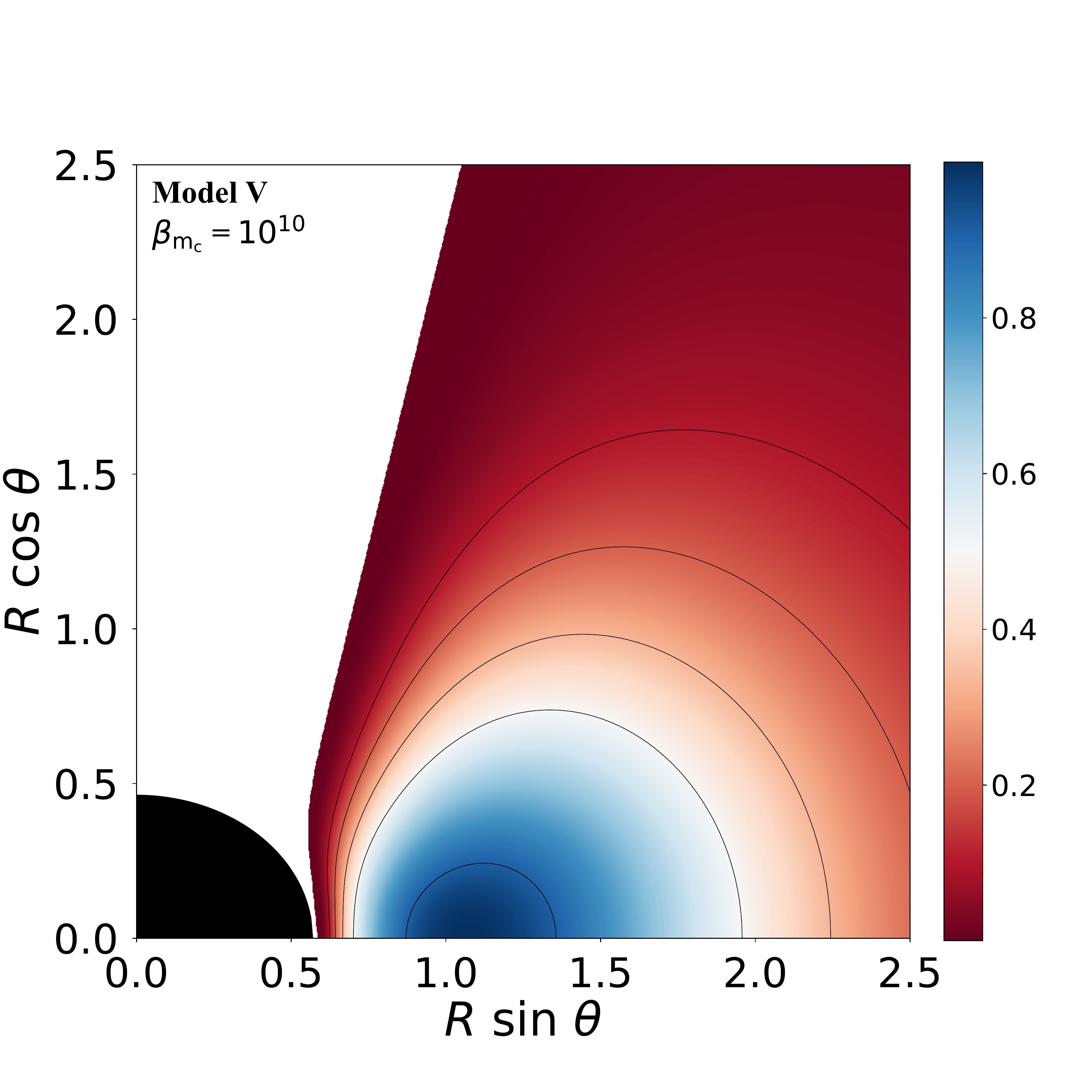}
\hspace{-0.3cm}
\includegraphics[scale=0.14]{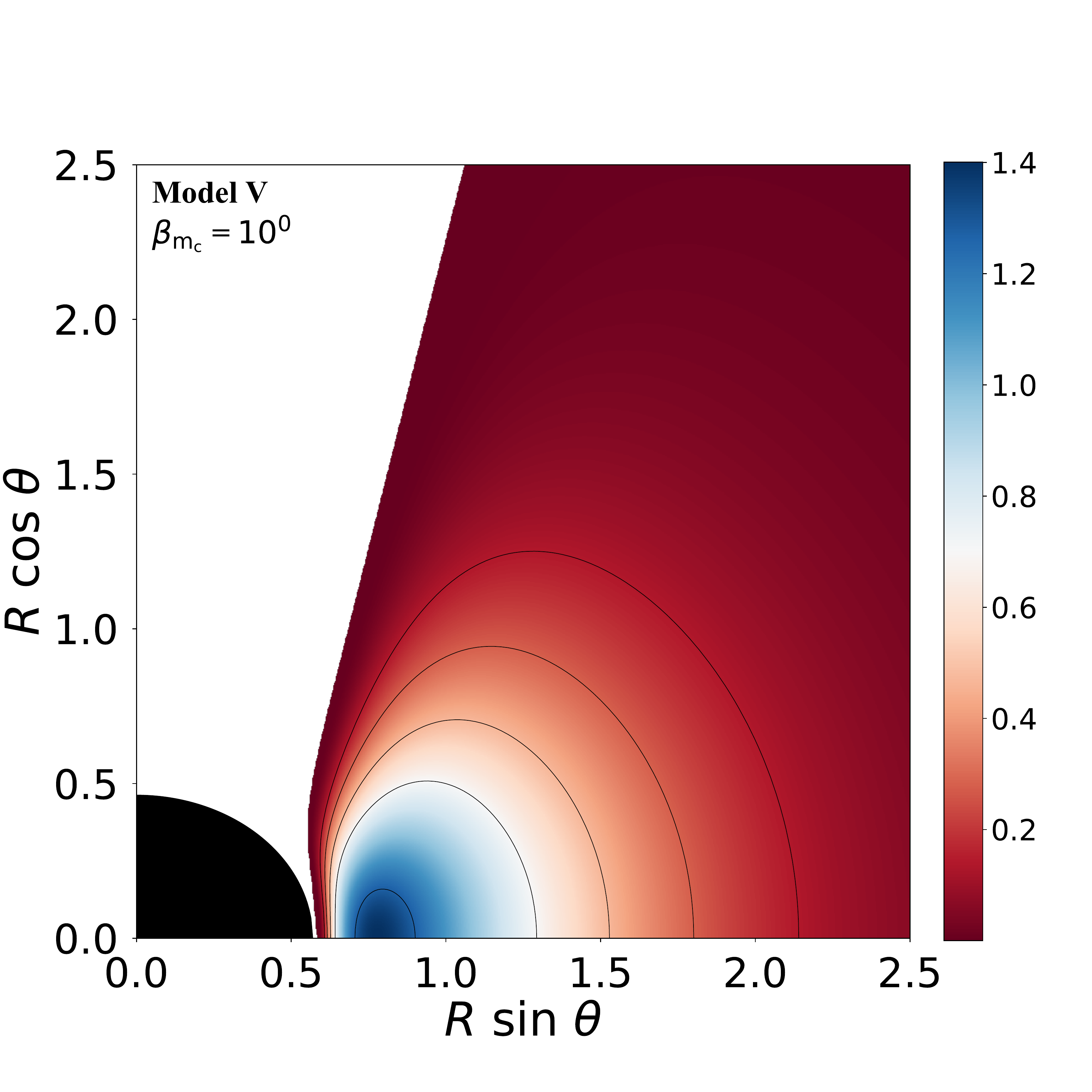}
\hspace{-0.2cm}
\includegraphics[scale=0.14]{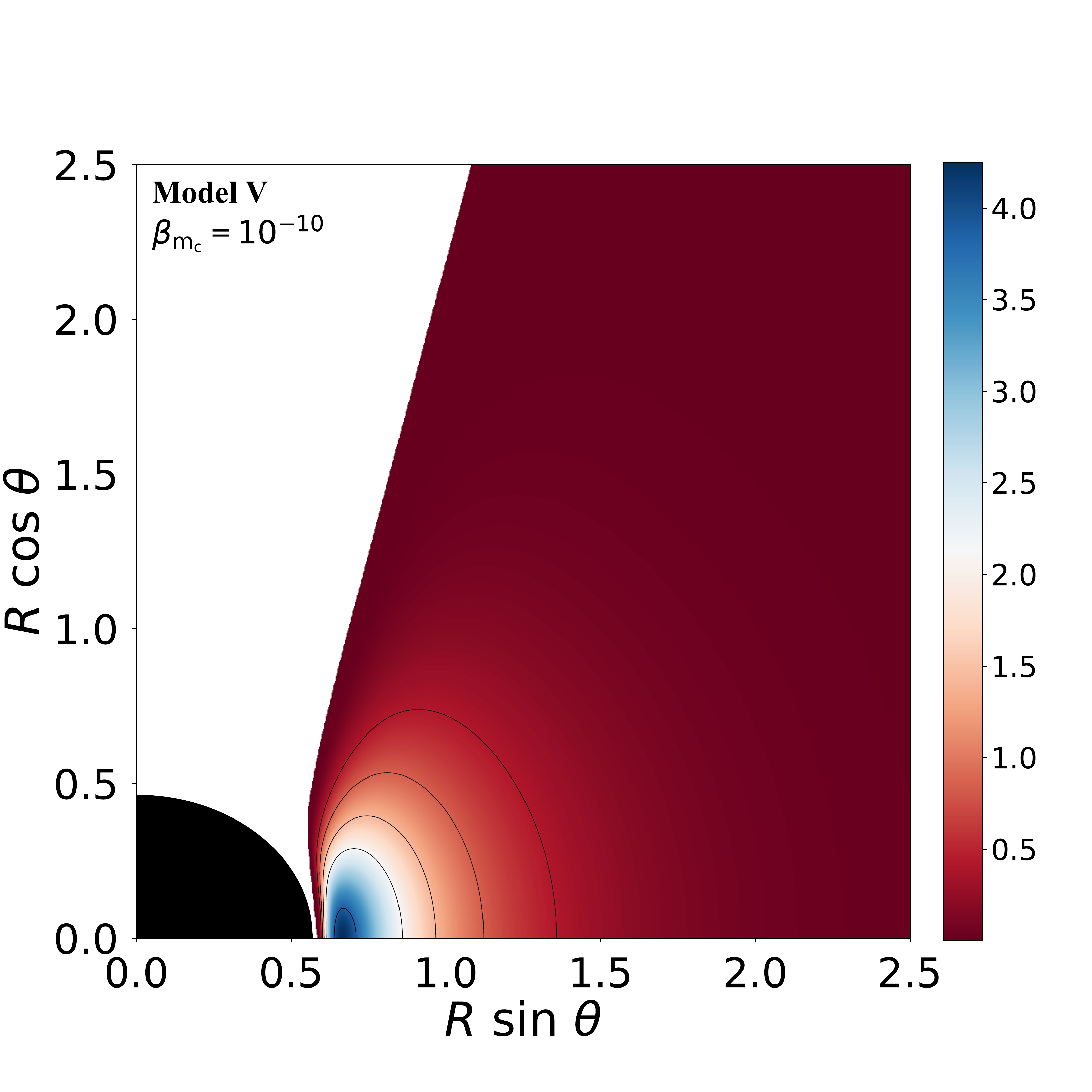}
\\
\includegraphics[scale=0.14]{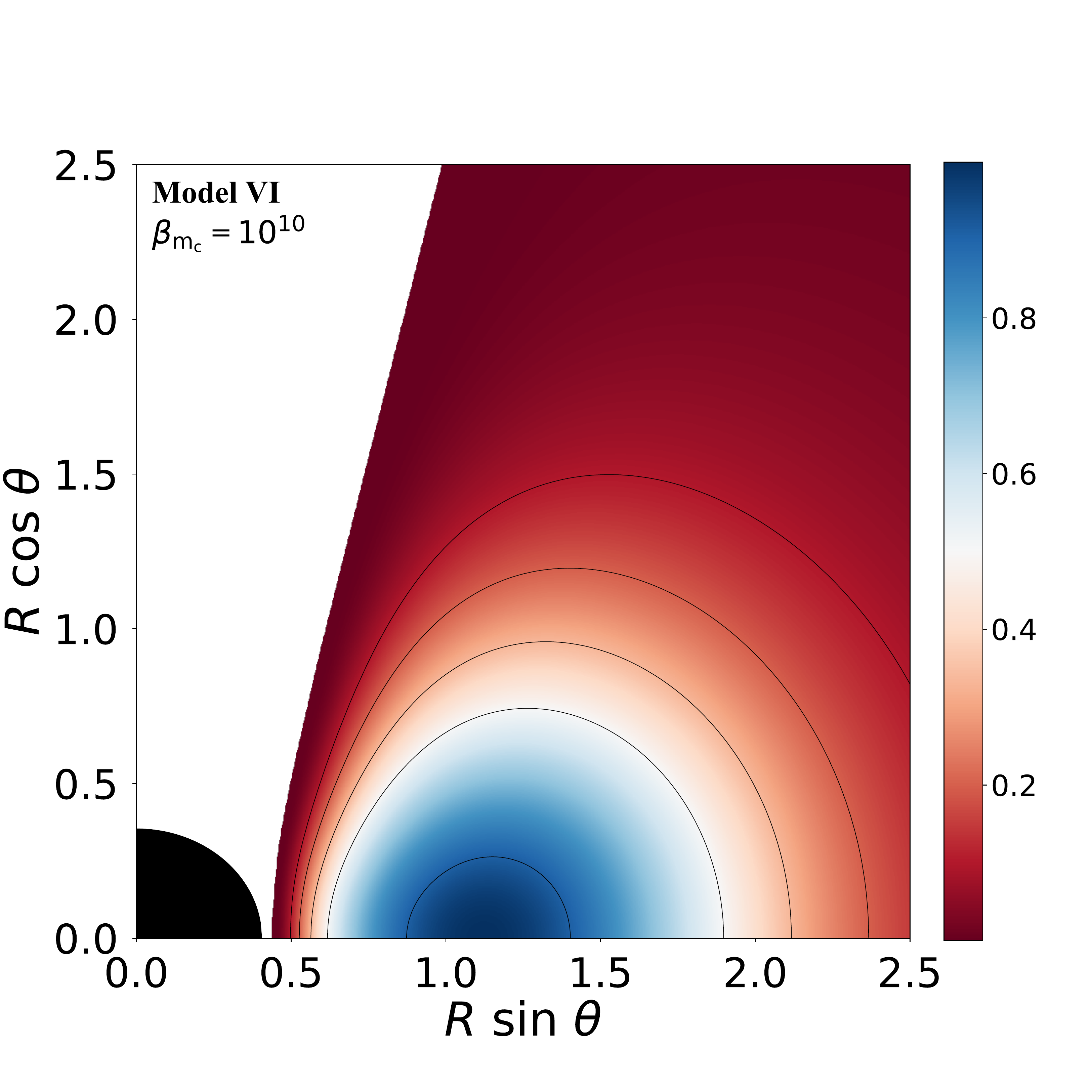}
\hspace{-0.3cm}
\includegraphics[scale=0.14]{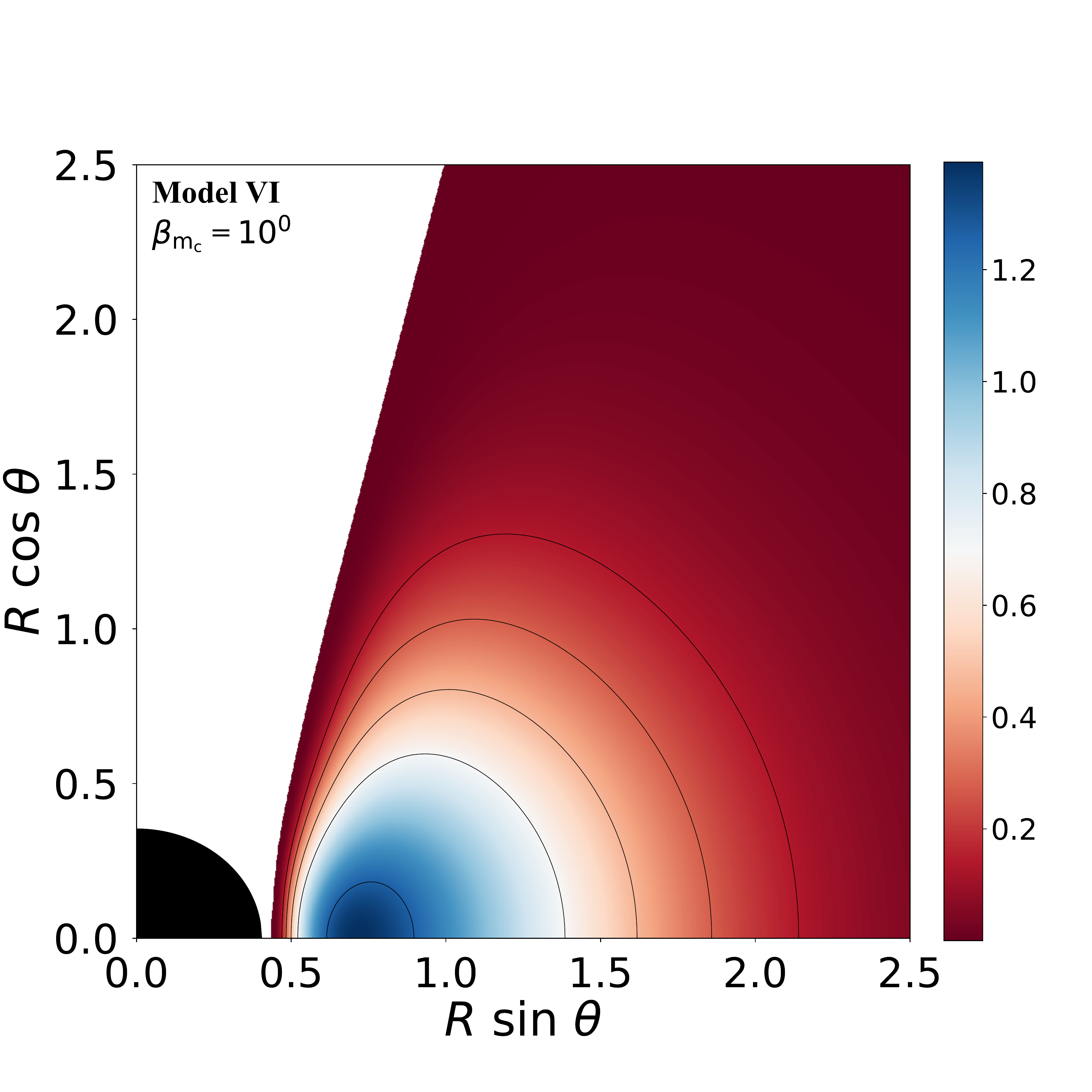}
\hspace{-0.2cm}
\includegraphics[scale=0.14]{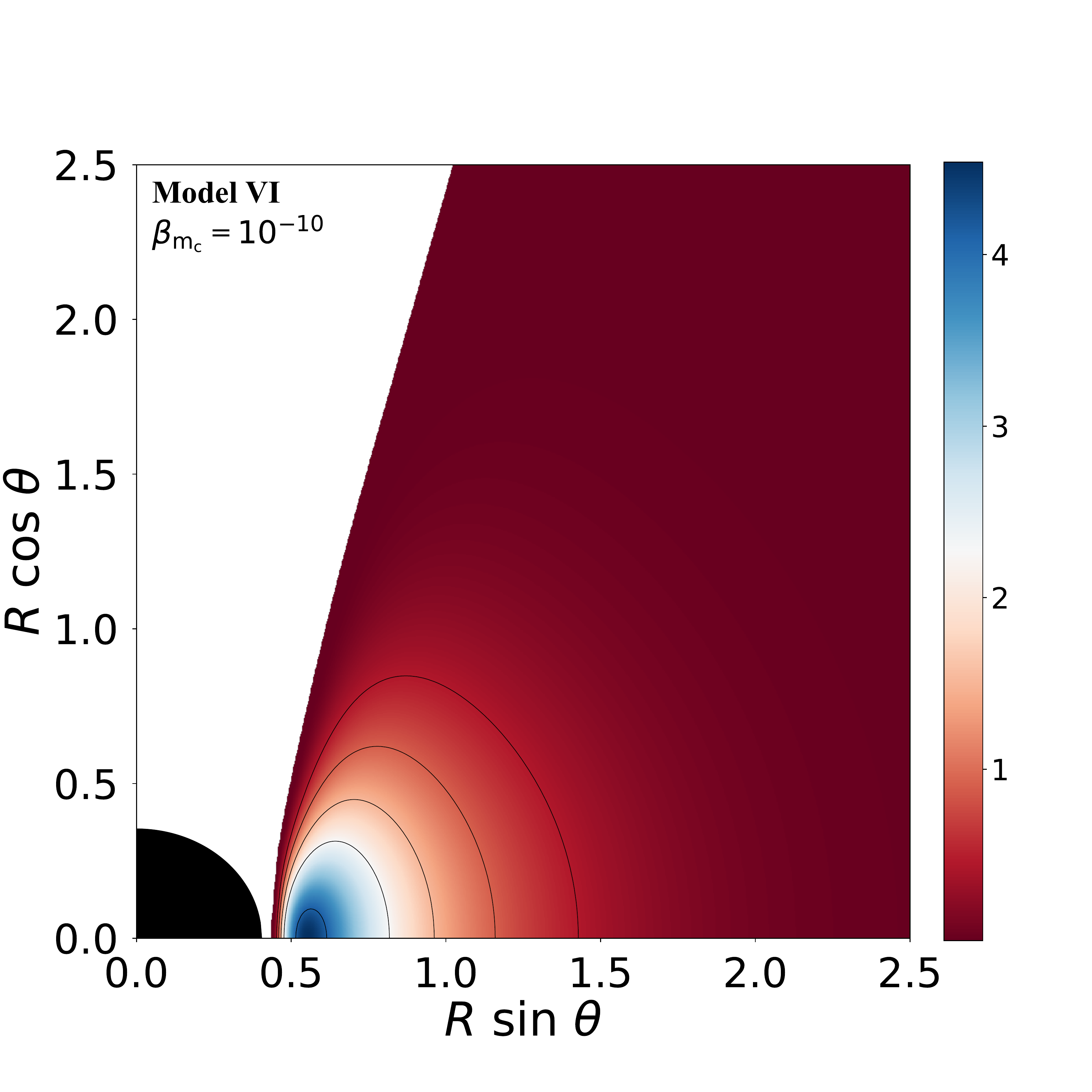}
\\
\includegraphics[scale=0.14]{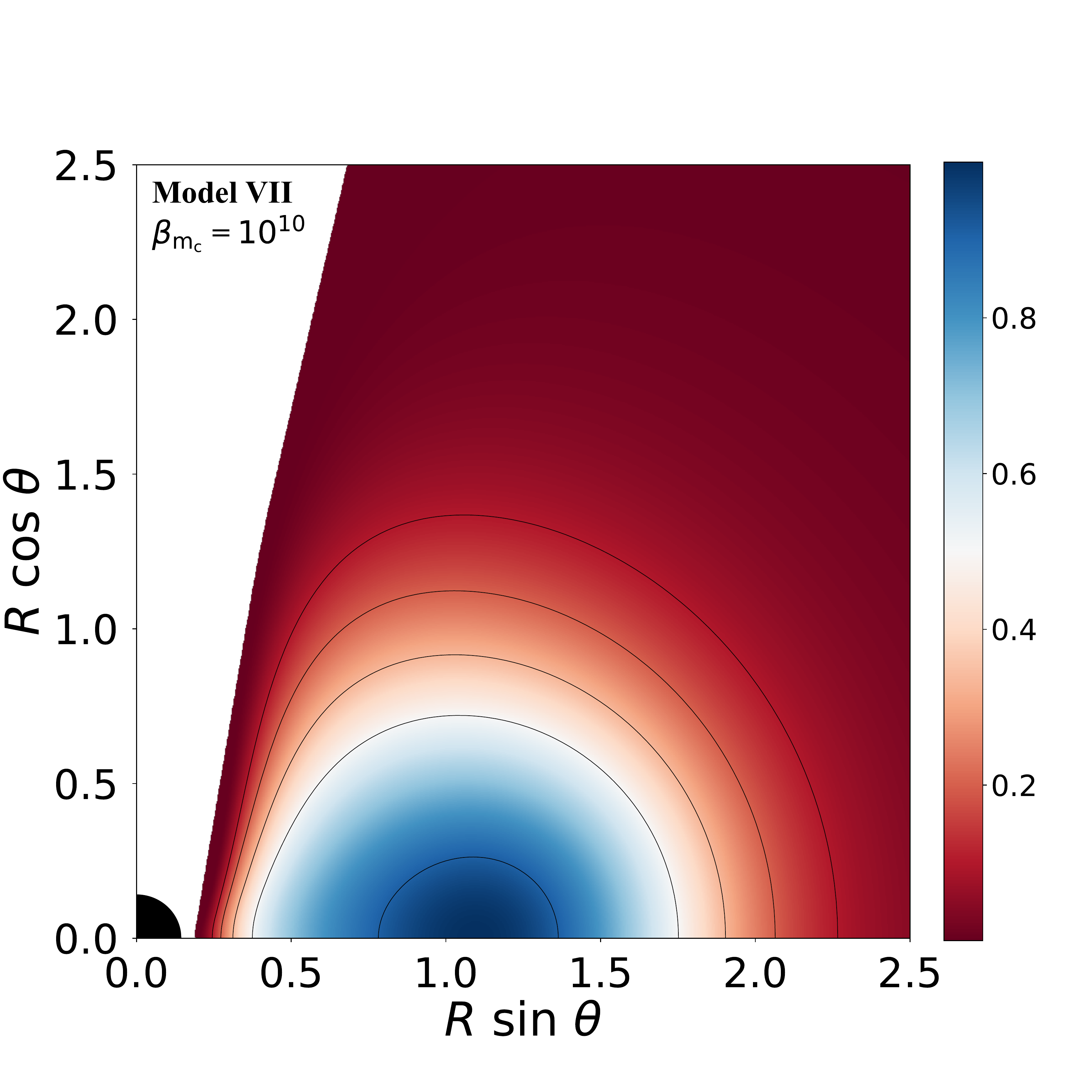}
\hspace{-0.3cm}
\includegraphics[scale=0.14]{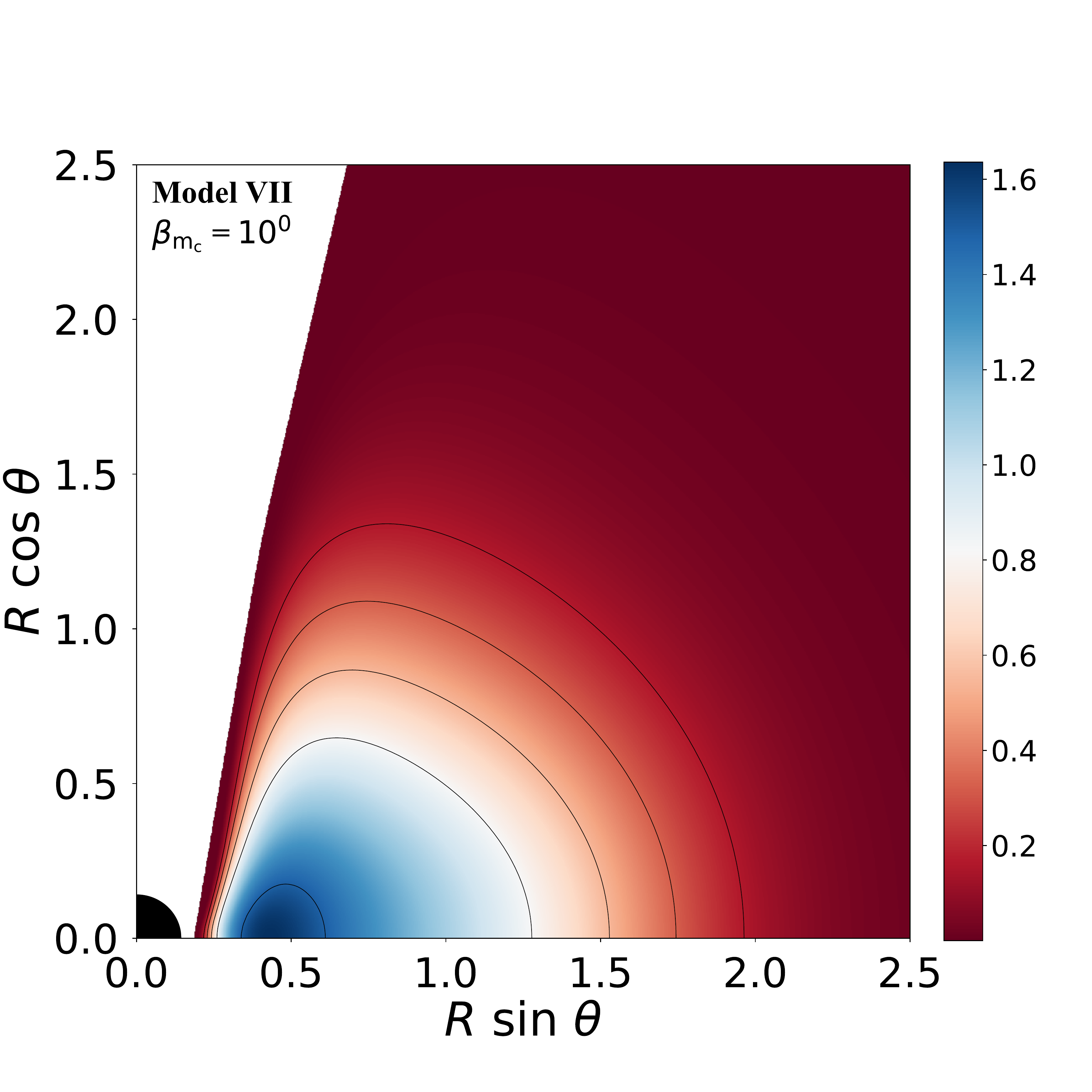}
\hspace{-0.2cm}
\includegraphics[scale=0.14]{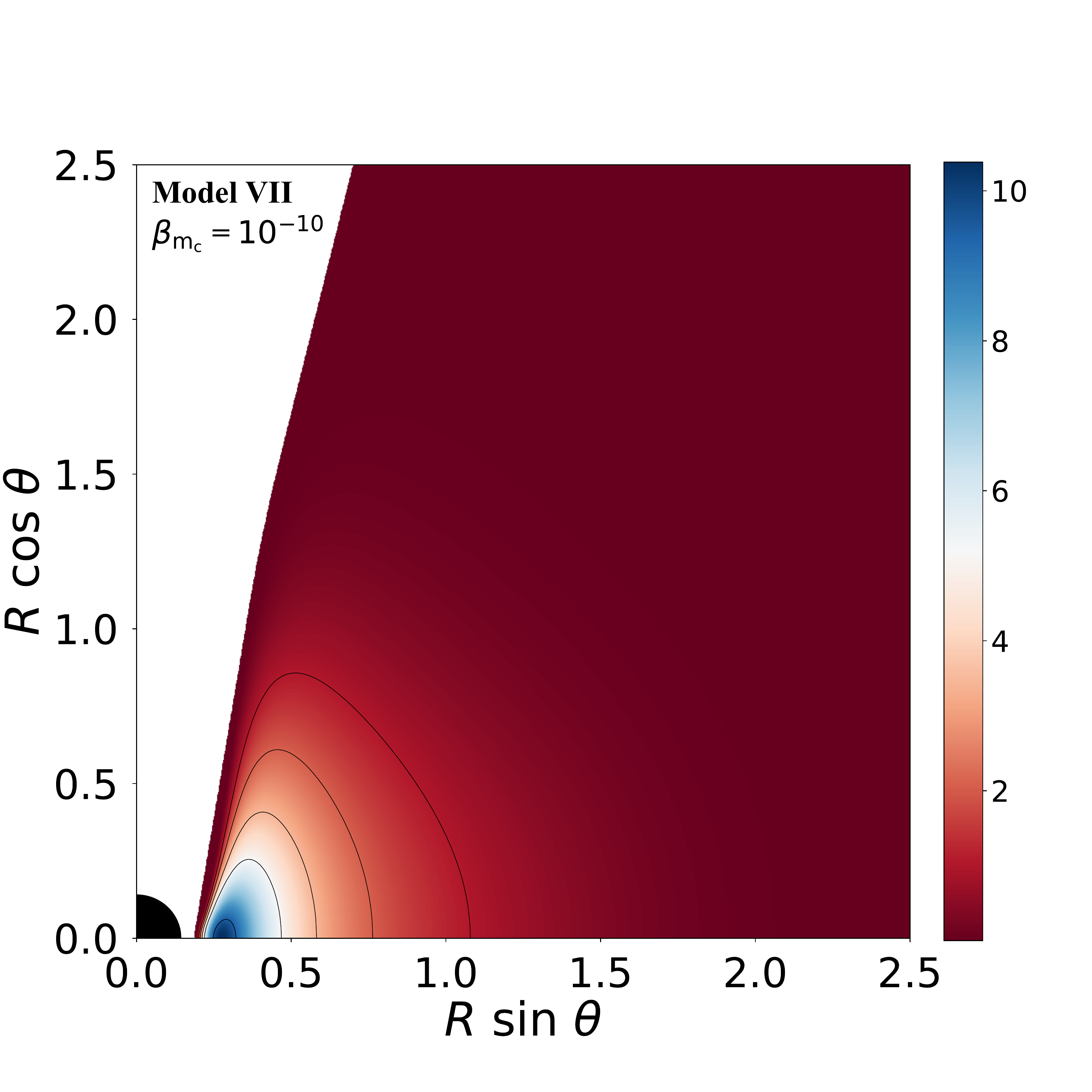}
\hspace{-0.2cm}
\caption{Same as Fig.~\ref{models_II} but using the perimeteral radial coordinate $R$.}
\label{models_peri_II}
\end{figure*}

\section{Methodology}
\label{procedure}

We now turn to describe the numerical methodology to build the disks. From the discussion in the preceding section it becomes apparent that the number of parameters defining the disk models is fairly large. In order to reduce the sample, in this work we set the mass of the scalar field to $\mu = 1$, the  azimuthal harmonic index to $m = 1$, the exponents of the polytropic EOS to $q = \Gamma = 4/3$, the density at the centre of the disk to $\rho_{\mathrm{c}} = 1$, the specific angular momentum to $l = l_{\mathrm{mb}}$ and the inner radius of the disk to $r_{\mathrm{in}} = r_{\mathrm{mb}}$. Thus, we leave the magnetization at the centre, $\beta_{\mathrm{m_c}}$, as the only free parameter for each model of KBHsSH. With this information we can compute all relevant physical quantities.

In particular, our choice of specific angular momentum and inner radius is made to allow disks to have a cusp and a centre. These disks are marginally stable, as they completely fill their Roche lobe, and a small perturbation can trigger accretion onto the BH. In addition, the thermodynamical quantities of the disks reach their maxima for this particular choice of parameters, as they are related to the total potential well $|\Delta W|$. Our choice also implies that the resulting disks will be semi-infinite (they are closed at infinity) but this is not a source of concern, as the external layers of the disk have extremely low density.

Before building the models, it is important to note that we need a sufficiently fine numerical grid to fully capture the behaviour of the physical magnitudes at the innermost regions of the disk. For this reason, we use a non-uniform $(r, \theta)$ grid with a typical domain given by $[r_{\mathrm{H}}, 199.2] \times [0, \pi/2]$ and a typical number of points $N_r \times N_{\theta} = 2500 \times 300$. Those numbers are only representative as the actual numbers depend on the horizon radius $r_{\mathrm{H}}$ and on the specific model. The spacetime metric data on this grid is interpolated from the original data obtained by~\cite{Herdeiro:2015b}. The original grid in~\cite{Herdeiro:2015b} is a uniform ($x$, $\theta$) grid (where $x$ is a compactified radial coordinate) with a domain $[0, 1] \times [0, \pi/2]$ and a number of points of $N_x \times N_{\theta} = 251 \times 30$~\footnote{Some samples are presented in~\cite{grav_web}}. To obtain our grid, we use the coordinate transformation provided in~\cite{grav_web} and interpolate the initial grid using cubic splines interpolation. 

To build the disks we first need to find $l_{\mathrm{mb}}$ and $r_{\mathrm{mb}}$ as the minimum of Eq.~\eqref{eq:mb_ang_mom} and the location of said minimum in terms of the radial coordinate respectively. Once this is done, we can compute the total potential distribution as
\begin{equation}
W(r,\theta) \equiv \ln |u_t| = \frac{1}{2} \ln \left| \frac{g_{t\phi}^2-g_{tt}g_{\phi\phi}}{g_{\phi\phi}+2g_{t\phi}l+g_{tt}l^2} \right|.
\end{equation}
With the total potential distribution, we can compute the location of the cusp $r_{\mathrm{cusp}}$ and the centre $r_{\mathrm{c}}$ as the extrema (maximum and minimum respectively) of the total potential in the equatorial plane. Also, we set $r_{\mathrm{in}} = r_{\mathrm{cusp}}$. For our choice of angular momentum distribution, this also means $W_{\mathrm{in}} = 0$. Having the total potential distribution and the characteristic radii of the disk, we can start to compute the thermodynamical quantities in the disk. First of all, we compute the polytropic constant $K$ by evaluating Eq.~\eqref{eq:final} at the centre
\begin{multline}
\label{eq:to_solve_K}
W - W_{\mathrm{in}} + \ln \left(1 + \frac{\Gamma K}{\Gamma -1}\rho_{\mathrm{c}}^{\Gamma -1}\right) 
\\
+ \frac{q}{q-1} \frac{K\rho_{\mathrm{c}}^{\Gamma}}{\beta_{\mathrm{m_c}} \left(\rho_{\mathrm{c}} + \frac{K\Gamma\rho_{\mathrm{c}}^{\Gamma}}{\Gamma-1}\right)} =0\,,
\end{multline}
where we have used the definition of magnetic pressure and the definition of the magnetization parameter $\beta$. Using their corresponding definitions, we can also compute $h_{\mathrm{c}}$, $p_{\mathrm{c}}$, $p_{\mathrm{m_c}}$ and the constant of the magnetic EOS $K_{\rm m}$. With both $K$ and $K_{\rm m}$ obtained, we can now compute the thermodynamical quantities in all our numerical domain. For points with $W(r, \theta) > 0$ we set $\rho = p = p_{\mathrm{m}} = 0$ and for points with $W_{\mathrm{c}} < W(r, \theta) < 0$, we write Eq.~\eqref{eq:final} as
\begin{multline}
\label{eq:to_solve_rho}
W - W_{\mathrm{in}} + \ln \left(1 + \frac{\Gamma K}{\Gamma -1}\rho^{\Gamma -1}\right) 
\\
+ \frac{q}{q-1}K_{\rm m}\left(\mathcal{L}\left(\rho + \frac{K\Gamma \rho^{\Gamma}}{\Gamma - 1}\right)\right)^{q-1}=0\,,
\end{multline}
to compute the rest-mass density $\rho$ of said point. Then, we can use again Eqs.~\eqref{eq:eos_fluid} and~\eqref{eq:eos_mag_tilde} and the definition of the specific enthalpy to compute the distribution of $p$, $p_{\mathrm{m}}$ and $h$.

It is relevant to note that Eqs.~\eqref{eq:to_solve_K} and~\eqref{eq:to_solve_rho} are trascendental equations and that Eq.~\eqref{eq:to_solve_rho} in particular must be solved at each point of our numerical grid. To solve these equations we use the bisection method. To ensure the accuracy of our computations (particularly the accuracy of the maximum and central quantities we report) we choose our grid to have a difference between two adjacent points of $\Delta r (r \simeq r_{\mathrm{c}}) \simeq 0.001$ in the equatorial plane.

\section{Results}
\label{results}

\begin{table*}[t]
\caption{Values of the relevant physical magnitudes of our models of magnetized, equilibrium tori around KBHsSH. All reported radii correspond to the perimeteral coordinate. For all cases,  $R_{\mathrm{in}} = R_{\mathrm{mb}}$ and $l = l_{\mathrm{mb}}$. From left to right the columns report: the specific angular momentum, $l$, the potential at the centre of the disk, $W_{\mathrm{c}}$, the inner radius of the disk, $R_{\mathrm{in}}$, its centre, $R_{\mathrm{c}}$, the value of the magnetization parameter at the centre, $\beta_{\mathrm{m_{\mathrm{c}}}}$, the maximum specific enthalpy, $h_{\mathrm{max}}$, density,  $\rho_{\mathrm{max}}$, thermal pressure, $p_{\mathrm{max}}$, and magnetic pressure, $p_{\mathrm{m, max}}$, and the location of the maximum of the thermal pressure and magnetic pressure, $R_{\mathrm{max}}$ and $R_{\mathrm{m, max}}$, respectively.}       
\label{HBH_disk_parameters}      
\centering          
\begin{tabular}{c c c c c  c c c c c c c}
\hline\hline       
 Model & $l$ & $W_{\mathrm{c}}$ & $R_{\mathrm{in}}$ & $R_{\mathrm{c}}$ &  $\beta_{\mathrm{m_{\mathrm{c}}}}$ & $h_{\mathrm{max}}$ & $\rho_{\mathrm{max}}$ & $p_{\mathrm{max}}$ & $p_{\mathrm{m, max}}$ & $R_{\mathrm{max}}$ & $R_{\mathrm{m, max}}$\\ 
\hline           
I & $0.934$ & $-0.188$ & $0.81$ & $1.14$ & $10^{10}$ & $1.21$ & $1.00$ & $5.16 \times 10^{-2}$ & $5.50 \times 10^{-12}$ & $1.14$ & $1.26$\\ 

 &  &  &  &  & $1$ & $1.10$ & $1.17$ & $3.11 \times 10^{-2}$ & $2.68 \times 10^{-2}$ & $1.01$ & $1.06$\\ 

 &  &  &  &  & $10^{-10}$ & $1.00$ & $1.90$ & $1.10 \times 10^{-11}$ & $7.80 \times 10^{-2}$ & $0.93$ & $0.96$\\ 

II & $0.933$ & $-0.205$ & $0.75$ & $1.18$ & $10^{10}$ & $1.23$ & $1.00$ & $5.69 \times 10^{-2}$ & $6.14 \times 10^{-12}$ & $1.18$ & $1.36$\\ 

 &  &  &  &  & $1$ & $1.12$ & $1.19$ & $3.50 \times 10^{-2}$ & $2.97 \times 10^{-2}$ & $1.00$ & $1.07$\\ 

 &  &  &  &  & $10^{-10}$ & $1.00$ & $2.01$ & $1.30 \times 10^{-11}$ & $8.99 \times 10^{-2}$ & $0.91$ & $0.94$ \\ 

III & $1.060$ & $-0.362$ & $0.84$ & $1.07$ & $10^{10}$ & $1.44$ & $1.00$ & $1.09 \times 10^{-1}$ & $1.21 \times 10^{-11}$ & $1.07$ & $1.22$\\ 

 &  &  &  &  & $1$ & $1.23$ & $1.28$ & $7.22 \times 10^{-2}$ & $5.76 \times 10^{-2}$ & $0.95$ & $0.99$ \\ 

 &  &  &  &  & $10^{-10}$ & $1.00$ & $2.74$ & $3.48 \times 10^{-11}$ & $2.06 \times 10^{-1}$ & $0.89$ & $0.91$\\ 
 
IV & $1.160$ & $-0.547$ & $0.67$ & $1.06$ & $10^{10}$ & $1.72$ & $1.00$ & $1.82\times 10^{-1}$ & $2.09 \times 10^{-11}$ & $1.06$ & $1.34$ \\ 

 &  &  &  &  & $1$ & $1.38$ & $1.37$ & $1.29 \times 10^{-1}$ & $9.76 \times 10^{-2}$ & $0.85$ & $0.91$\\ 

 &  &  &  &  & $10^{-10}$ & $1.00$ & $3.70$ & $7.83 \times 10^{-11}$ & $4.08\times 10^{-1}$ & $0.76$ & $0.78$ \\ 

V & $1.200$ & $-0.685$ & $0.58$ & $1.07$ & $10^{10}$ & $1.98$ & $1.00$ & $2.46 \times 10^{-1}$ & $2.76 \times 10^{-11}$ & $1.07$ & $1.31$\\ 

 &  &  &  &  & $1$ & $1.51$ & $1.40$ & $1.78 \times 10^{-1}$ & $1.32\times 10^{-1}$ & $0.78$ & $0.87$ \\ 

 &  &  &  &  & $10^{-10}$ & $1.00$ & $4.26$ & $1.18 \times 10^{-10}$ & $5.79\times 10^{-1}$ & $0.67$ & $0.69$ \\ 
  
VI & $1.200$ & $-0.832$ & $0.43$ & $1.12$ & $10^{10}$ & $2.30$ & $1.00$ & $3.24 \times 10^{-1}$ & $3.52 \times 10^{-11}$ & $1.12$ & $1.32$ \\ 

 &  &  &  &  & $1$ & $1.66$ & $1.39$ & $2.28 \times 10^{-1}$ & $1.69\times 10^{-1}$ & $0.72$ & $0.86$ \\ 

 &  &  &  &  & $10^{-10}$ & $1.00$ & $4.54$ & $1.57 \times 10^{-10}$ & $7.40\times 10^{-1}$ & $0.55$ & $0.59$ \\ 
 
VII & $0.920$ & $-1.236$ & $0.18$ & $1.10$ & $10^{-10}$ & $3.44$ & $1.00$ & $6.10 \times 10^{-1}$ & $6.46 \times 10^{-11}$ & $1.10$ & $1.25$ \\ 

 &  &  &  &  & $1$ & $2.25$ & $1.64$ & $5.10 \times 10^{-1}$ & $3.22\times 10^{-1}$ & $0.43$ & $0.62$\\ 

 &  &  &  &  & $10^{-10}$ & $1.00$ & $10.42$ & $7.03 \times 10^{-10}$ & $0.24 \times 10^{-1}$ & $0.28$ & $0.30$\\ 
\hline\hline
\end{tabular}
\end{table*}

\subsection{2D Morphology}

We start presenting the morphological distribution of the models in the ($r\sin\theta, r\cos\theta$) plane in figures~\ref{models_I} and~\ref{models_II}. These figures show the rest-mass density distribution for all our KBHsSH models for 3 different values of the magnetization parameter at the centre of the disks, $\beta_{\mathrm{m_c}}$, namely $10^{10}$ (unmagnetized, left column), $1$ (mildly magnetized, middle column) and $10^{-10}$ (strongly magnetized, right column). 

\begin{figure*}
\centering
\includegraphics[scale=0.1267]{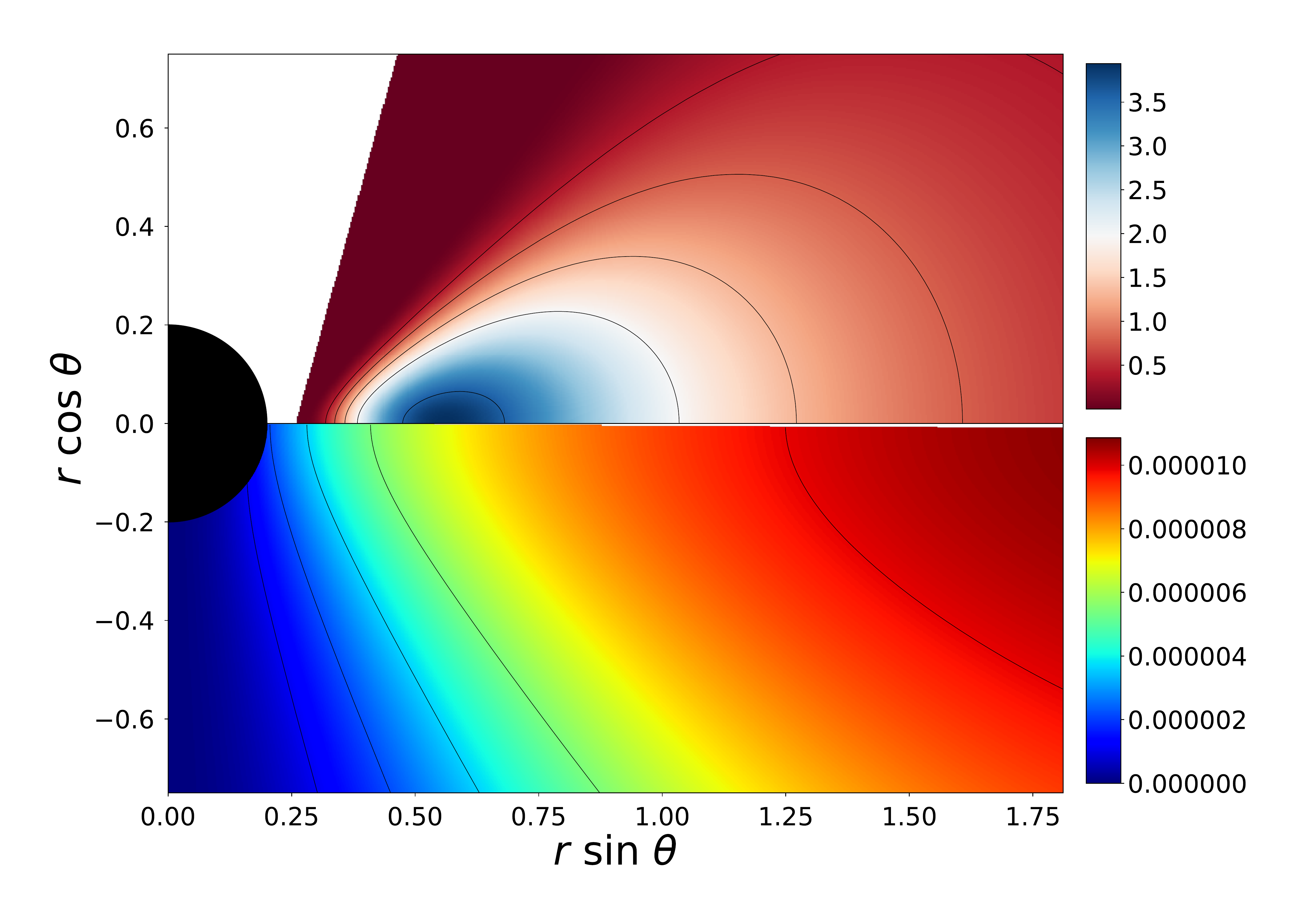}
\hspace{-0.3cm}
\includegraphics[scale=0.12]{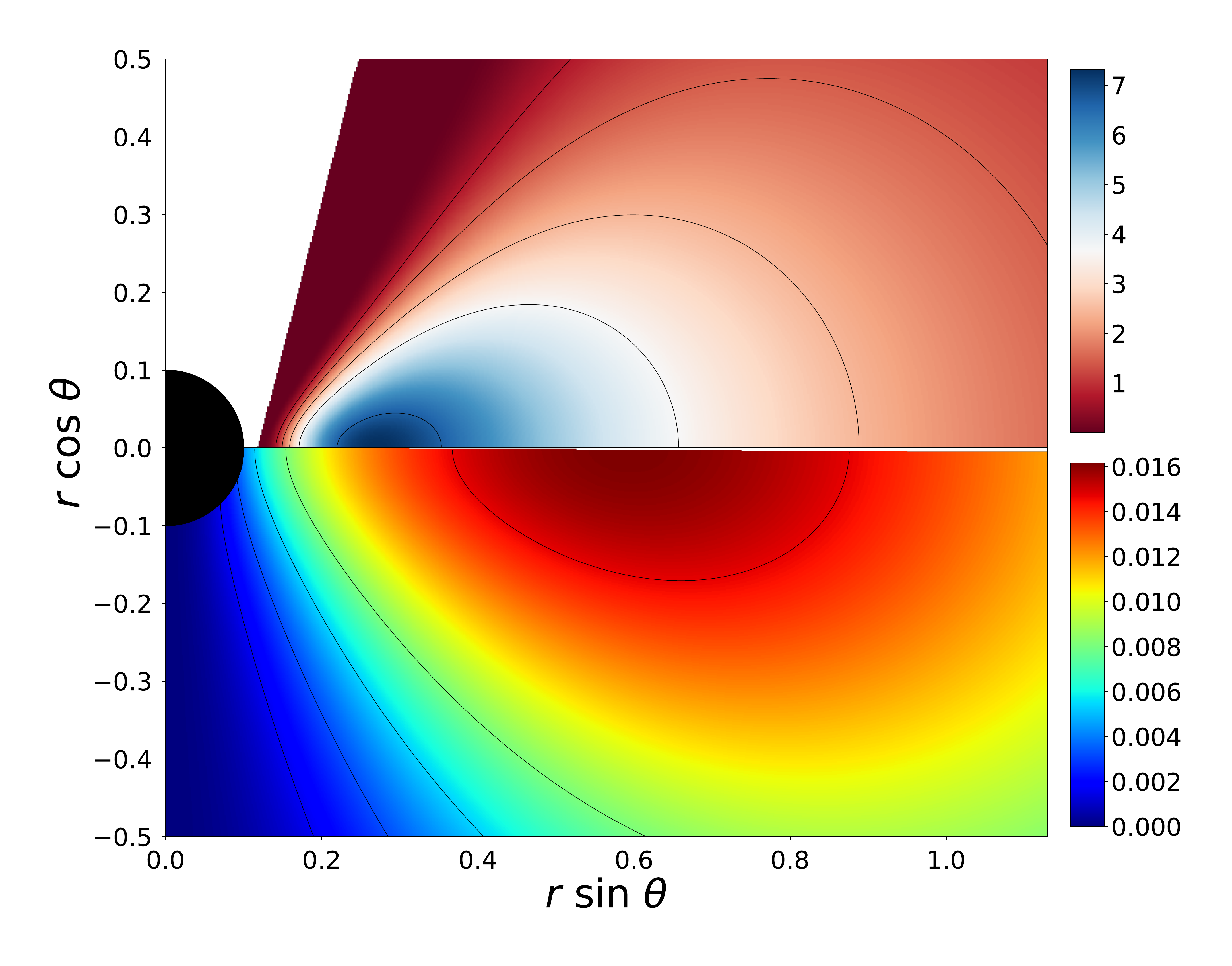}
\hspace{-0.3cm}
\includegraphics[scale=0.1267]{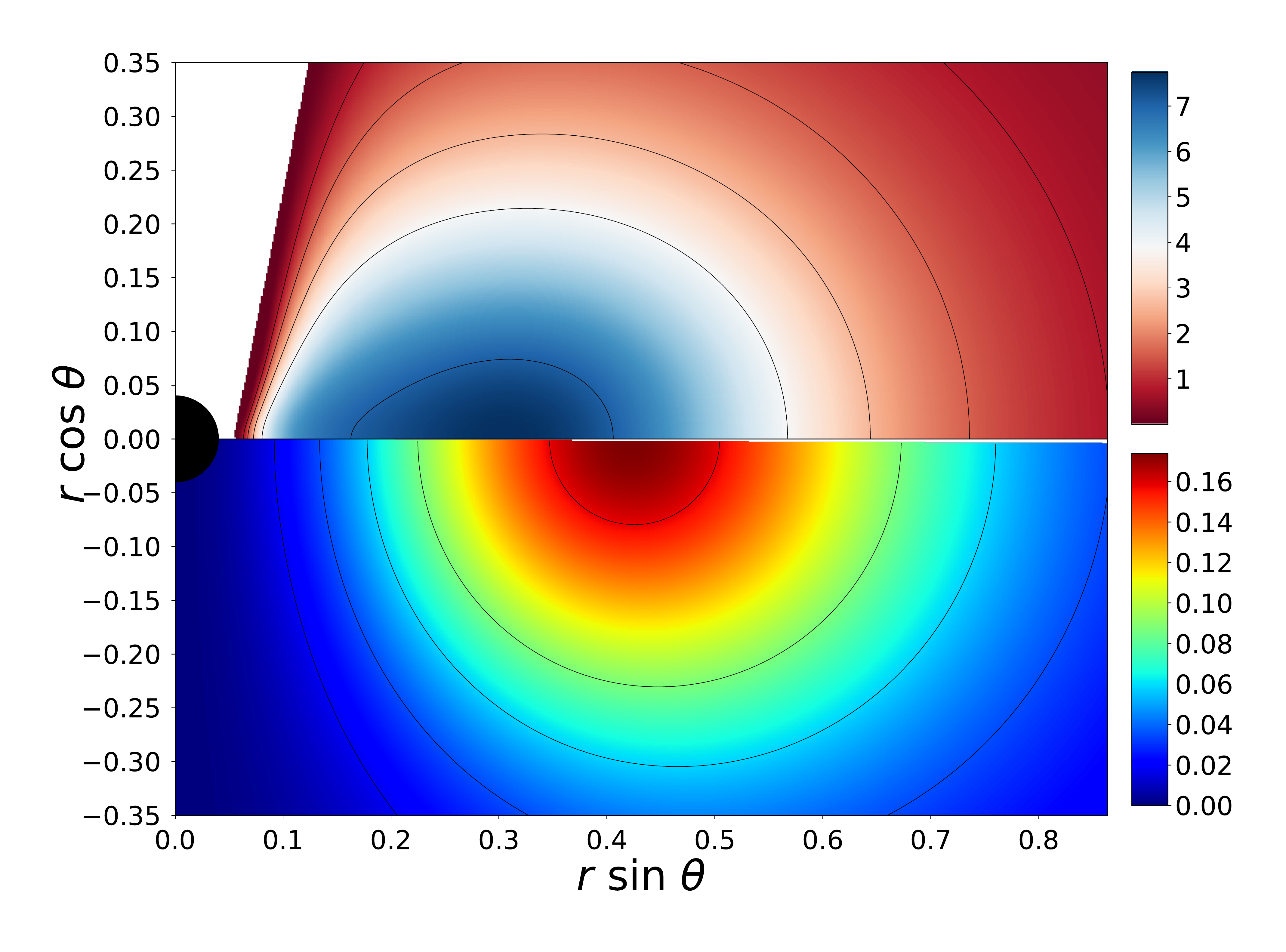}
\hspace{-0.2cm}
\\
\includegraphics[scale=0.1267]{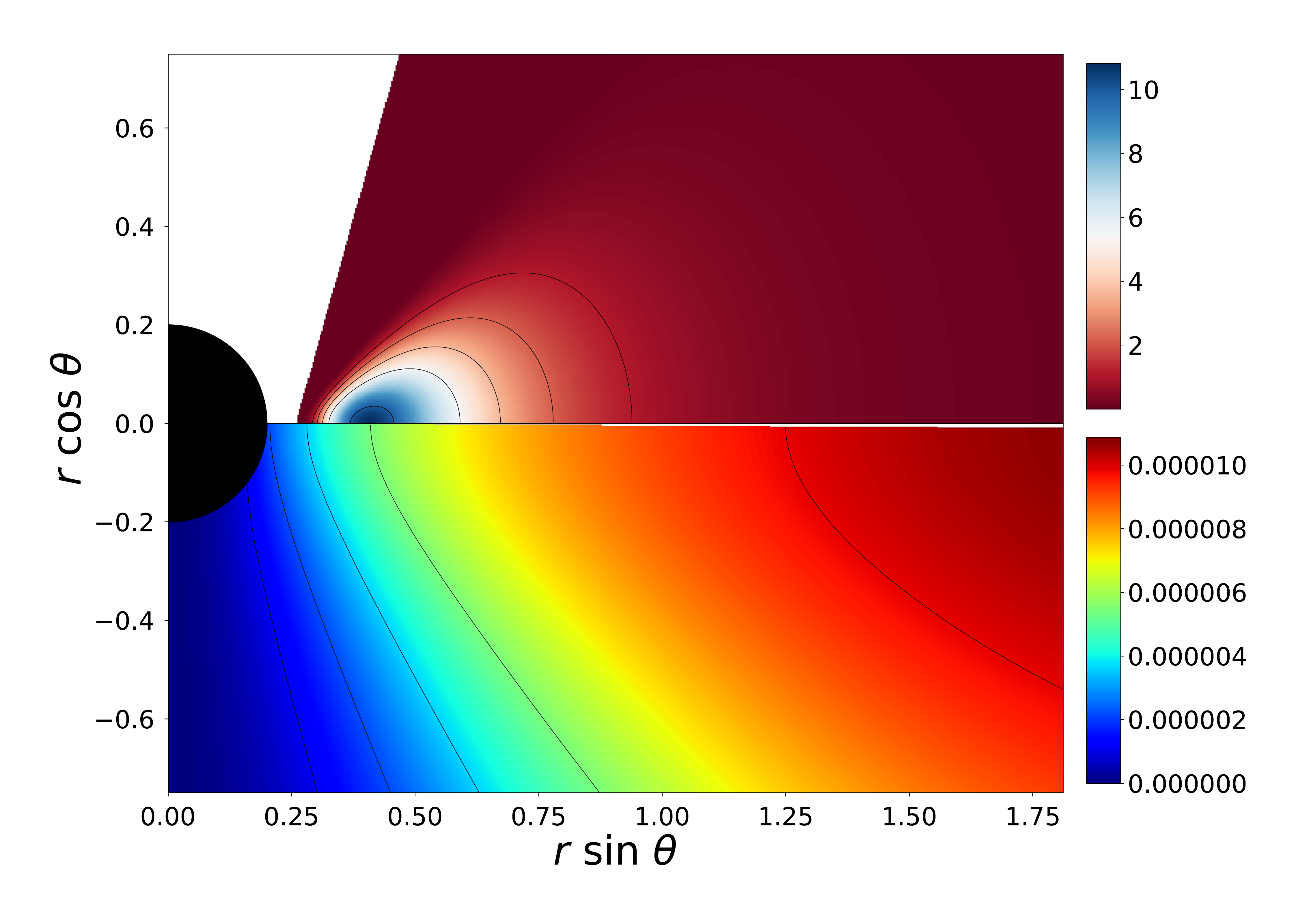}
\hspace{-0.3cm}
\includegraphics[scale=0.12]{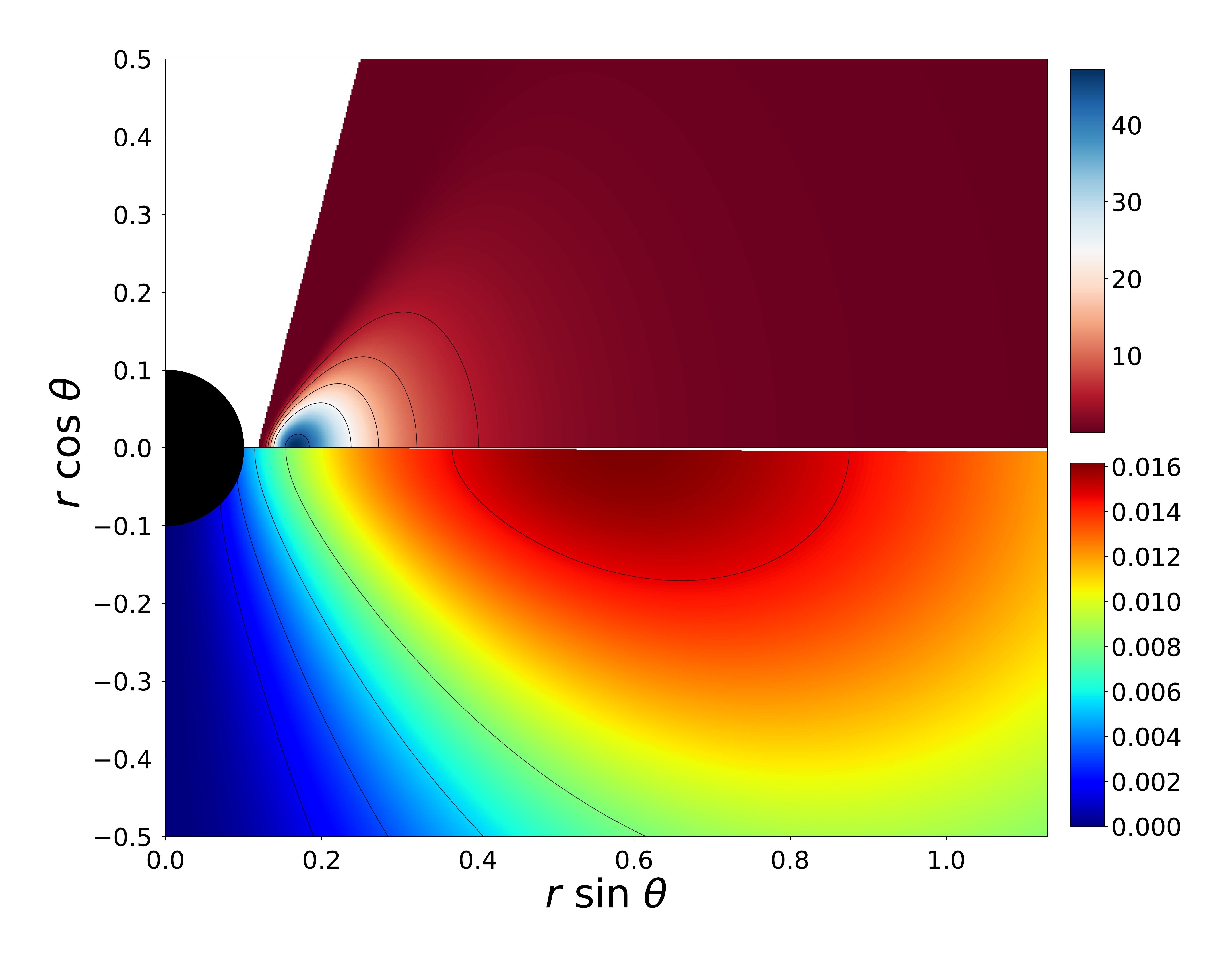}
\hspace{-0.3cm}
\includegraphics[scale=0.1267]{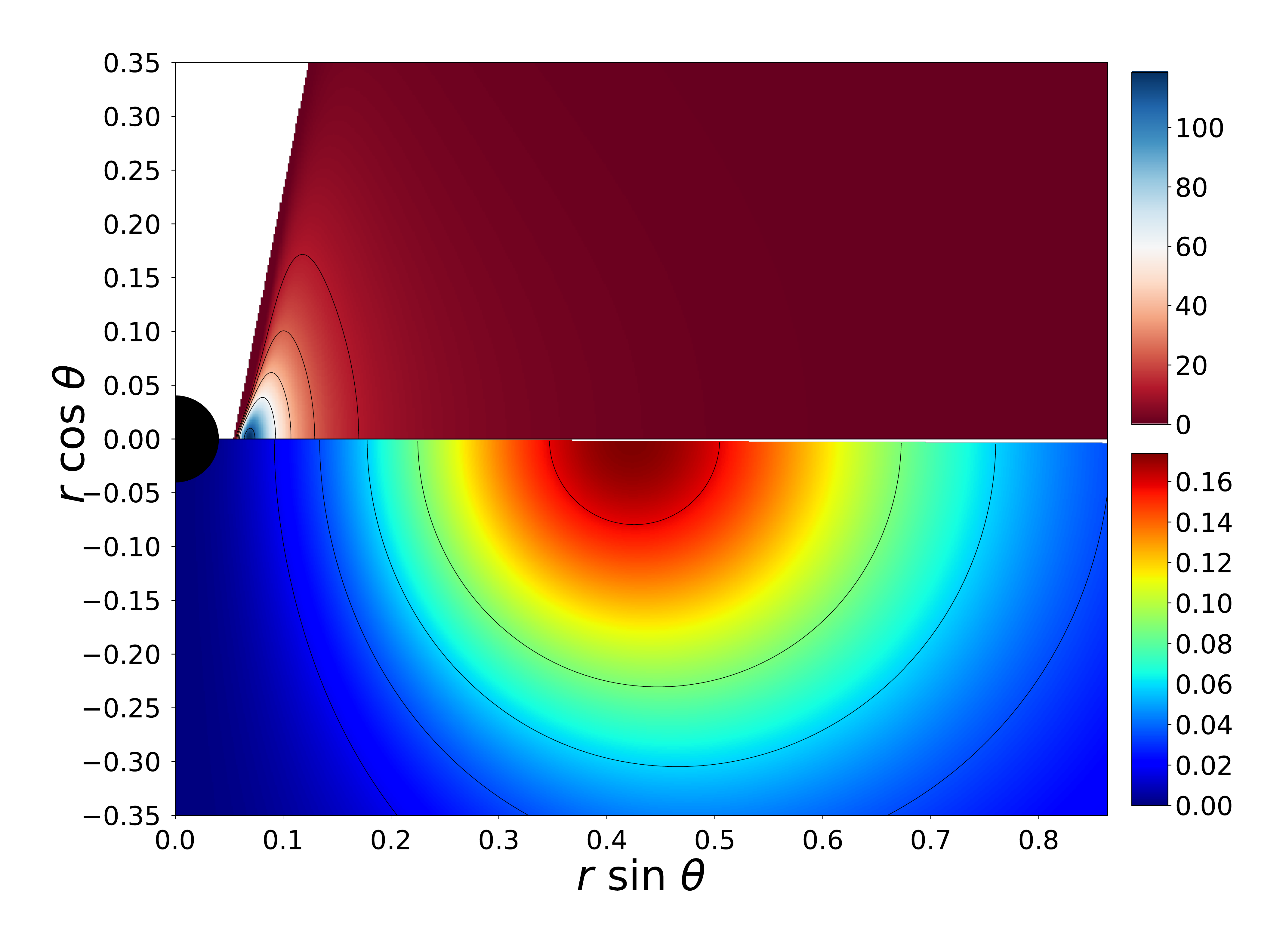}
\hspace{-0.2cm}
\caption{Energy density distribution for the torus $\rho_{\mathrm{T}}$ (upper half of the images) and for the scalar field $\rho_{\mathrm{SF}}$ (lower half). From left to right the columns correspond to models I, IV, and VII. The top row corresponds to non-magnetized models ($\beta_{\mathrm{m}_{\mathrm{c}}} = 10^{10}$) and the bottom row to strongly magnetized models ($\beta_{\mathrm{m}_{\mathrm{c}}} = 10^{-10}$).}
\label{comparison_mass_density}
\end{figure*}

The structure of the disks is similar for all values of $\beta_{\mathrm{m_c}}$ with the only quantitative differences being the location of the centre of the disk, which moves closer to the BH as the magnetization increases, and the range of variation of the isodensity contours, whose upper ends become larger with decreasing $\beta_{\mathrm{m_c}}$. This behaviour is in complete agreement with that found for Kerr BHs in~\cite{Gimeno-Soler:2017} irrespective of the BH spin. For the particular case of Model VII, the maximum of the rest-mass density for the strongly magnetized case is significantly larger than for the other models and the spatial extent of the disk is fairly small. 

The size of the disks can be best quantified by plotting the radial profiles of the rest-mass density on the equatorial plane. This is shown in the upper panels of Fig.~\ref{radial_profiles} for models I, IV and VII and for the same three values of the magnetization parameter shown in Figs.~\ref{models_I} and~\ref{models_II}. (The lower panels of this figure correspond to disks around Kerr BHs and will be discussed below.) From Fig.~\ref{radial_profiles} we see that model I disks are significantly larger than models IV and VII, i.e.~the hairier the models the more compact and smaller the disks become. We also note the presence of an extended region of high density in the unmagnetized model VII  (the mildly-magnetized case also shows this feature but to a lesser extent). This could be related to the existence of an extra gravitational well due to the scalar field distribution that overlaps with the matter distribution of the disk (as can be seen in the right panel of Fig.~\ref{comparison_mass_density} below).

In figures~\ref{models_peri_I} and~\ref{models_peri_II} we show the same morphological distribution of Figs.~\ref{models_I} and~\ref{models_II} but using, instead, a perimeteral radial coordinate $R$, related to the radial coordinate $r$ according to $R = e^{F_2} r$. This perimeteral coordinate represents the proper length along the azimuthal direction, which constitutes a geometrically meaningful direction since it runs along the orbits of the azimuthal Killing vector field. Therefore, the proper size of a full $\phi$ orbit is given by $2\pi R$, i.e.~$R$ is the perimeteral radius. The most salient feature of the morphologies shown in Figs.~\ref{models_peri_I} and~\ref{models_peri_II}, when comparing to those displayed in Figs.~\ref{models_I} and~\ref{models_II}, is the deformation of the disks in their innermost regions. In general, the deformations become larger the higher the horizon sphericity $\mathfrak{s}$ and the closer the disk is to the horizon. Model III is the one showing the largest deformation, as $(R_{\rm in}-R_{\rm H})/R_{\rm H}$ attains the smallest value for this model. It is also worth noticing that the shape of the BH also changes when using the perimeteral coordinate. While in the $r$ coordinate the horizon is spherical (cf.~Figs.~\ref{models_I} and~\ref{models_II}) in the perimeteral coordinate $R$ is not always so. Moreover, the larger the value of $v_{\rm H}$, the more elliptic the horizon becomes, which in our sample corresponds to model III (cf.~Table~\ref{models_list}, $\mathfrak{s}=1.489$). 

In addition, an interesting geometrical property of the perimeteral coordinates is that, for the Kerr metric, $R_{\mathrm{H}} = 2M$ irrespective of the value of the angular momentum. However, for the KBHsSH cases, $2M_{\mathrm{H}} < R_{\mathrm{H}} < 2M_{\mathrm{ADM}}$, and the quotient $R_{\mathrm{H}}/ 2M_{\mathrm{H}}$ increases as more mass and angular momentum is stored in the scalar field. 

Table~\ref{HBH_disk_parameters} reports the relevant physical quantities for all of our disk models around KBHsSH. It is worth mentioning that KBHsSH can violate the Kerr bound for the potential $\Delta W \equiv W_{\mathrm{in}} - W_{\mathrm{c}}$. As shown in~\cite{Abramowicz:1978}, constant angular momentum disks arround Kerr BHs exhibit a maximum for $|\Delta W|$ when the spin parameter $a\rightarrow 1$. This value is $\Delta W_{\mathrm{max}} = -\frac{1}{2} \ln 3 \simeq -0.549$. Models V, VI, and VII of our sample violate that bound. As a result, the maximum values of the fluid quantities for disks around KBHsSH are significantly larger than in the Kerr BH case. In both cases, these values increase as $|\Delta W|$ increases, irrespective of the magnetization, as shown in Table~\ref{HBH_disk_parameters}.

\begin{figure*}
\centering
\includegraphics[scale=0.14]{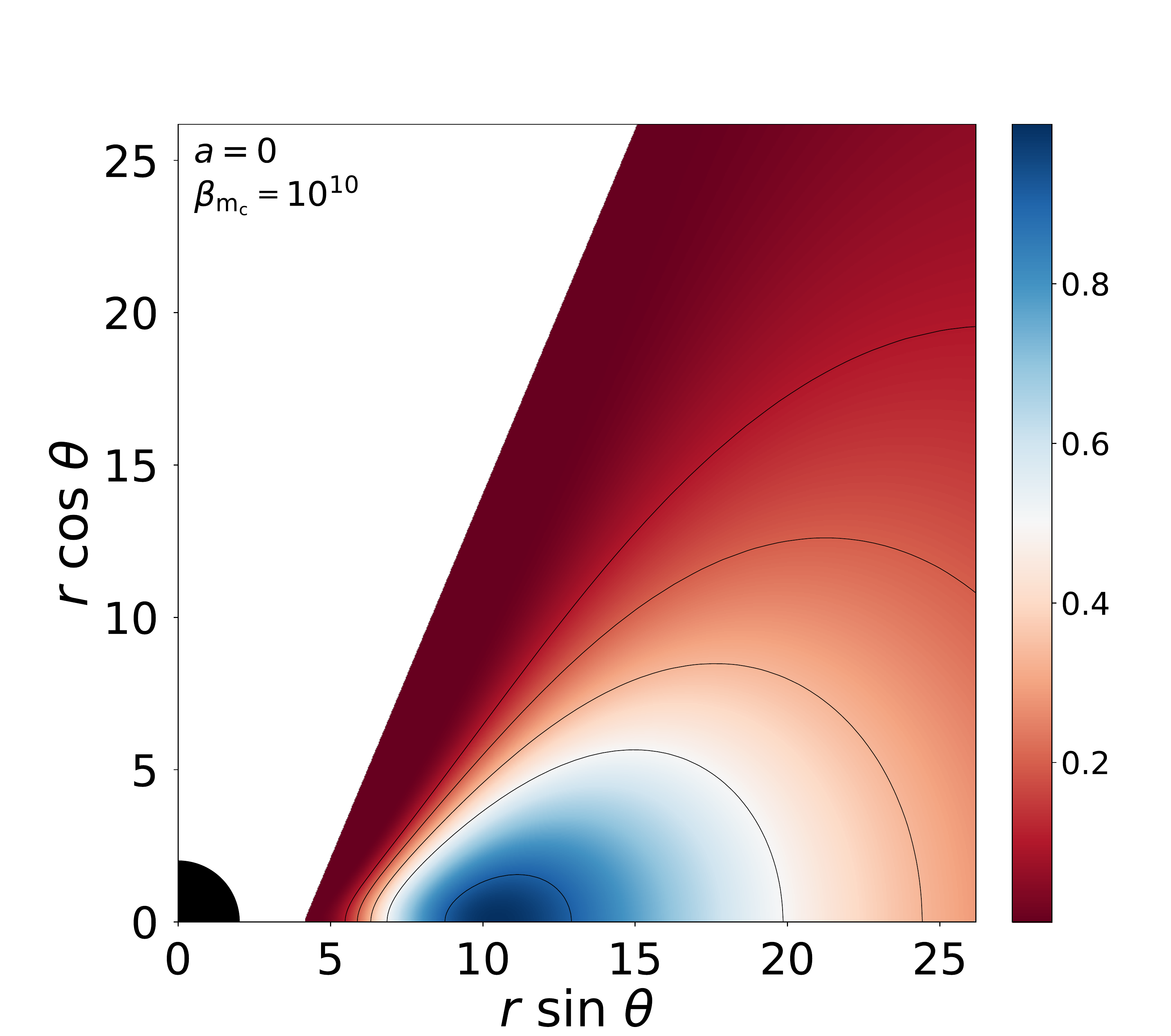}
\hspace{-0.3cm}
\includegraphics[scale=0.14]{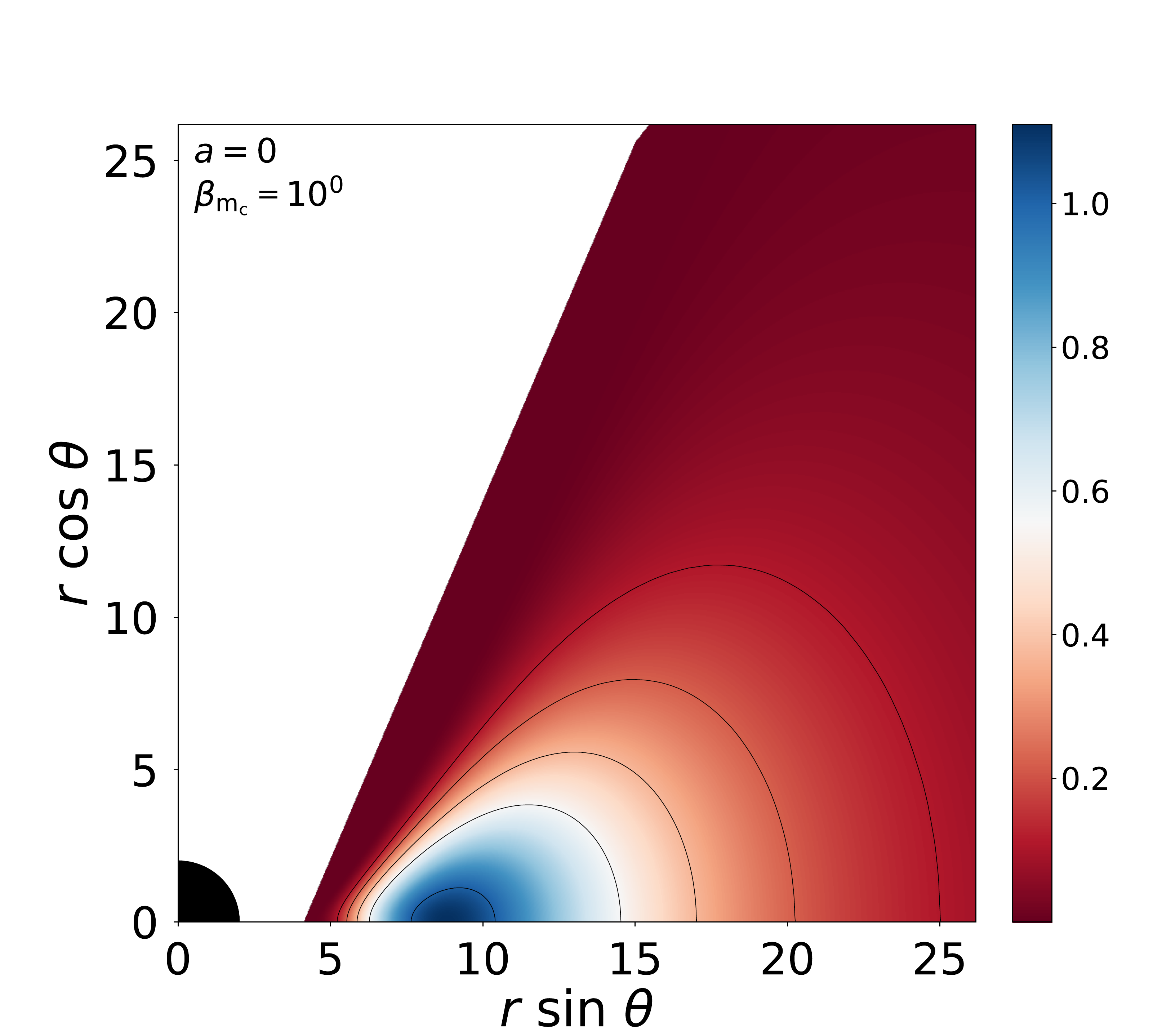}
\hspace{-0.2cm}
\includegraphics[scale=0.14]{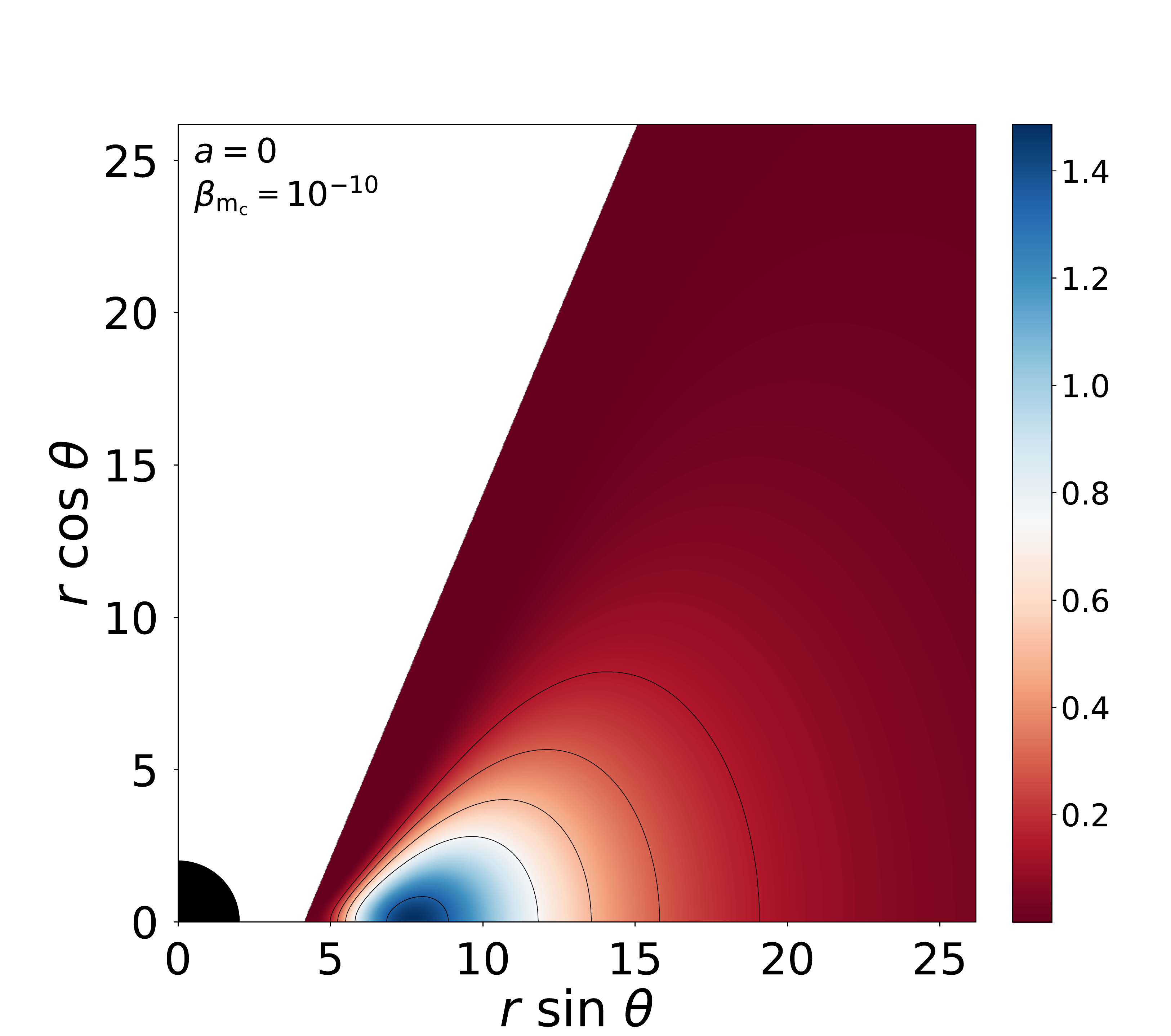}
\\
\includegraphics[scale=0.14]{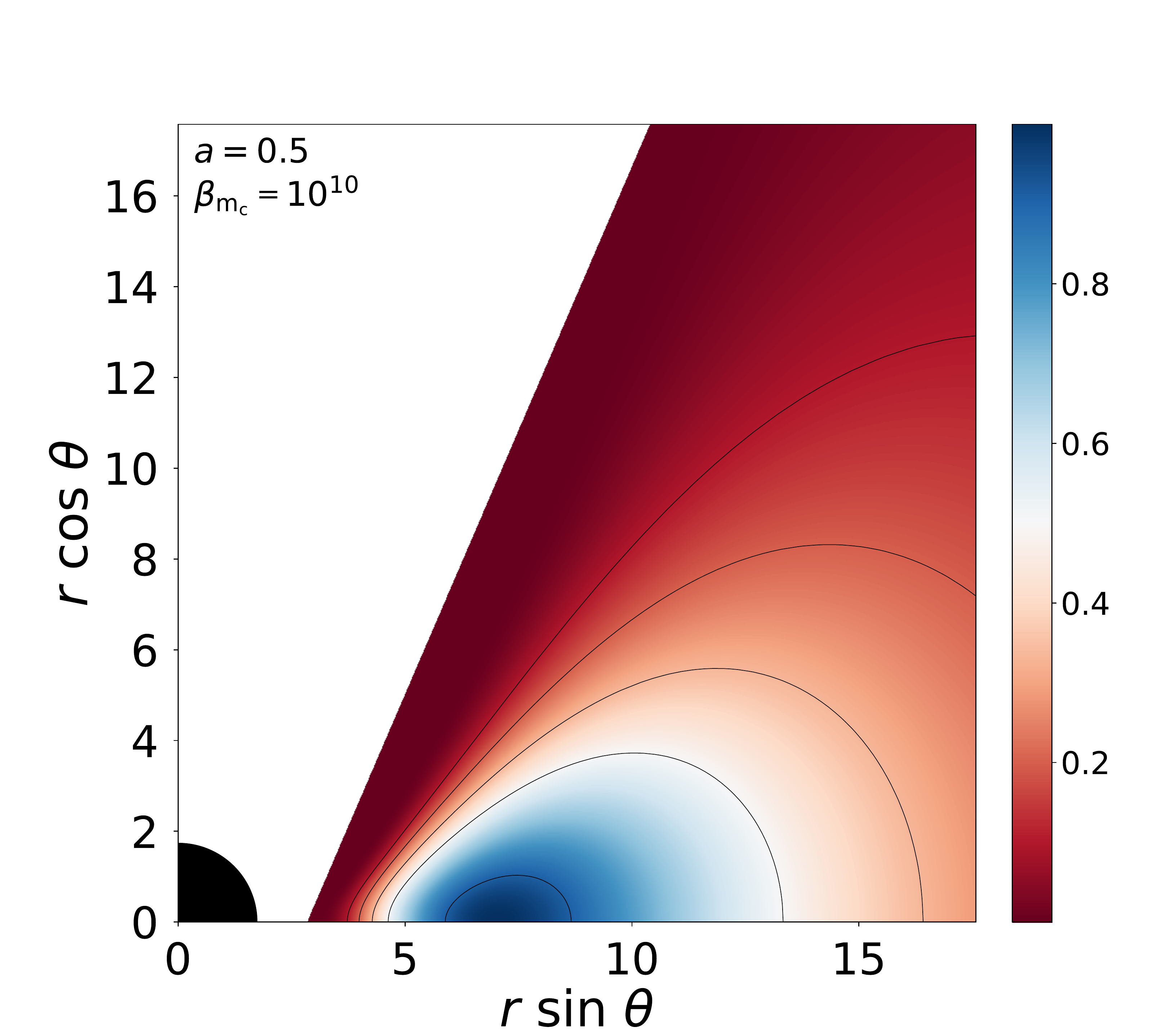}
\hspace{-0.3cm}
\includegraphics[scale=0.14]{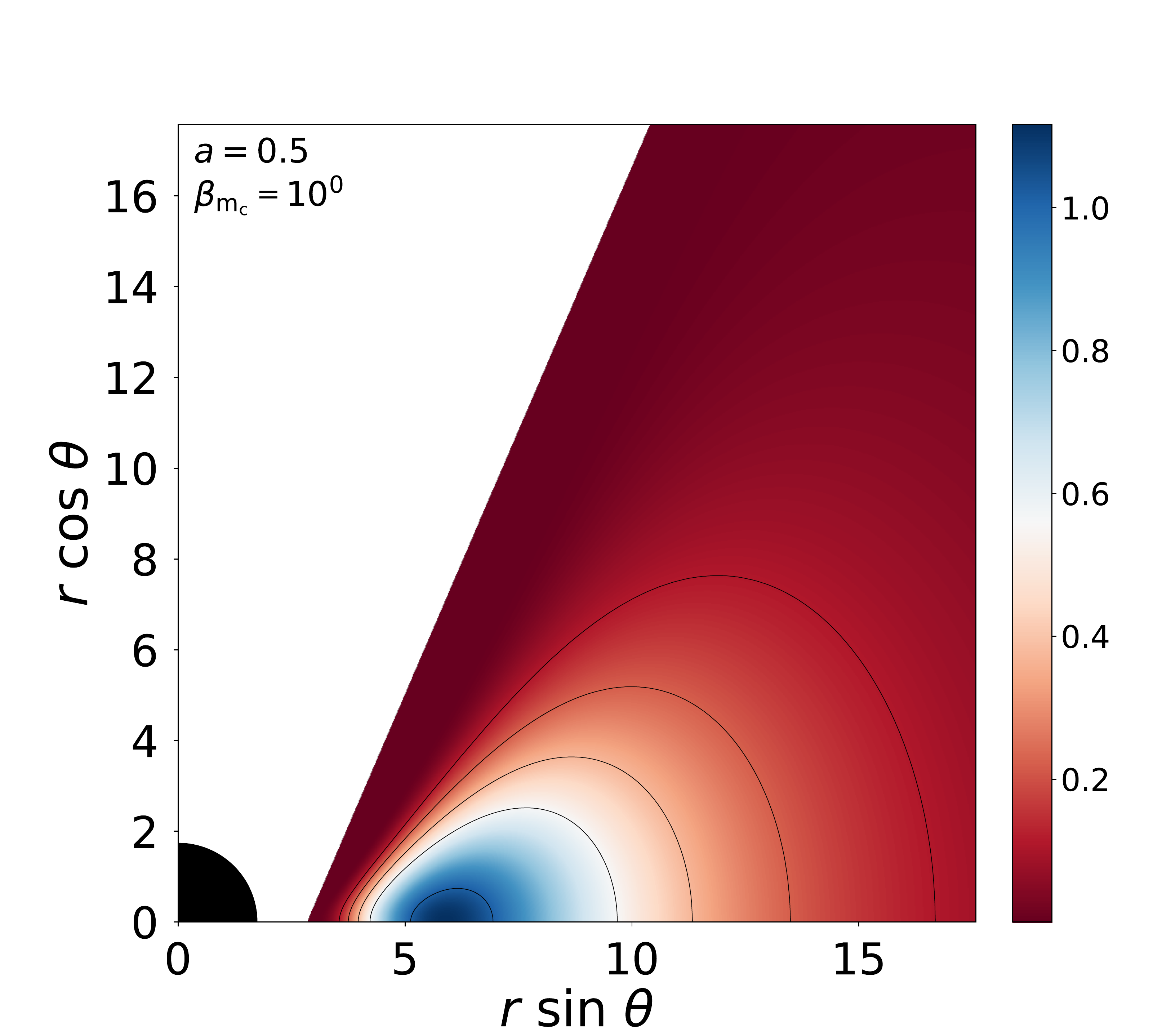}
\hspace{-0.2cm}
\includegraphics[scale=0.14]{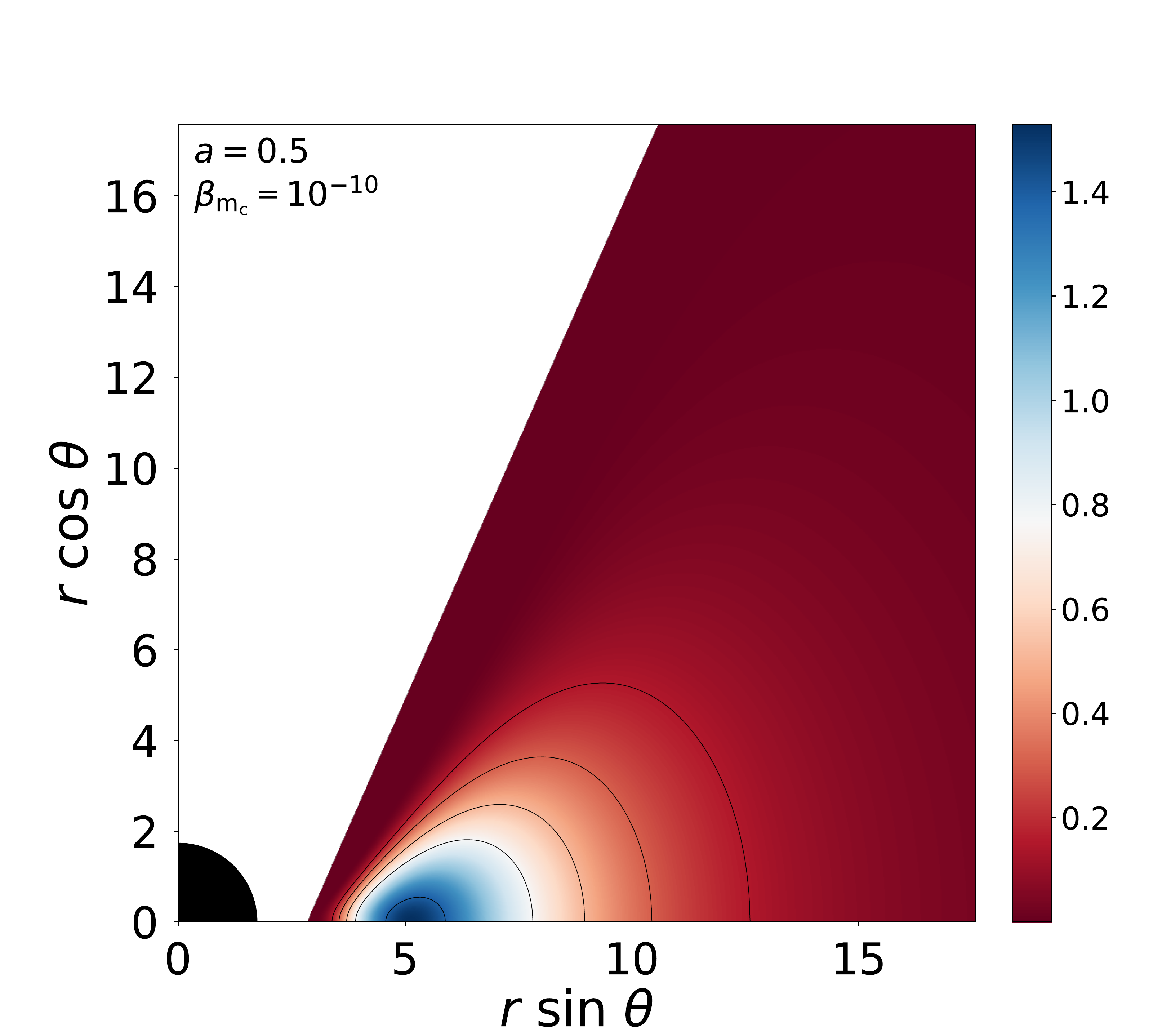}
\\
\includegraphics[scale=0.14]{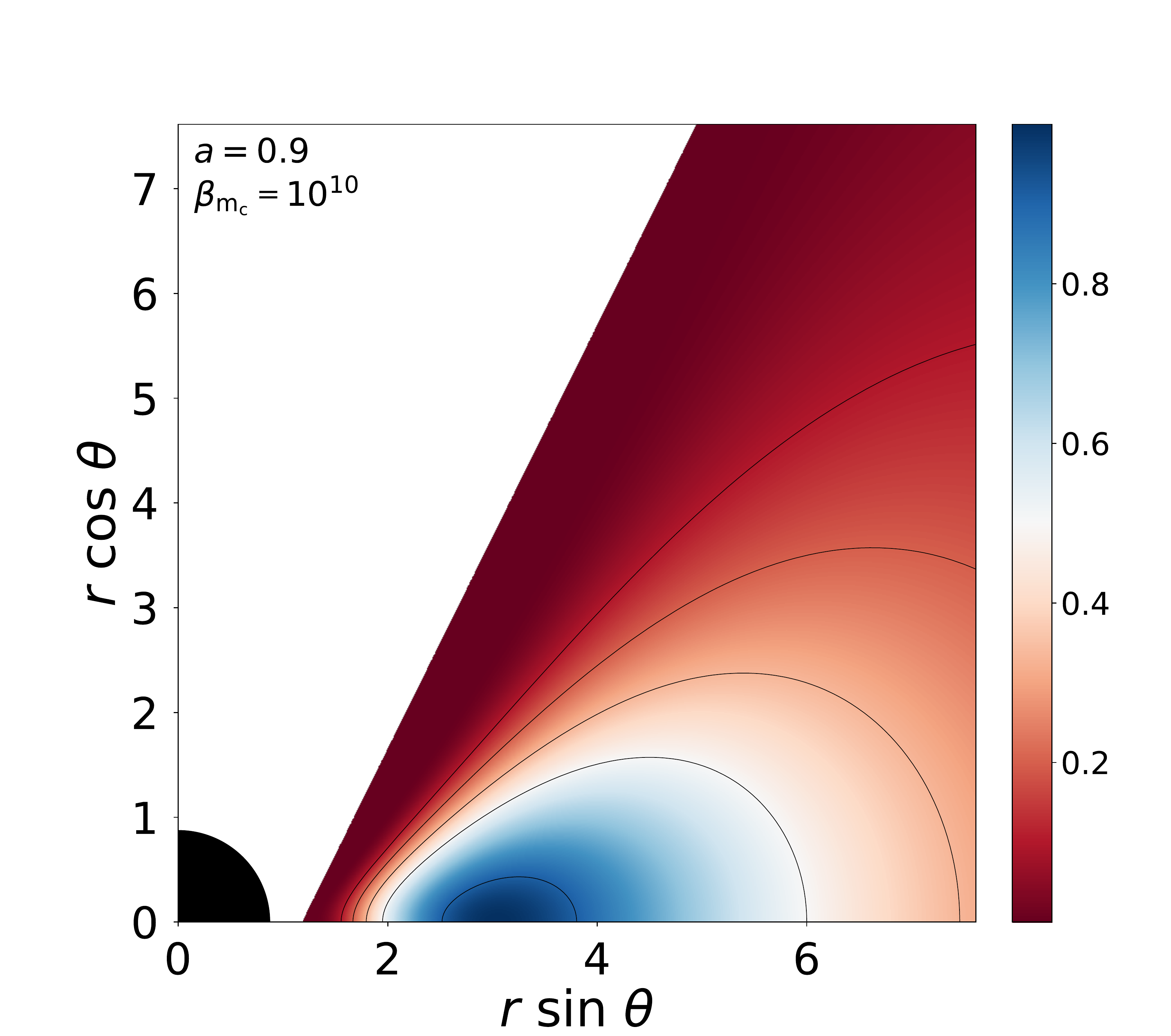}
\hspace{-0.3cm}
\includegraphics[scale=0.14]{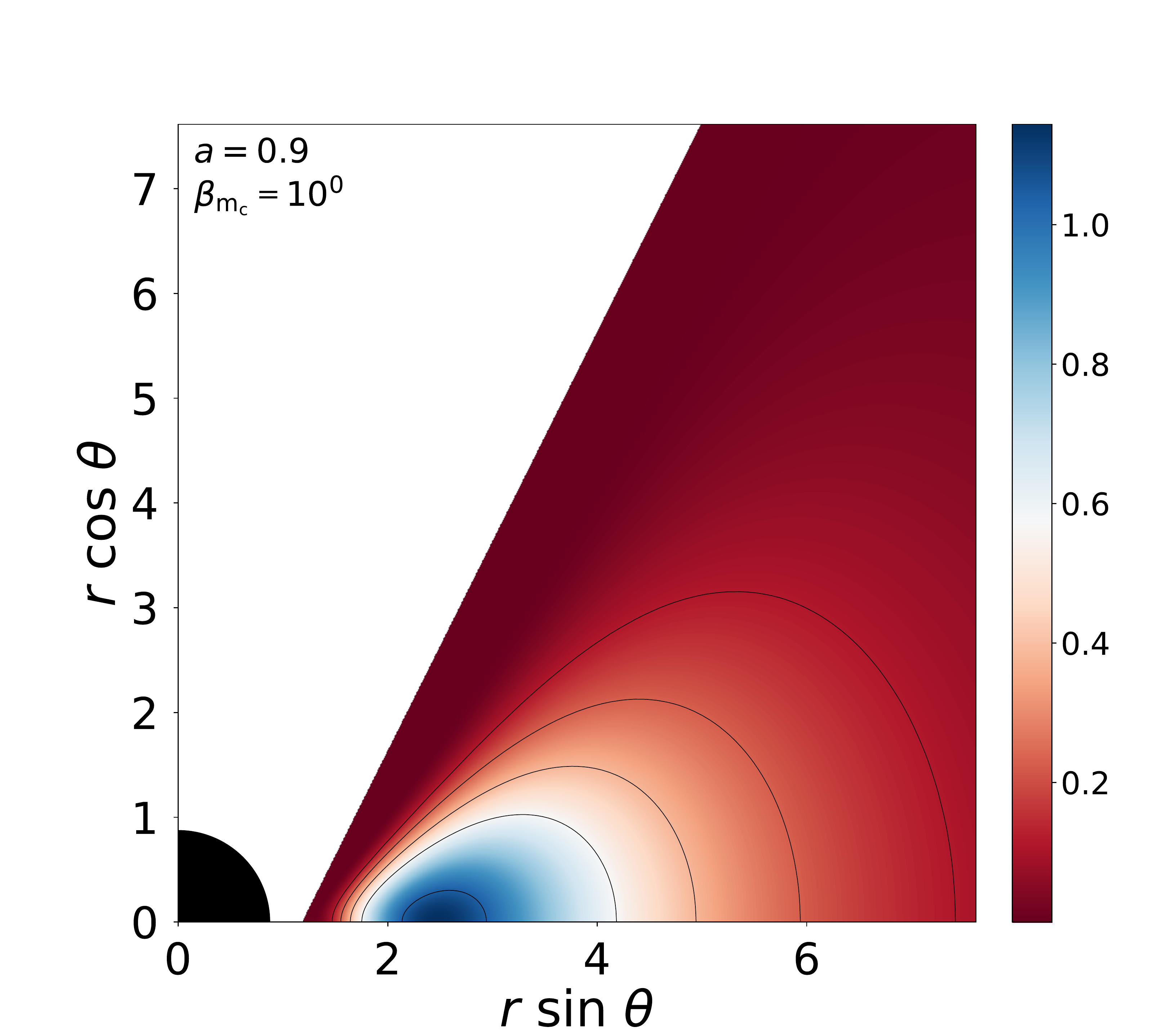}
\hspace{-0.2cm}
\includegraphics[scale=0.14]{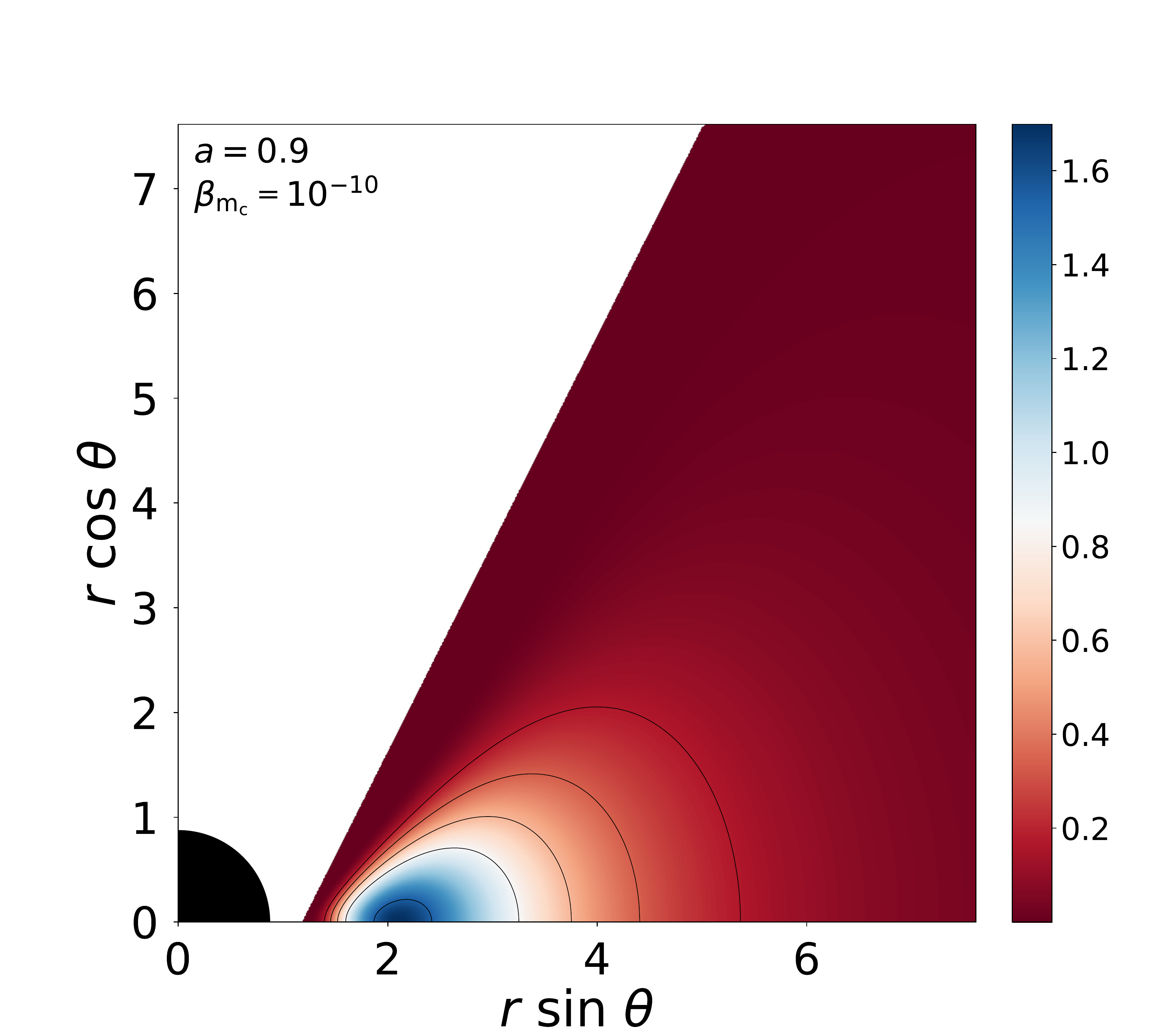}
\\
\includegraphics[scale=0.14]{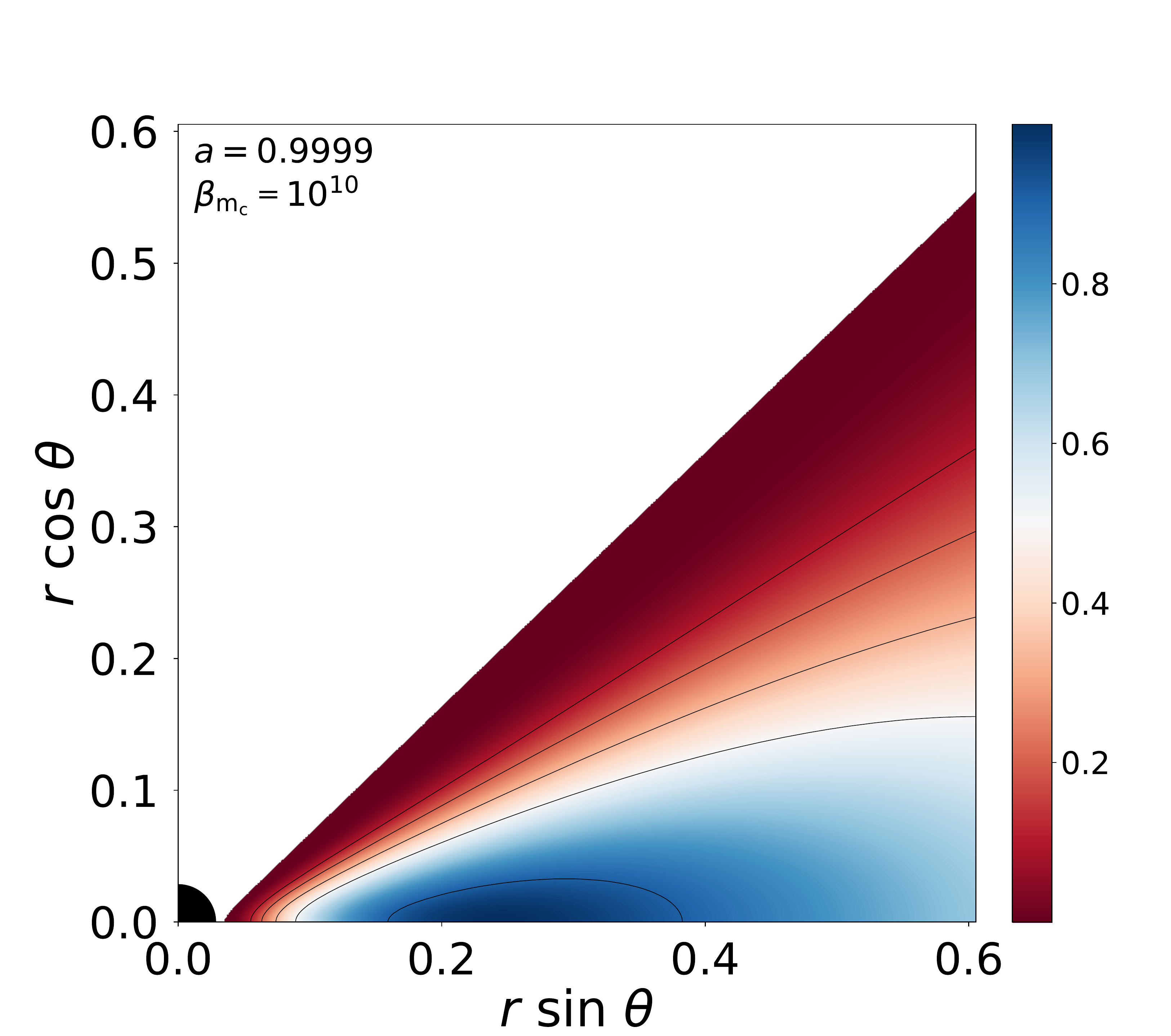}
\hspace{-0.3cm}
\includegraphics[scale=0.14]{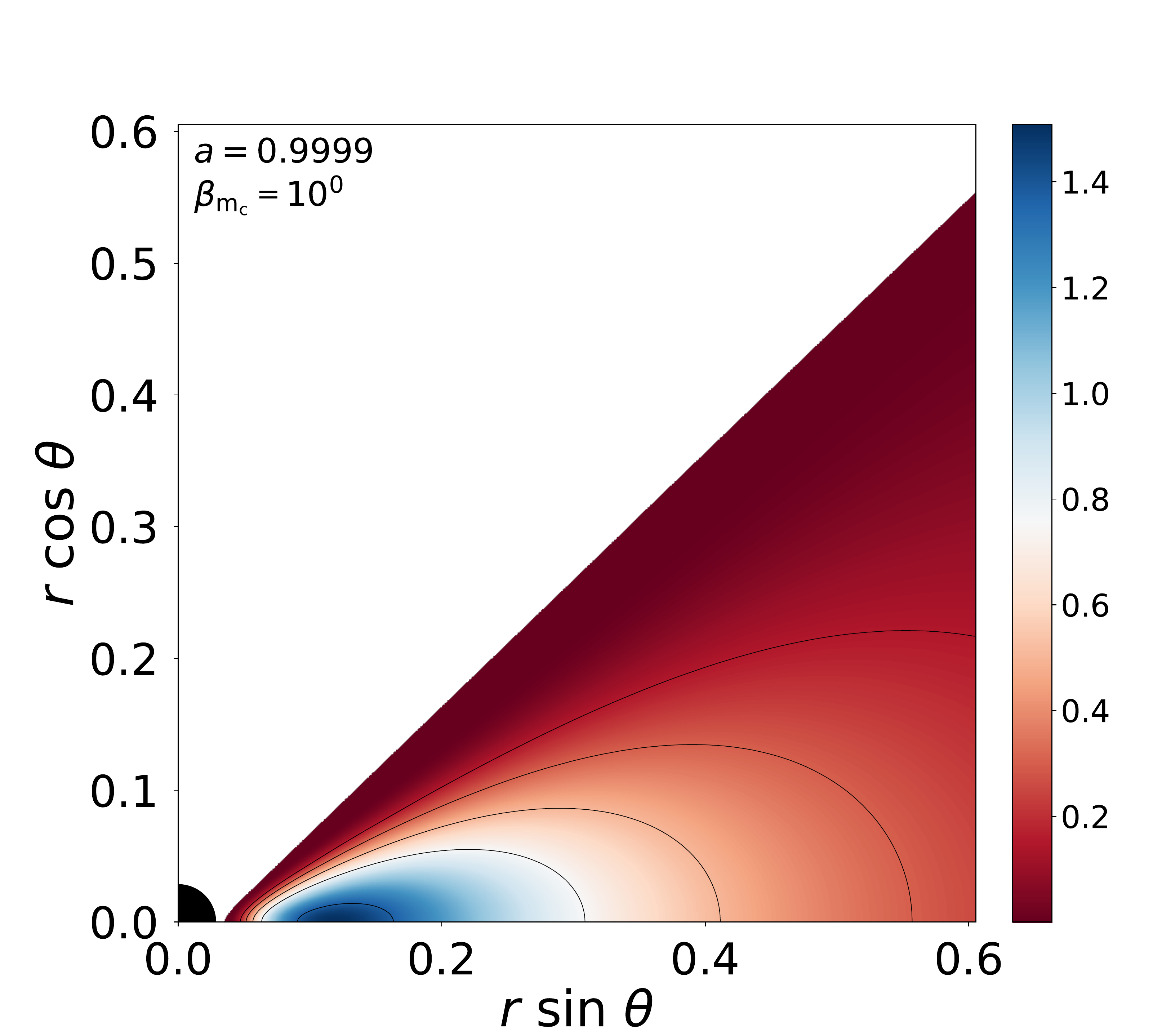}
\hspace{-0.2cm}
\includegraphics[scale=0.14]{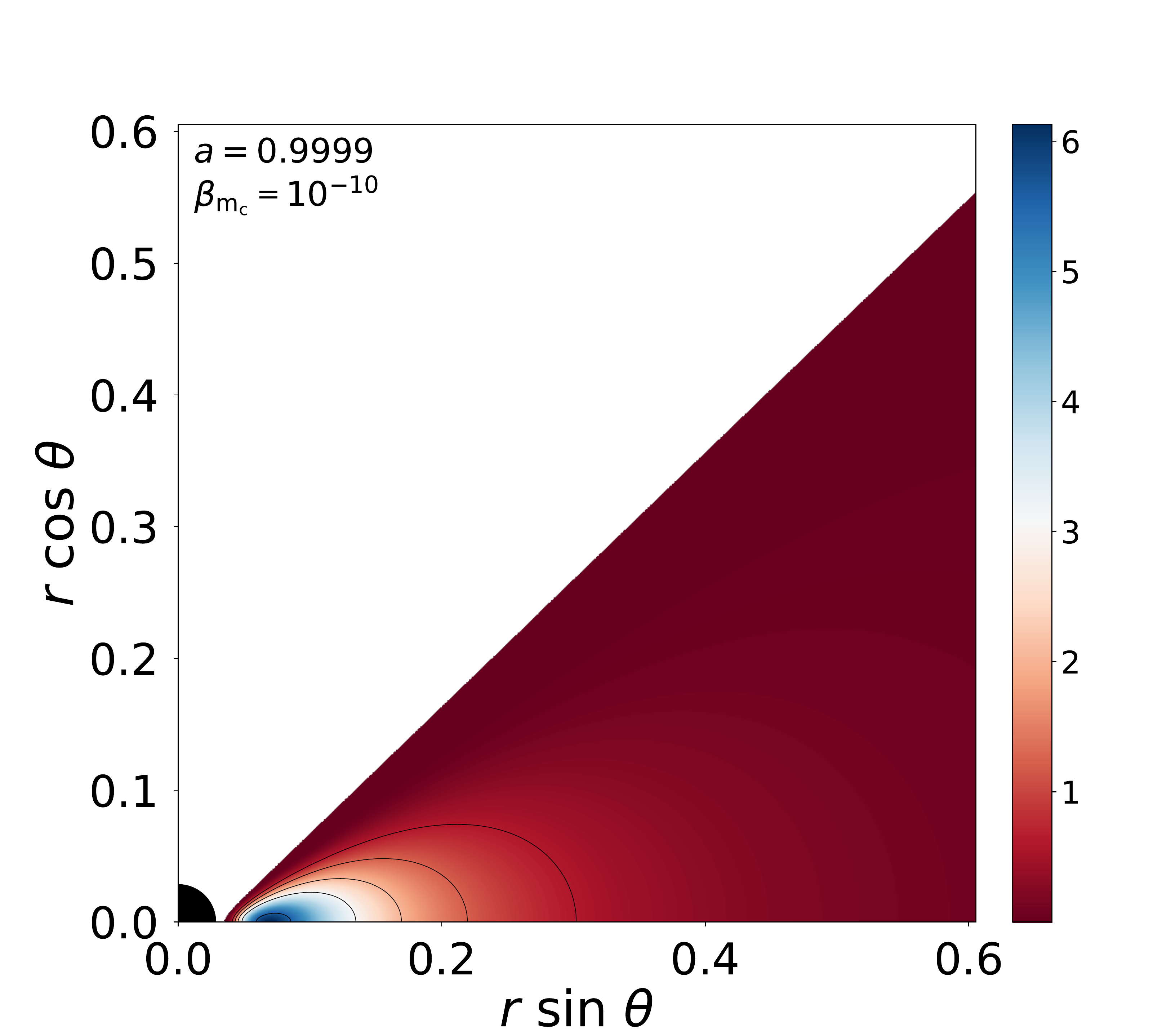}
\hspace{-0.2cm}
\caption{Rest-mass density distribution. From top to bottom the rows correspond to a sequence of Kerr BHs with increasing spin parameter $a$ (0, 0.5, 0.9 and 0.9999). From left to right the columns correspond to different values of the magnetization parameter, namely non-magnetized ($\beta_{\mathrm{m}_{\mathrm{c}}} = 10^{10}$), mildly magnetized ($\beta_{\mathrm{m}_{\mathrm{c}}} = 1$) and strongly magnetized ($\beta_{\mathrm{m}_{\mathrm{c}}} = 10^{-10}$)}
\label{models_Kerr}
\end{figure*}

\begin{figure*}
\centering
\includegraphics[scale=0.14]{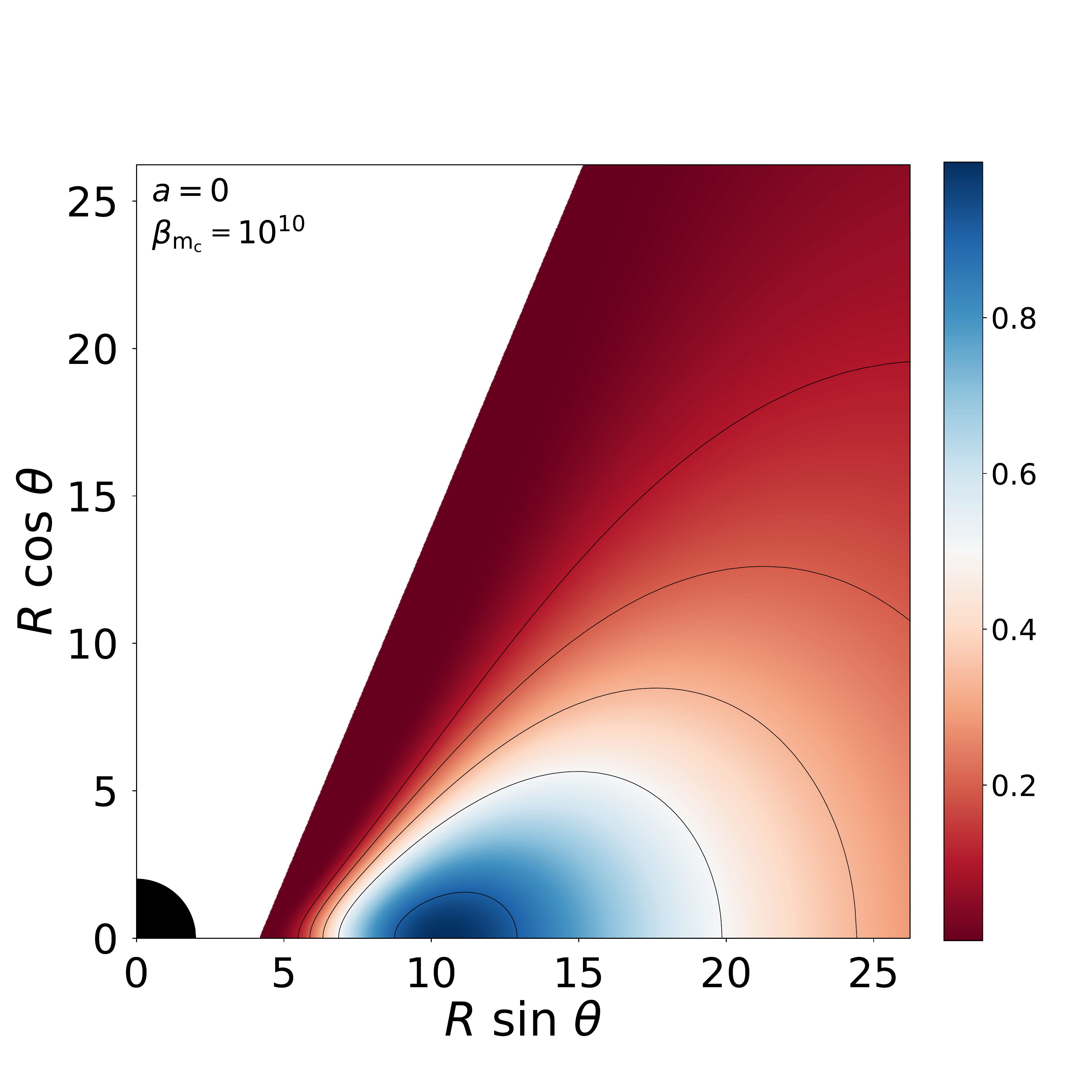}
\hspace{-0.3cm}
\includegraphics[scale=0.14]{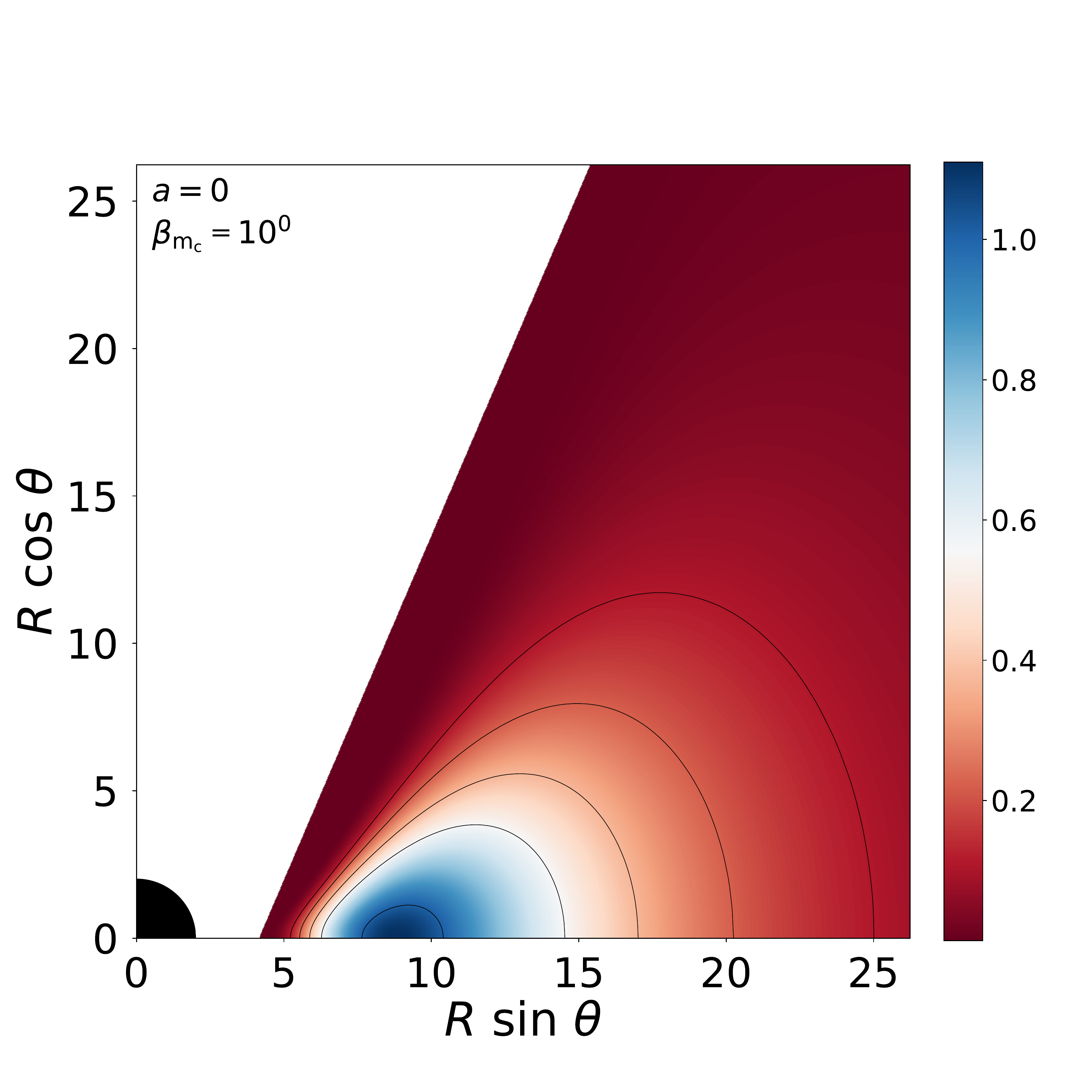}
\hspace{-0.2cm}
\includegraphics[scale=0.14]{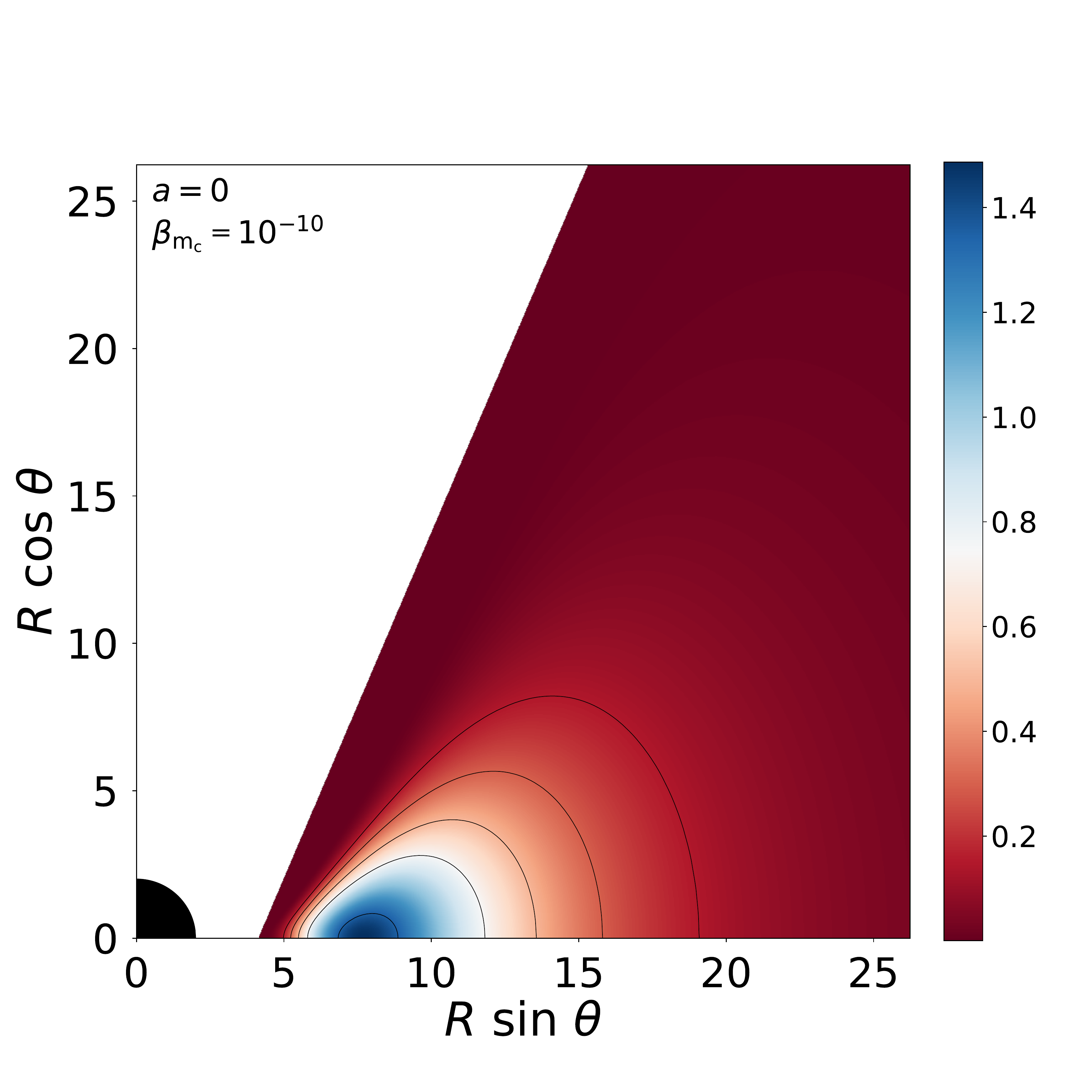}
\\
\includegraphics[scale=0.14]{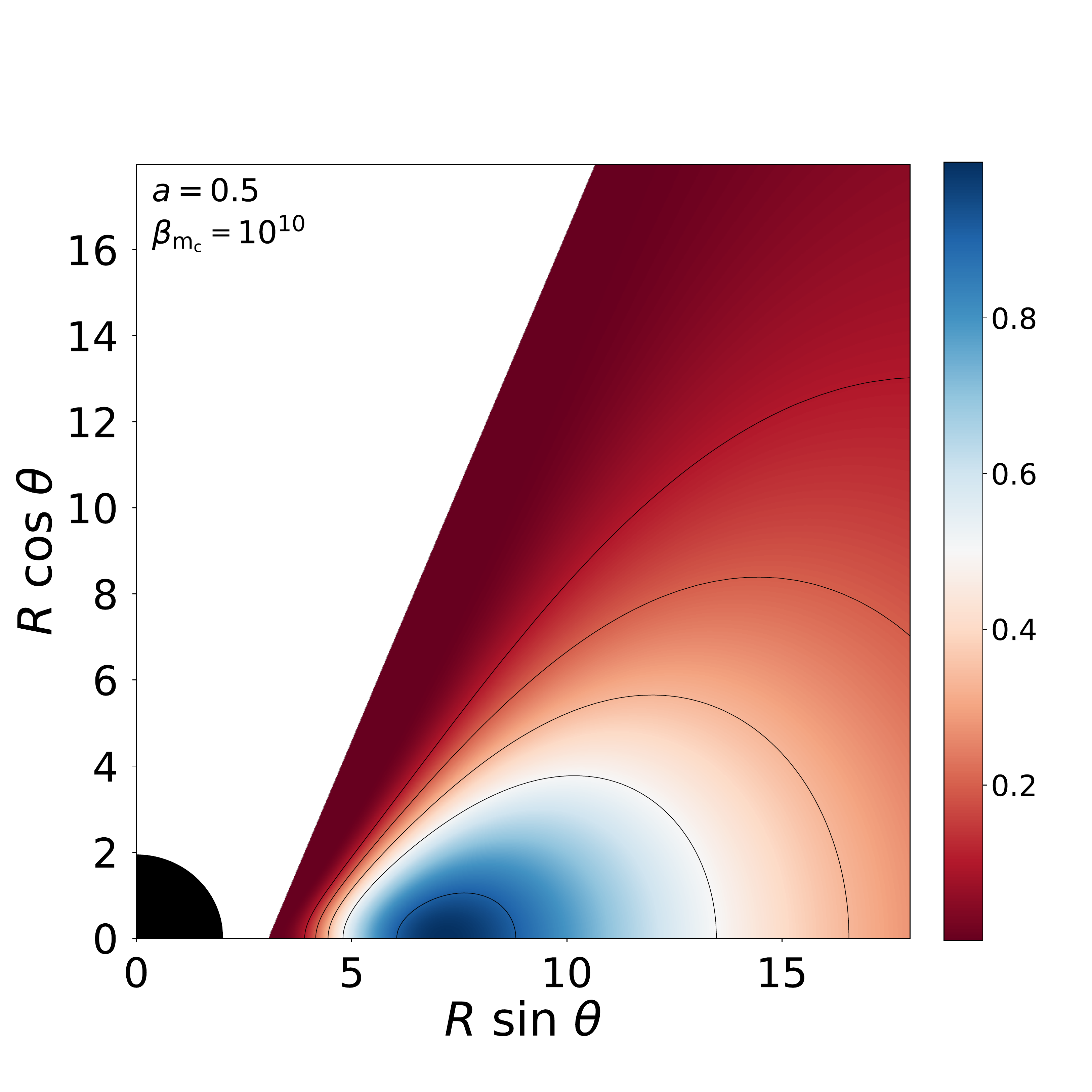}
\hspace{-0.3cm}
\includegraphics[scale=0.14]{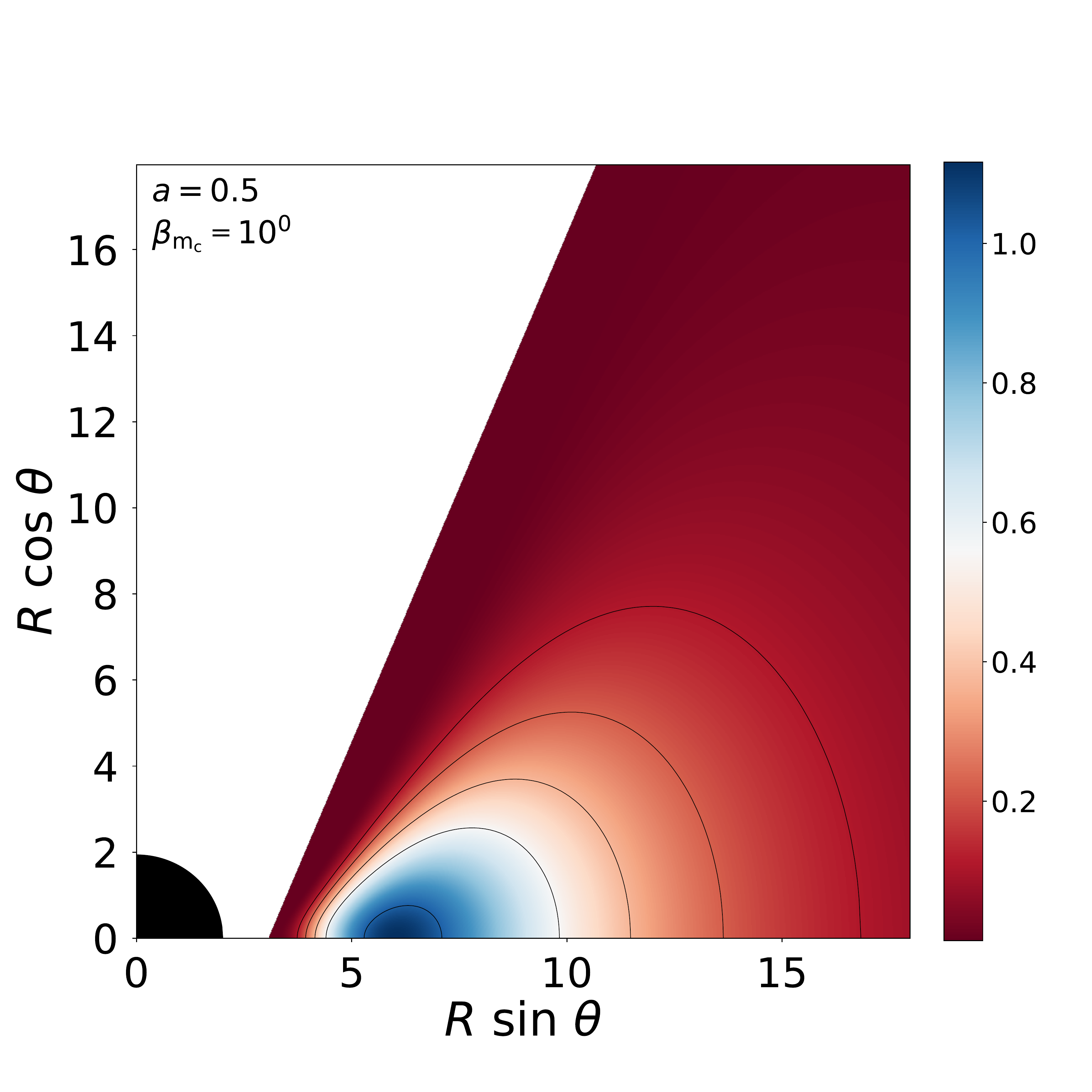}
\hspace{-0.2cm}
\includegraphics[scale=0.14]{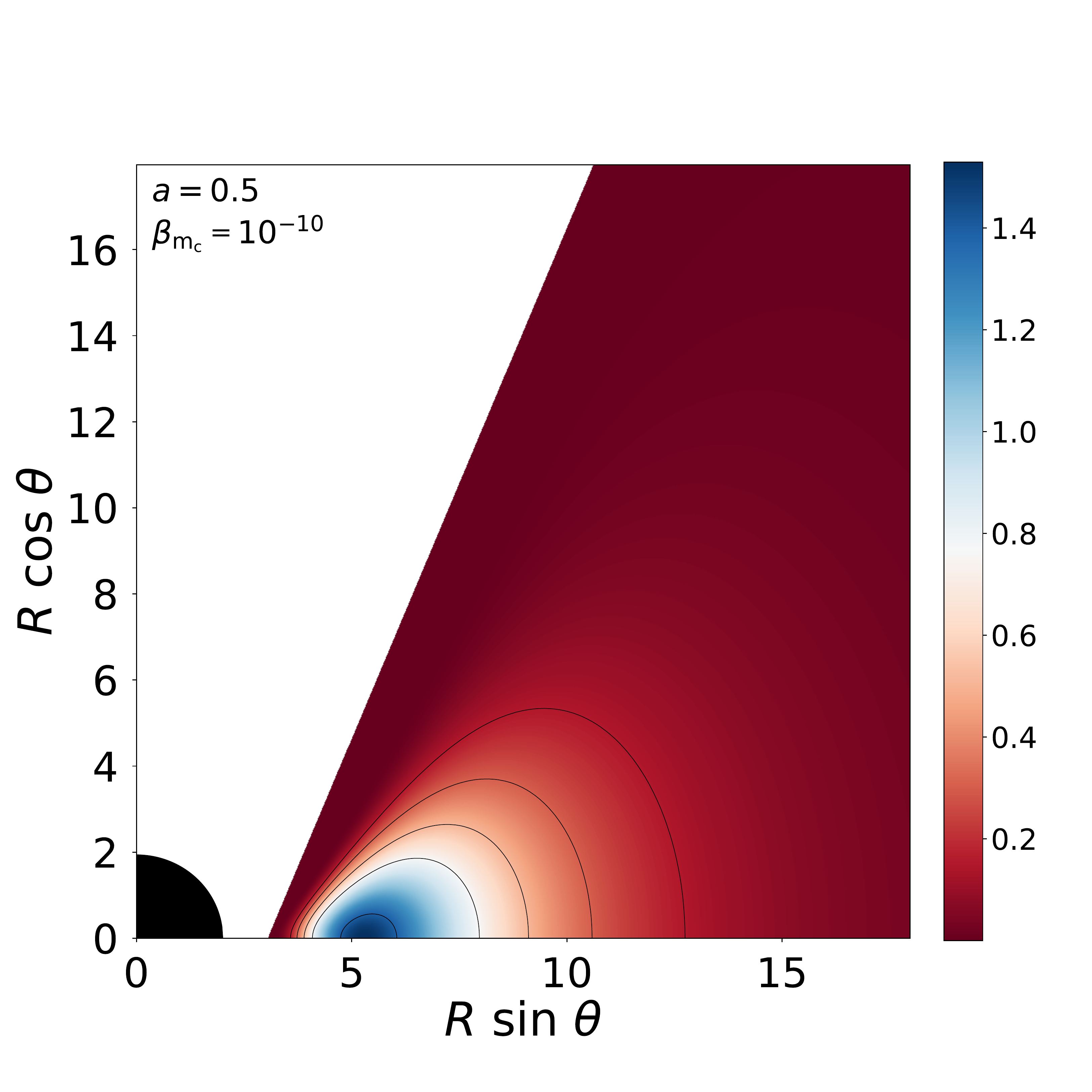}
\\
\includegraphics[scale=0.14]{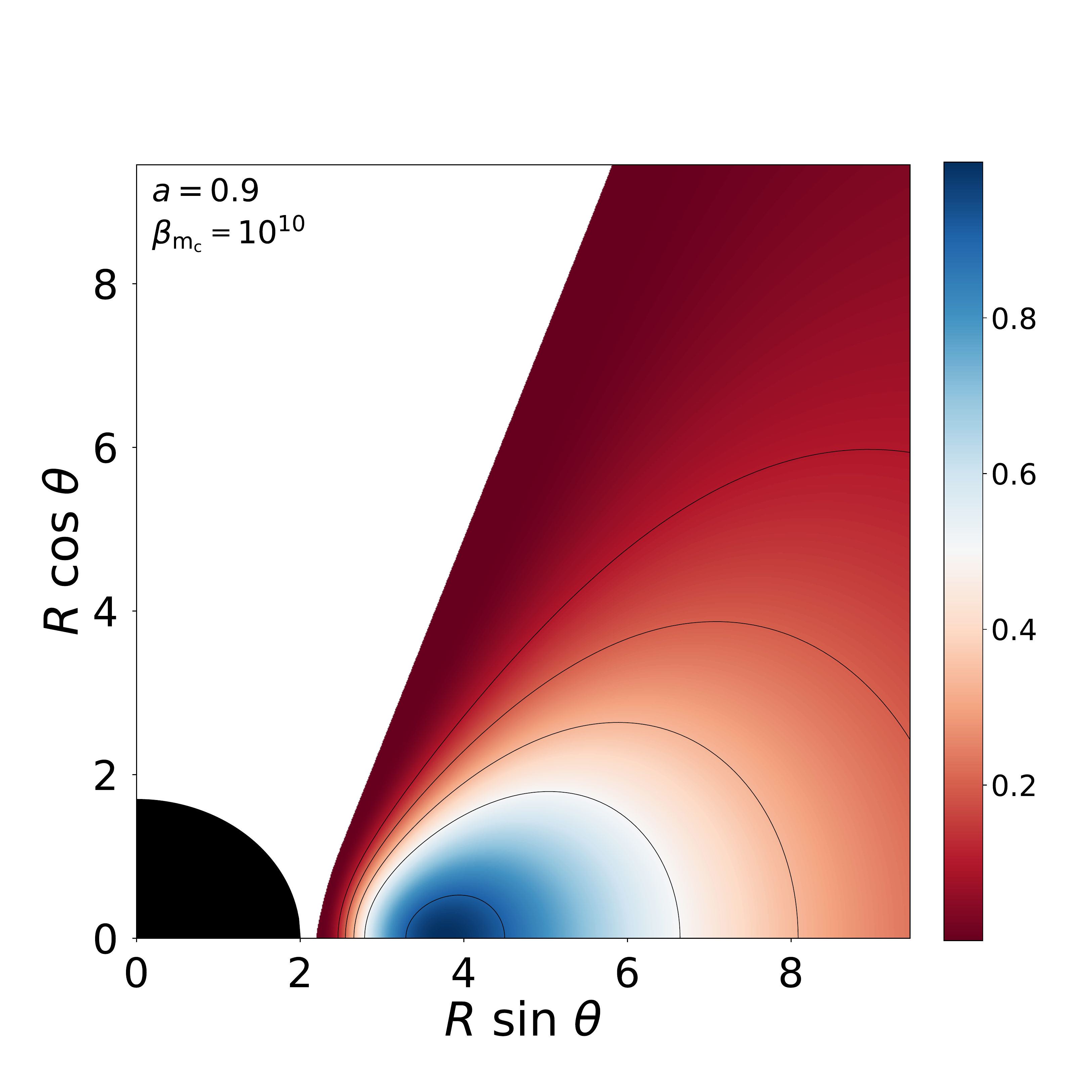}
\hspace{-0.3cm}
\includegraphics[scale=0.14]{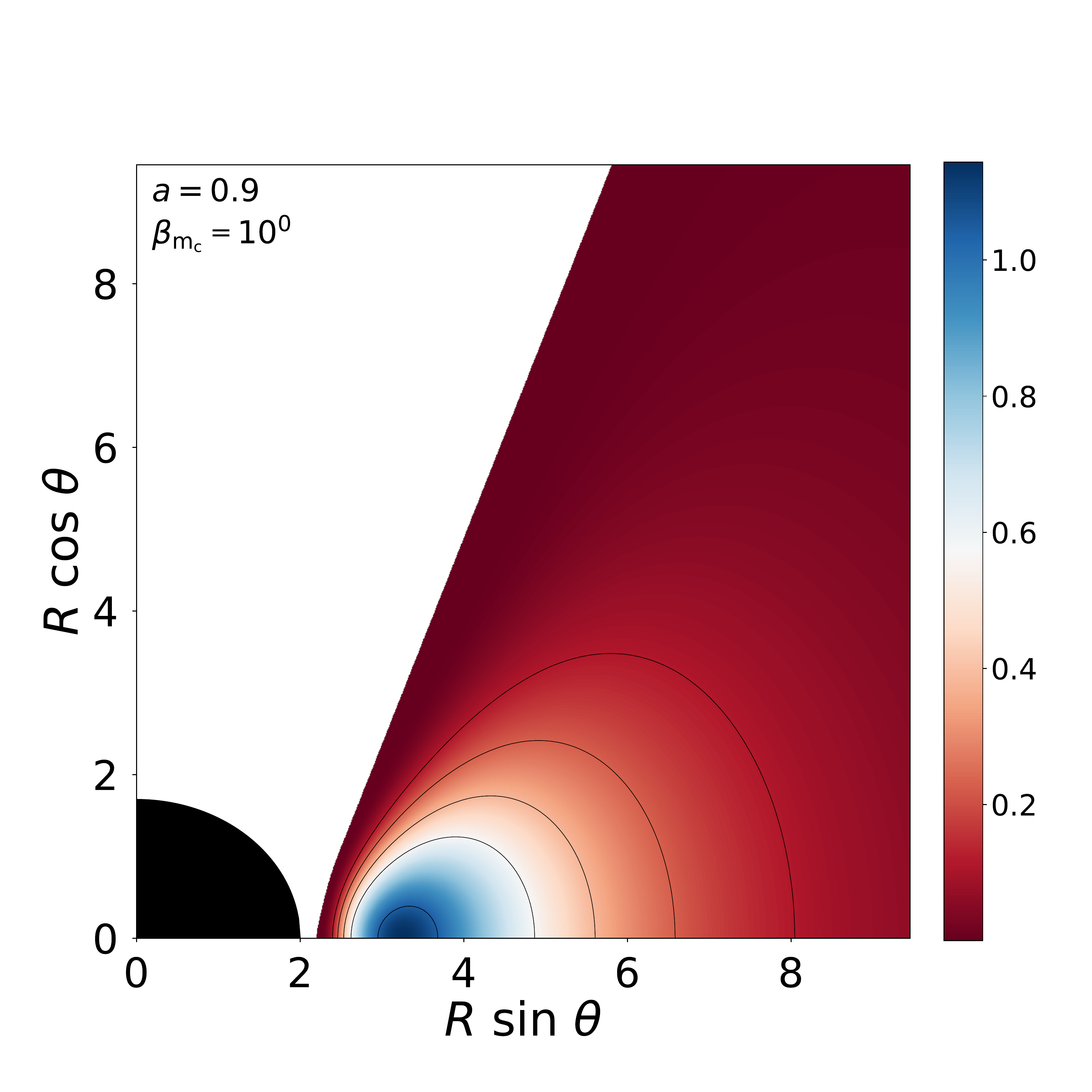}
\hspace{-0.2cm}
\includegraphics[scale=0.14]{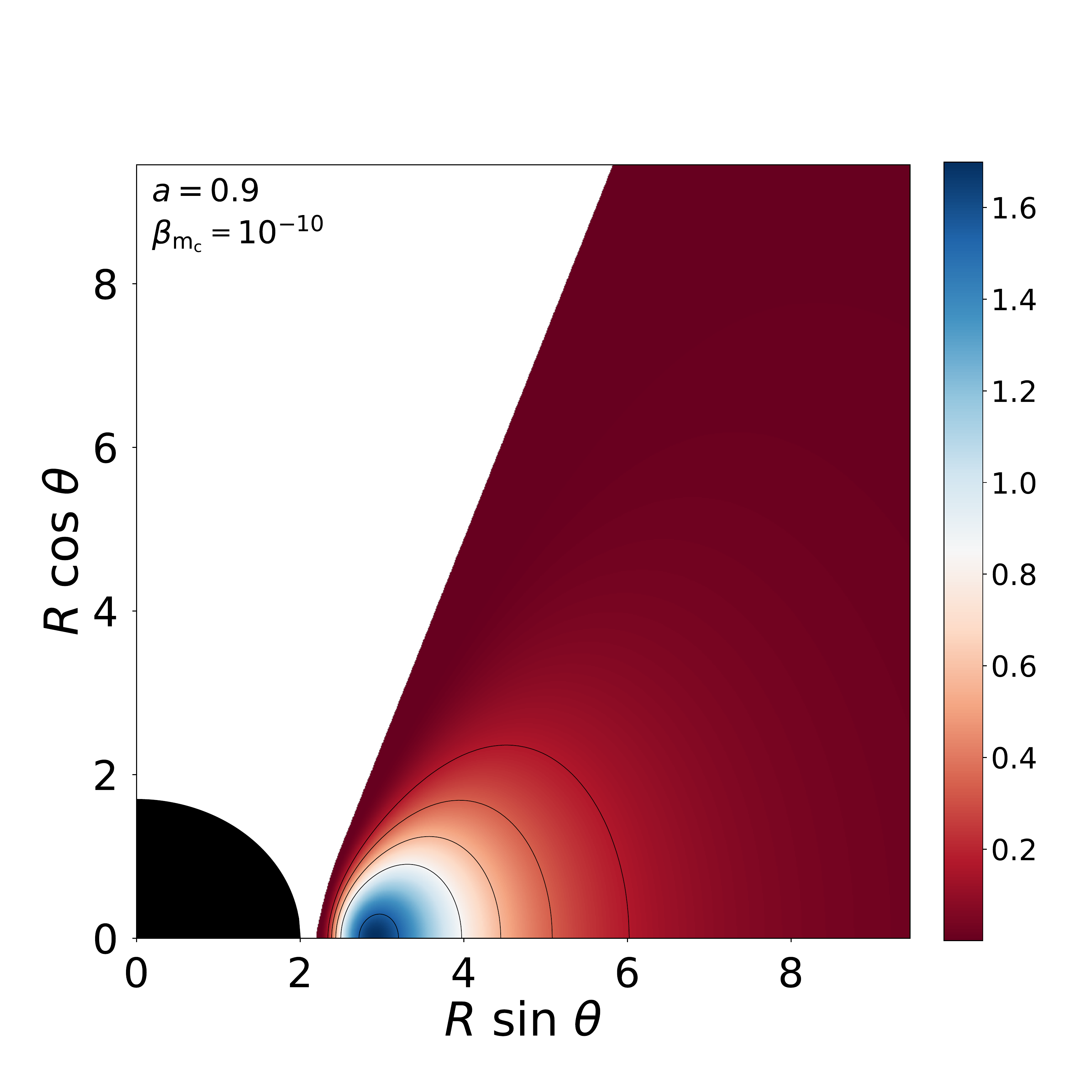}
\\
\includegraphics[scale=0.14]{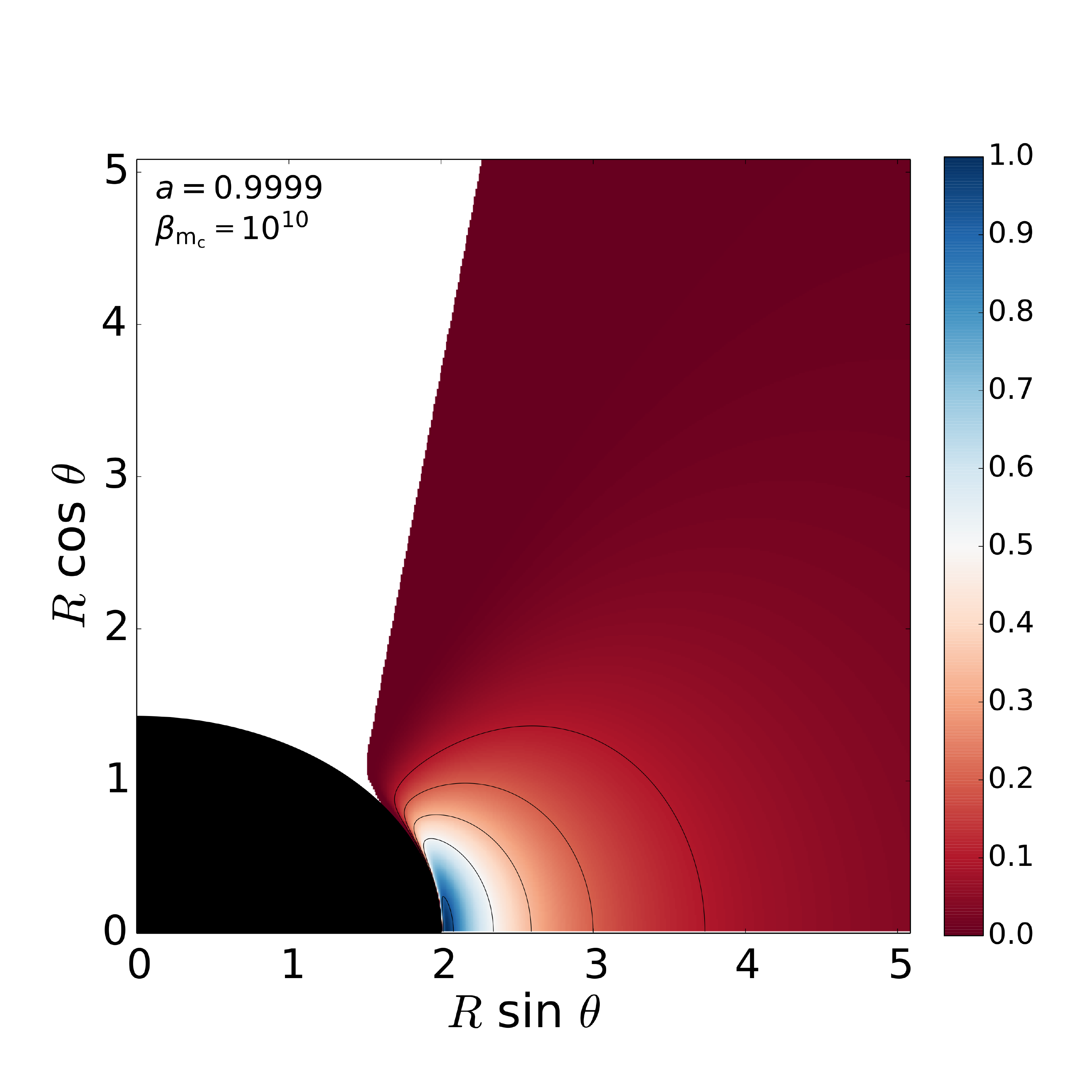}
\hspace{-0.3cm}
\includegraphics[scale=0.14]{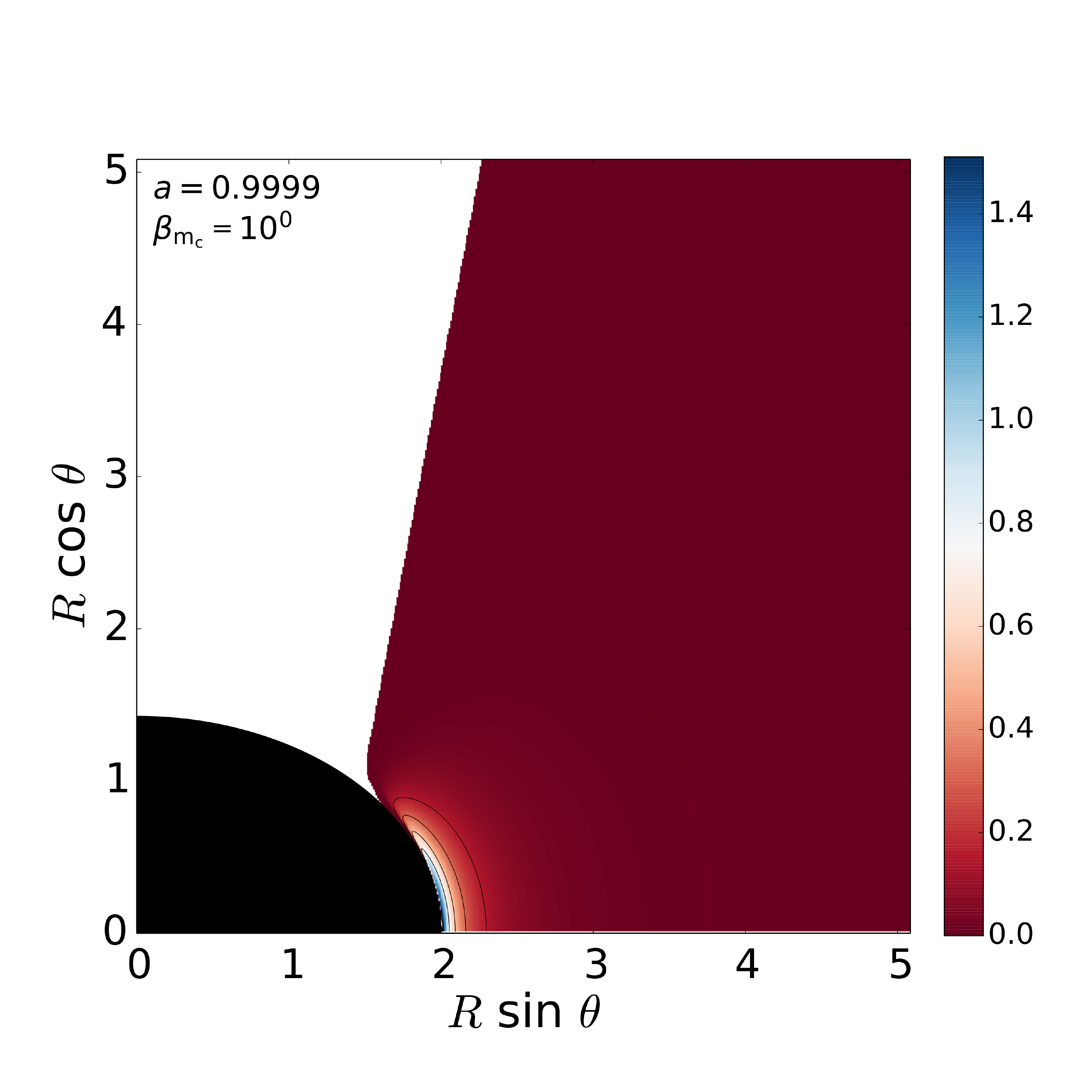}
\hspace{-0.2cm}
\includegraphics[scale=0.14]{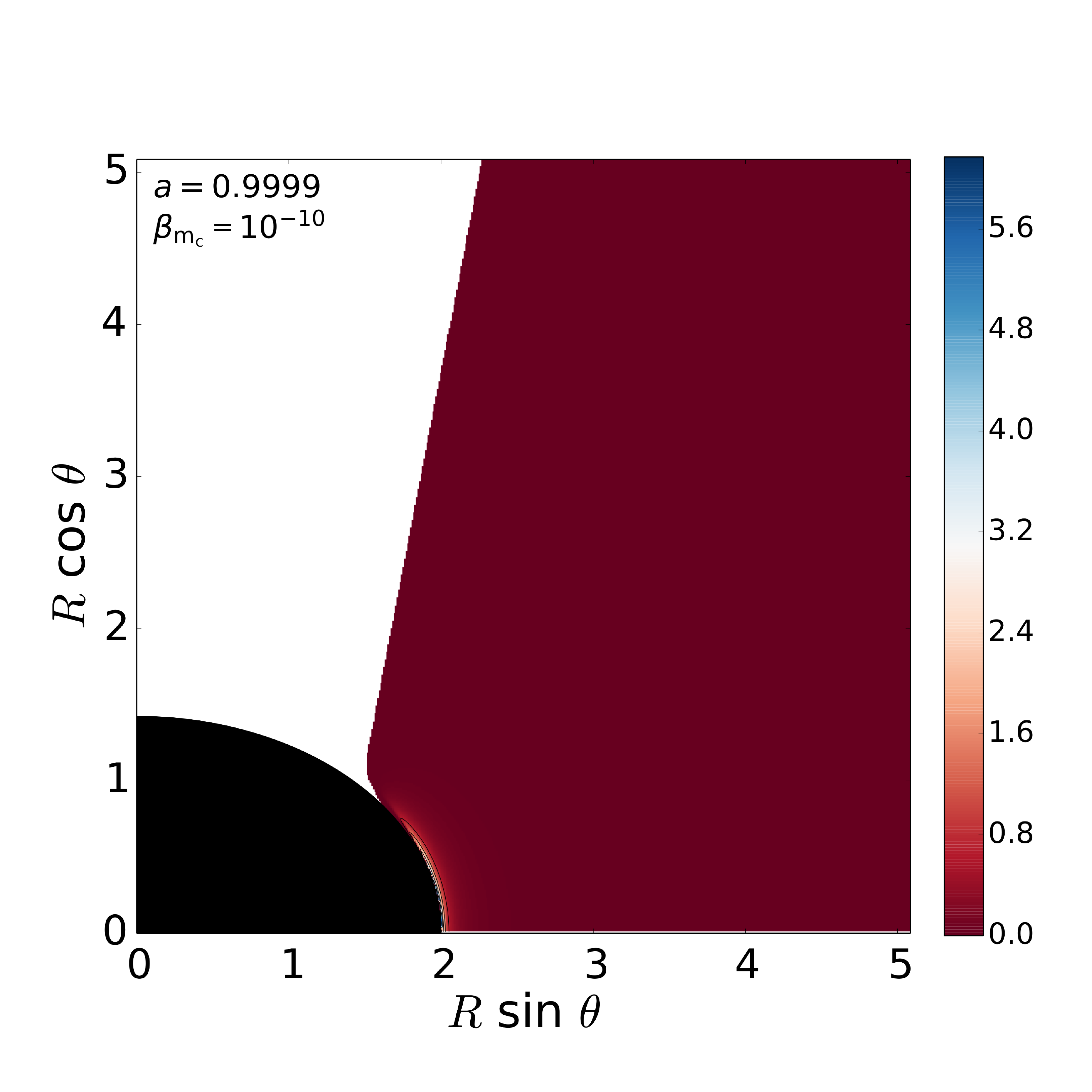}
\hspace{-0.2cm}
\caption{Rest-mass density distribution using perimeteral coordinates. From top to bottom the rows correspond to a sequence of Kerr BHs with increasing spin parameter $a$ (0, 0.5, 0.9 and 0.9999). From left to right the columns correspond to different values of the magnetization parameter, namely non-magnetized ($\beta_{\mathrm{m}_{\mathrm{c}}} = 10^{10}$), mildly magnetized ($\beta_{\mathrm{m}_{\mathrm{c}}} = 1$) and strongly magnetized ($\beta_{\mathrm{m}_{\mathrm{c}}} = 10^{-10}$)}
\label{models_Kerr_peri}
\end{figure*}

\begin{figure*}
\centering
\includegraphics[scale=0.2]{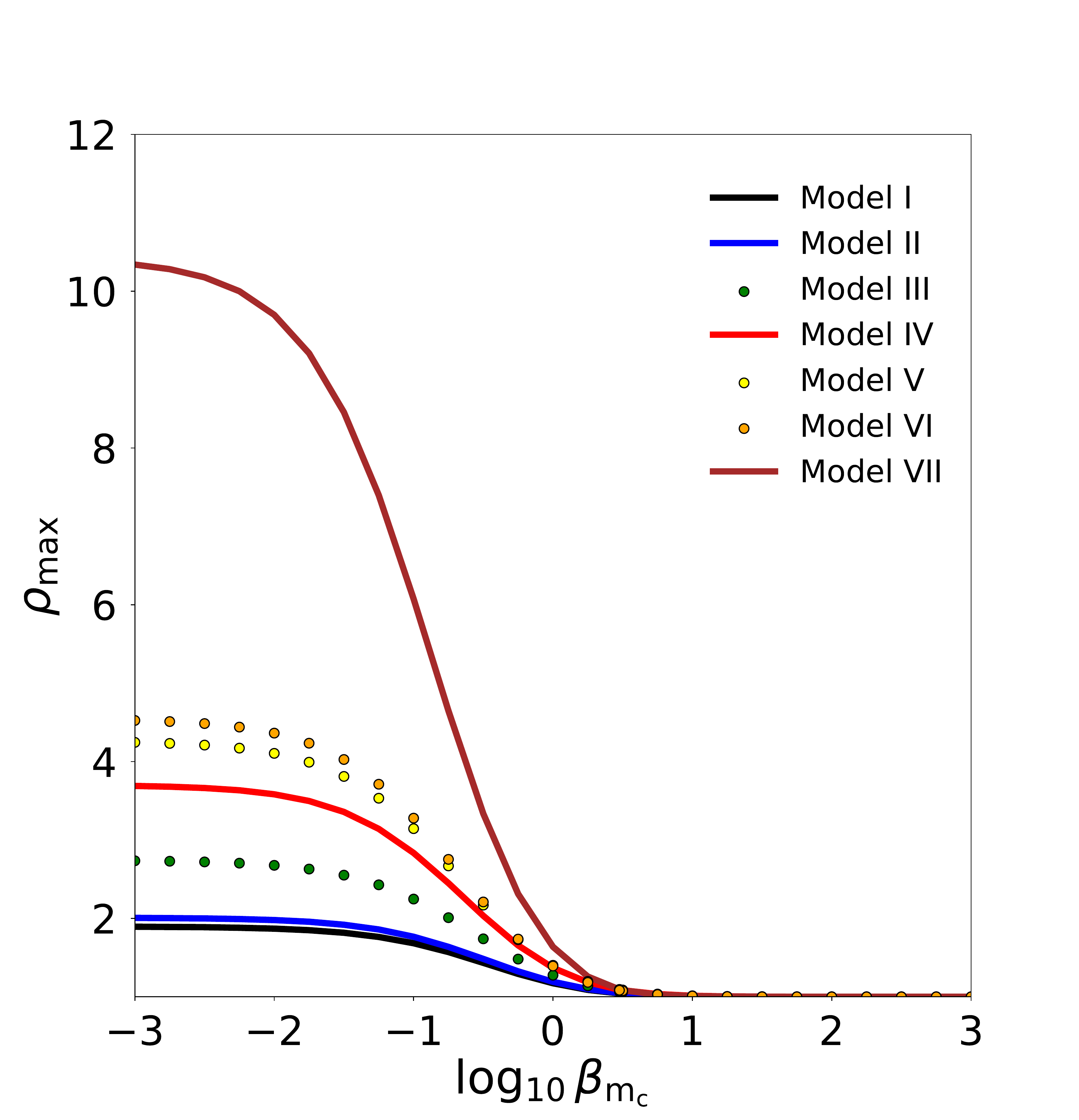}
\hspace{-0.cm}
\includegraphics[scale=0.2]{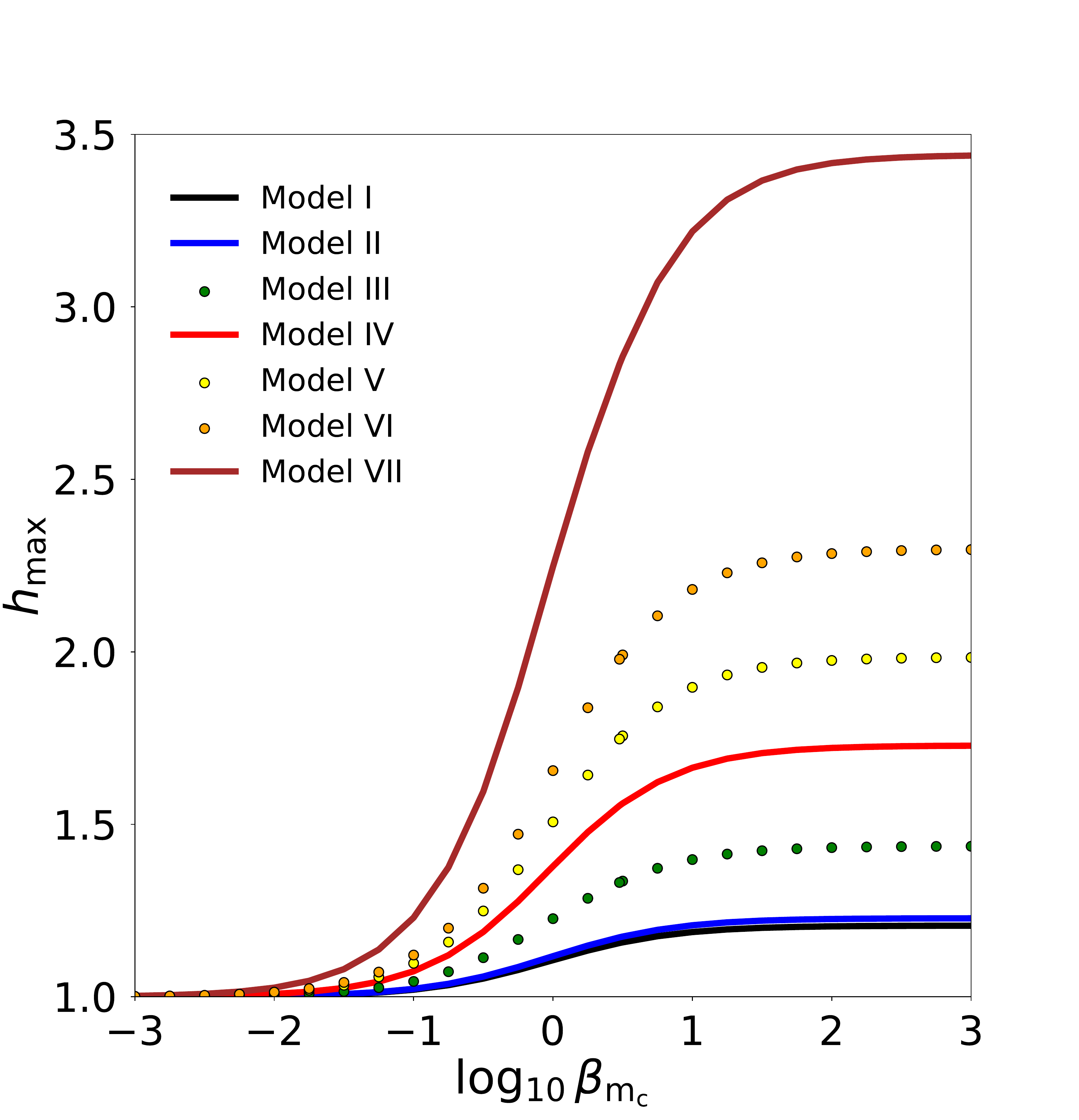}
\hspace{-0.cm}
\\
\includegraphics[scale=0.2]{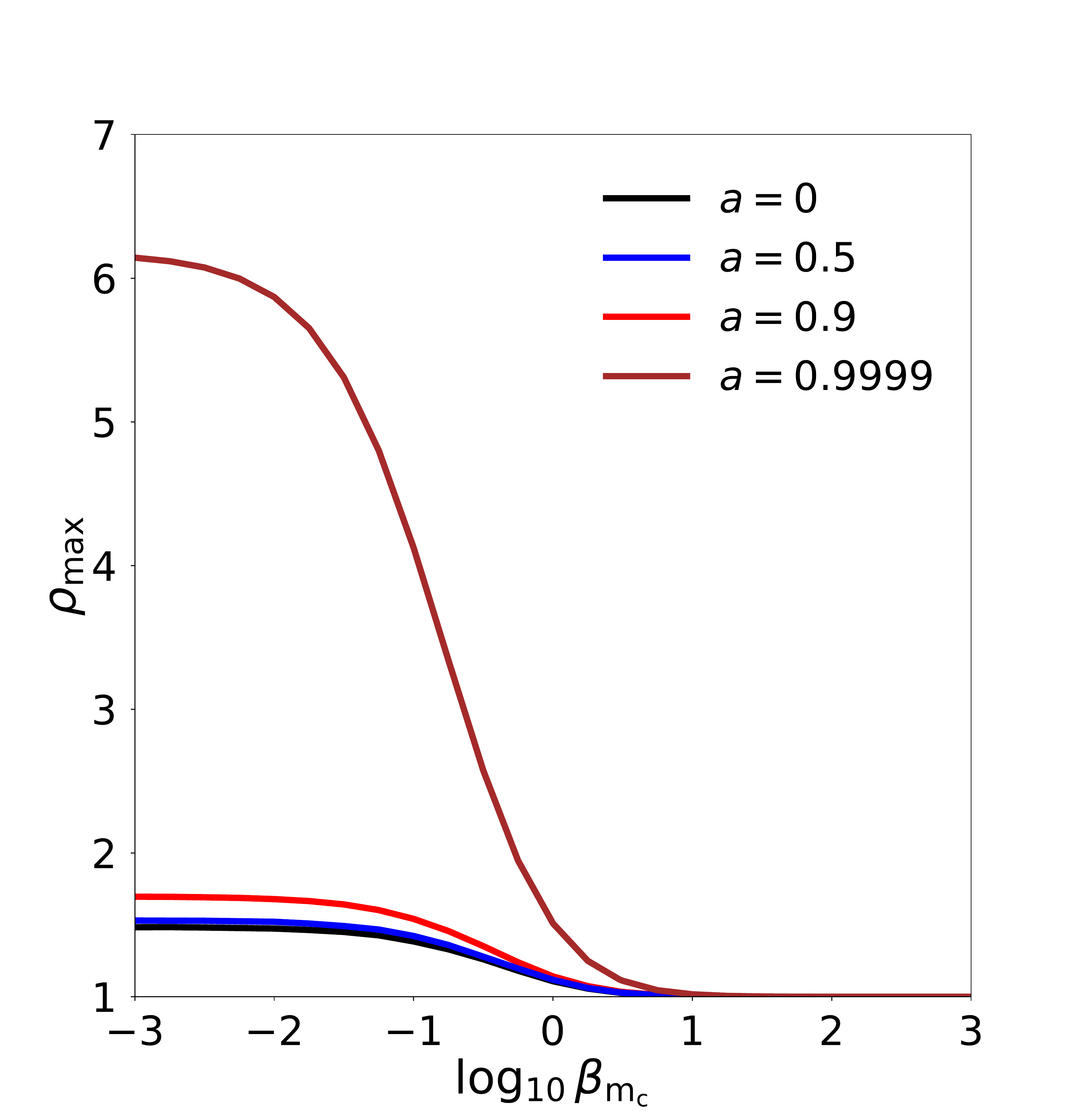}
\hspace{-0.cm}
\includegraphics[scale=0.2]{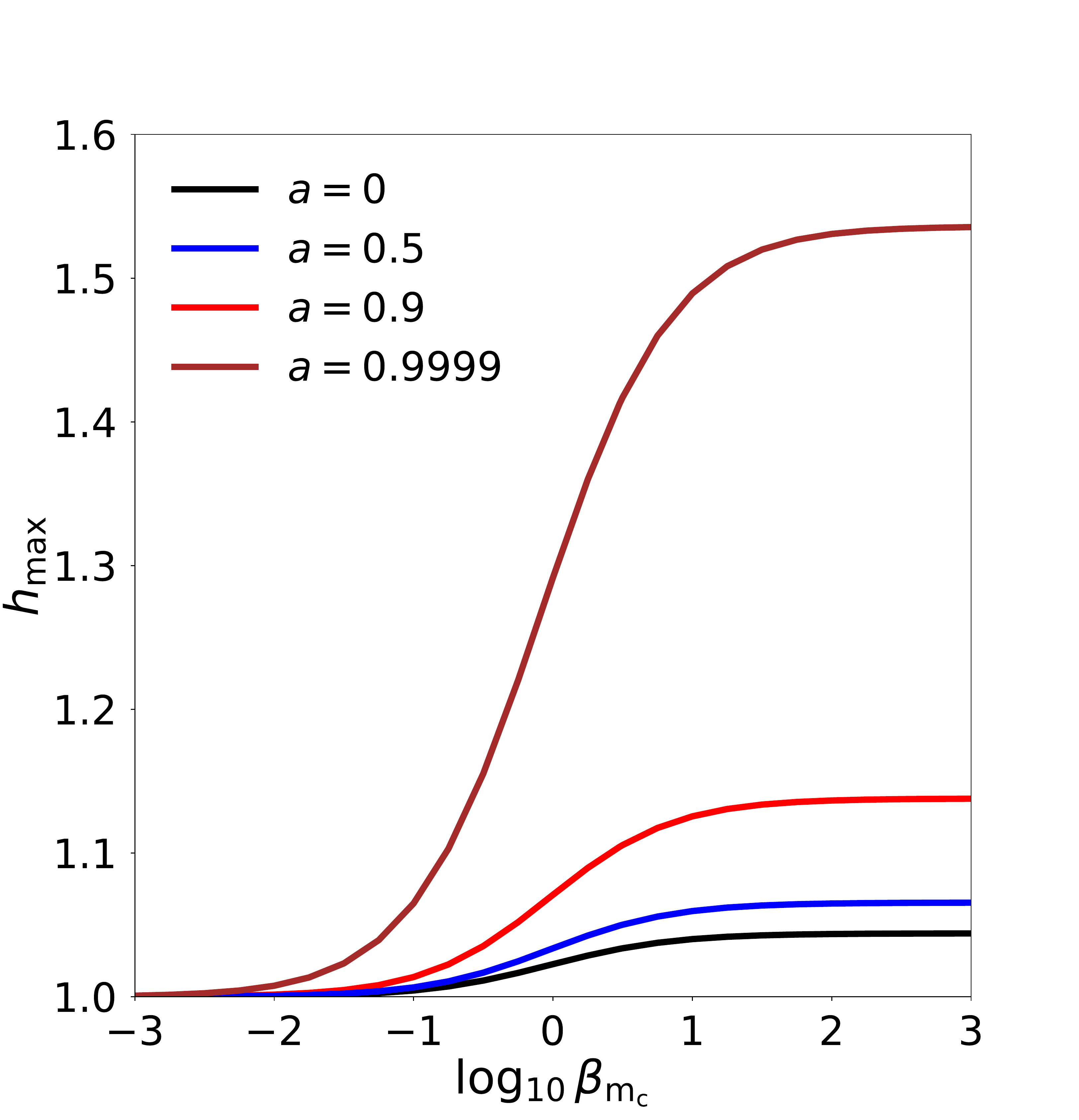}
\hspace{-0.cm}
\caption{Effects of the magnetization on the values for the maximum density (left) and enthalpy (right) of the disks. In the first row, we show this for all of our KBHsSH models. In the second row, we show this for a sequence of Kerr BHs with increasing spin parameter.}
\label{comparison_HBH_Kerr_dens_enth}
\end{figure*}

\begin{figure*}
\centering
\includegraphics[width=0.4\textwidth]{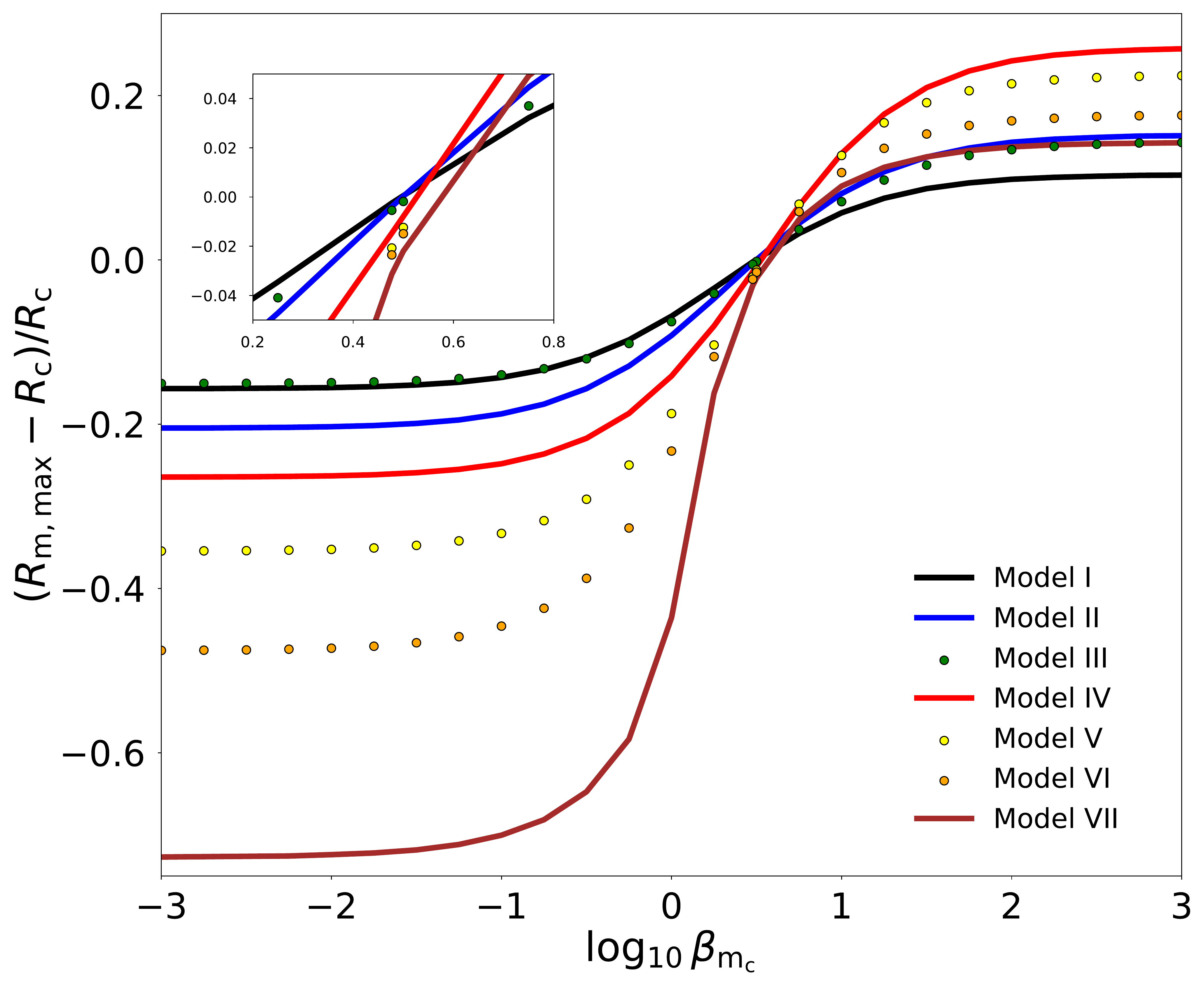}
\hspace{0.5cm}
\includegraphics[width=0.4\textwidth]{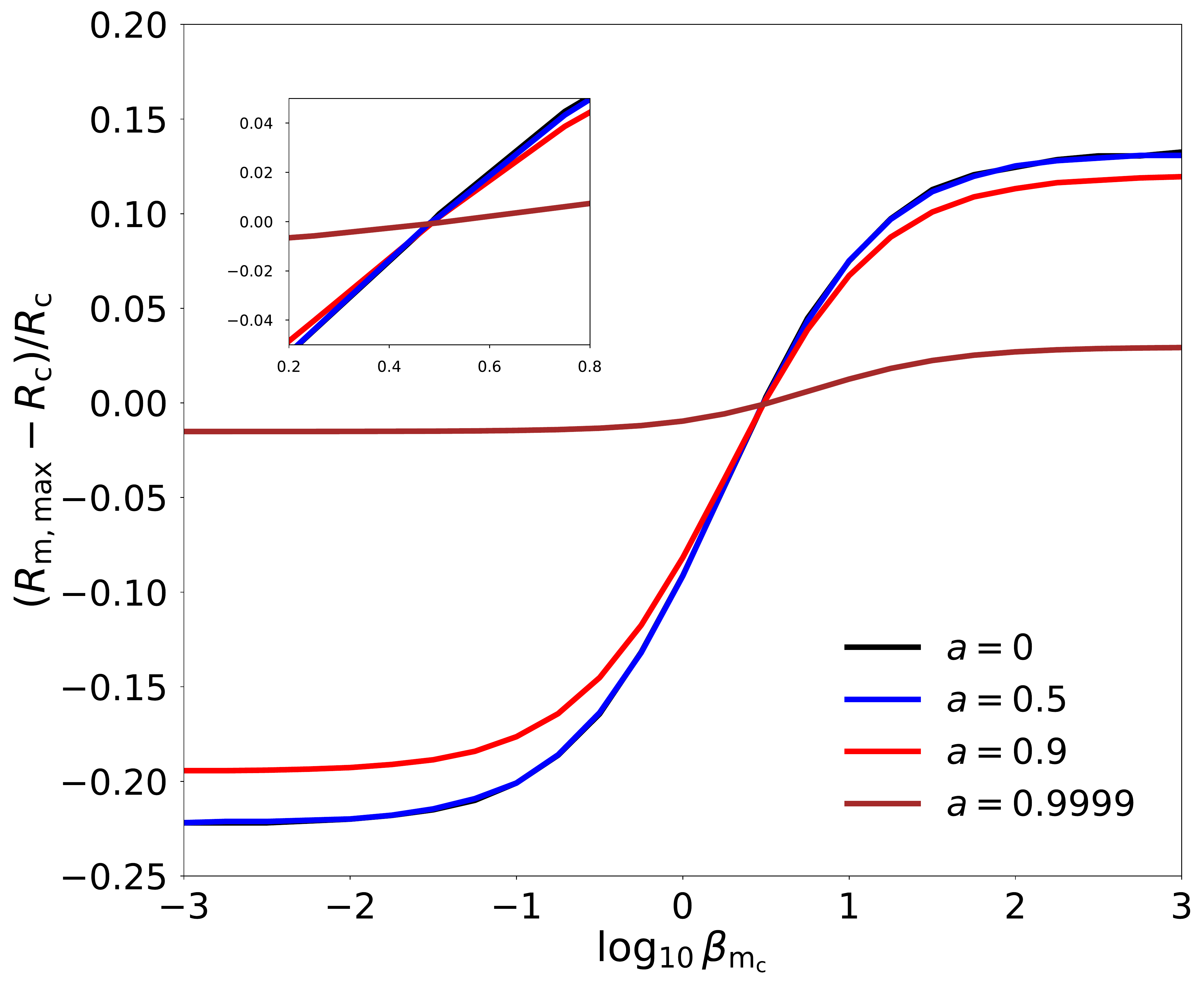}
\hspace{0.5cm}
\caption{Effects of the magnetization on the perimeteral location of the magnetic pressure maximum (divided by the the perimeteral radius of the centre), $R_{\mathrm{mag}, \mathrm{max}} - R_{\mathrm{c}})/ R_{\mathrm{c}}$). Left panel: KBHsSH models. Right panel: A sequence of Kerr BHs with increasing spin parameter.}
\label{comparison_HBH_Kerr_r_m_max}
\end{figure*}

\begin{table*}[t]
\caption{Disk parameters and values of their relevant physical magnitudes for the Kerr BH case. For all models, $R_{\mathrm{in}} = R_{\mathrm{mb}}$ , $l = l_{\mathrm{mb}}$ and $M_{\mathrm{BH}} = 1$. The meaning of the quantities reported is as in Table~\ref{HBH_disk_parameters}.}        
\label{KBH_disk_parameters}      
\centering          
\begin{tabular}{c c c c c c  c c c c c c c}
\hline\hline       
 $a$ & $\mathfrak{s}$ &  $l$ & $W_{\mathrm{c}}$ & $R_{\mathrm{in}}$ & $R_{\mathrm{c}}$ &  $\beta_{\mathrm{m_{\mathrm{c}}}}$ & $h_{\mathrm{max}}$ & $\rho_{\mathrm{max}}$ & $p_{\mathrm{max}}$ & $p_{\mathrm{m, max}}$ & $R_{\mathrm{max}}$ & $R_{\mathrm{m, max}}$\\ 
\hline           
$0$ & $1$ & $4.00$ & $-4.32 \times 10^{-2}$ & $4.00$ & $10.47$ & $10^{10}$ & $1.04$ & $1.00$ & $1.10 \times 10^{-2}$ & $1.15 \times 10^{-12}$ & $10.47$ & $11.86$\\ 

 &  &  &  &  &  & $1$ & $1.02$ & $1.11$ & $6.29 \times 10^{-3}$ & $5.69 \times 10^{-3}$ & $8.81$ & $9.52$\\ 

 &  &  & &  &  & $10^{-10}$ & $1.00$ & $1.48$ & $1.83 \times 10^{-12}$ & $1.48 \times 10^{-2}$ & $7.70$ & $8.14$\\ 

 $0.5$ & $1.053$ & $3.41$ & $-6.35 \times 10^{-2}$ & $2.99$ & $7.12$ & $10^{10}$ & $1.07$ & $1.00$ & $1.64 \times 10^{-2}$ & $1.72 \times 10^{-12}$ & $7.19$ & $8.14$\\ 

 &  &  & &  &  & $1$ & $1.03$ & $1.12$ & $9.43 \times 10^{-3}$ & $8.47 \times 10^{-3}$ & $6.05$ & $6.53$ \\ 
 
 &  &  & &  &  & $10^{-10}$ & $1.00$ & $1.53$ & $2.81 \times 10^{-12}$ & $2.23 \times 10^{-2}$ & $5.29$ & $5.59$\\ 
 
$0.9$ & $1.276$ & $2.63$ & $-0.129$ & $2.18$ & $3.78$ & $10^{10}$ & $1.14$ & $1.00$ & $1.64 \times 10^{-2}$ & $3.65 \times 10^{-12}$ & $3.78$ & $4.23$\\ 

 &  &  &  &  &  & $1$ & $1.07$ & $1.14$ & $2.03 \times 10^{-2}$ & $1.78 \times 10^{-2}$ & $3.25$ & $3.47$\\ 

 &  &  &  &  &  & $10^{-10}$ & $1.00$ & $1.70$ & $6.54 \times 10^{-12}$ & $4.92 \times 10^{-2}$ & $2.92$ & $3.04$ \\ 

$0.9999$ & $1.629$ & $2.02$ & $-0.429$ & $2.00015$ & $2.034$ & $10^{10}$ & $1.54$ & $1.00$ & $1.34 \times 10^{-1}$ & $1.61 \times 10^{-11}$ & $2.034$ & $2.094$ \\ 

 &  &  &  &  &  & $1$ & $1.29$ & $1.51$ & $1.10 \times 10^{-1}$ & $7.52 \times 10^{-2}$ & $2.0075$ & $2.014$\\ 

 &  &  &  &  &  & $10^{-10}$ & $1.00$ & $6.17$ & $1.22 \times 10^{-10}$ & $4.91 \times 10^{-1}$ & $2.0021$ & $2.0030$ \\ 
\hline\hline
\end{tabular}
\end{table*}

\begin{table*}[t]
\caption{Central density for the different models. A value of $r_{\mathrm{in}}$ such that $\Delta W = 0.9 \Delta W_{\mathrm{Total}} \equiv W_{\mathrm{cusp}} - W_{\mathrm{c}}$ is chosen and a torus gravitational mass of $M_{\mathrm{T}} = 0.1 M_{\mathrm{ADM}}$ is assumed. In the third column, the value of the central density is reported in geometrized units (`g.u.') while in the fourth column this value is reported in cgs units. The fifth and sixth columns provide those values but for tori built around Kerr BHs with the same ADM quantities as the KBHsSH models. Finally, the last column reports the sphericity of the Kerr BH models.}

\label{density_table}      
\centering          
\begin{tabular}{c c c c c c c}
\hline\hline       
 Model  & $\beta_{\mathrm{m_{\mathrm{c}}}}$ & $\rho_{\mathrm{c}}$ & $\rho_{\mathrm{c}}$ & $\rho^{\rm K}_{\mathrm{c}}$ & $\rho^{\rm K}_{\mathrm{c}}$ & $\mathfrak{s}^{\rm K}$
 \\
 & & [g.u.] & $[\mathrm{g} \, \rm{cm}^{-3}]$ & [g.u.] & $[\mathrm{g} \, \rm{cm}^{-3}]$ & 
  \\ 
\hline           
I  & $10^{10}$ & $6.818 \times 10^{-4}$ & $1.739 \times 10^{13}$  & $3.752 \times 10^{-3}$ &  $9.567 \times 10^{13}$ & $1.589$ \\ 

  & $1$ & $2.185 \times 10^{-3}$ & $5.503 \times 10^{13}$  & $7.942 \times 10^{-3}$ & $2.025 \times 10^{14}$ & \\ 

  & $10^{-10}$ & $3.227 \times 10^{-3}$ & $8.229 \times 10^{13}$  & $7.641 \times 10^{-3}$ & $1.948 \times 10^{14}$ & \\ 

II  & $10^{10}$ & $3.216 \times 10^{-4}$ & $8.201 \times 10^{12}$ & $-$ & $-$ & $-$ \\ 
 
  & $1$ & $1.651 \times 10^{-3}$ & $4.210 \times 10^{13}$ & $-$ & $-$ & $-$ \\ 

  & $10^{-10}$ & $3.026 \times 10^{-3}$ & $7.716 \times 10^{13}$ & $-$ & $-$ & $-$ \\ 

III & $10^{10}$ & $8.120 \times 10^{-4}$ & $2.071 \times 10^{13}$  & $6.683 \times 10^{-5}$ & $1.704 \times 10^{12}$ & $1.278$ \\ 

  & $1$ & $3.497 \times 10^{-3}$ & $8.917 \times 10^{13}$ & $2.075 \times 10^{-4}$ & $5.291 \times 10^{12}$ & \\ 

  & $10^{-10}$ & $5.452 \times 10^{-3}$ & $1.390 \times 10^{14}$  & $3.265 \times 10^{-4}$ & $8.325 \times 10^{12}$ & \\ 

IV & $10^{10}$ & $1.197 \times 10^{-3}$ & $3.052 \times 10^{13}$  & $3.001 \times 10^{-5}$ & $7.652 \times 10^{11}$ & $1.219$ \\ 

  & $1$ & $3.421 \times 10^{-3}$ & $8.723 \times 10^{13}$  & $9.512 \times 10^{-5}$ & $2.425 \times 10^{12}$ & \\ 

  & $10^{-10}$ & $5.135 \times 10^{-3}$ & $1.309 \times 10^{14}$ & $1.533 \times 10^{-4}$ & $3.909 \times 10^{12}$ & \\ 
 
V  & $10^{10}$ & $1.792 \times 10^{-3}$ & $4.569 \times 10^{13}$  & $3.152 \times 10^{-5}$ & $8.037 \times 10^{11}$ & $1.227$\\ 

  & $1$ & $3.883 \times 10^{-3}$ & $9.901 \times 10^{13}$  & $9.942 \times 10^{-5}$ & $2.535 \times 10^{12}$ & \\ 

  & $10^{-10}$ & $5.435 \times 10^{-3} $ & $1.386 \times 10^{14}$  & $1.596 \times 10^{-4}$ & $4.070 \times 10^{12}$ & \\ 

VI & $10^{10}$ & $2.348 \times 10^{-3}$ & $5.987 \times 10^{13}$  & $3.232 \times 10^{-5}$ & $8.241 \times 10^{11}$ & $1.234$\\ 

  & $1$ & $4.106 \times 10^{-3}$ & $1.047 \times 10^{14}$  & $1.019 \times 10^{-4}$ & $2.598 \times 10^{12}$ & \\ 

  & $10^{-10}$ & $5.685 \times 10^{-3}$ & $1.450 \times 10^{14}$ & $1.632 \times 10^{-4}$ & $4.161 \times 10^{12}$ & \\ 

VII  & $10^{10}$ & $3.737 \times 10^{-3}$ & $9.529 \times 10^{13}$ & $4.114 \times 10^{-5}$ & $1.049 \times 10^{12}$ & $1.268$\\ 

  & $1$ & $5.356 \times 10^{-3}$ & $1.366 \times 10^{14}$ & $1.280 \times 10^{-4}$ & $3.264 \times 10^{12}$ & \\ 

  & $10^{-10}$ & $7.598 \times 10^{-3}$ & $1.937 \times 10^{14}$ & $2.021 \times 10^{-4}$ & $5.153 \times 10^{12}$ & \\ 
 \hline\hline   
\end{tabular}
\end{table*}

In figure~\ref{comparison_mass_density} we show the total energy density of the torus $\rho_{\mathrm{T}}$ (upper half of each image) and the total energy density of the scalar field $\rho_{\mathrm{SF}}$ (lower half) for models I, IV and VII and two  values of the magnetization parameter at the centre ($10^{10}$, top row, and $10^{-10}$, bottom row). This figure shows that, for non-magnetized disks, the maximum of the total energy density of the disk $\rho_{\mathrm{T}}$ is closer to the maximum of the total energy density of the scalar field $\rho_{\mathrm{SF}}$ for increasing hair. This trend disappears with increasing magnetization, as the disk moves closer to the horizon in such case.

\subsection{Comparison with Kerr BHs}

For the sake of comparison we also build equilibrium sequences of magnetized disks around four Kerr BHs of the same mass ($M_{\mathrm{BH}} = 1$) and varying spins, from $a=0$ to $a=0.9999$. These models are more general than the corresponding ones presented in~\cite{Gimeno-Soler:2017} as the $h=1$ assumption is now relaxed. Our numerical approach can handle BH spins as large as $|a-1|=10^{-7}$ without modifying the resolution of our numerical grid. However, for higher values of the spin parameter, we would need to increase our resolution (especially the resolution along the polar angle $\theta$ for the most highly magnetized case) but such extreme cases do not add further relevant information to our discussion. Table~\ref{KBH_disk_parameters} reports a summary of the values of the main physical quantities of these disks, whose morphology is displayed in Figs.~\ref{models_Kerr} and~\ref{models_Kerr_peri}. As for the disks built around KBHsSH, the maximum values of the enthalpy, density, pressure and magnetic pressure increase with increasing $|\Delta W|$, which, in the Kerr BH case, also means with increasing values of $a$. It can be seen that both the cusp and the centre move closer to the horizon with increasing $a$, i.e.~the disks reduce their size and approach the BH as the spin parameter increases. (Note that, as we mentioned before, in the Kerr case the radial location of the horizon at the equatorial plane in perimeteral coordinates is $R_{\mathrm{H}} = 2M$ irrespective of the value of the BH spin.) 

The comparison of the values of the physical quantities shows that, even for highly rotating Kerr BHs, the maximum values for $h$, $p$ and $p_{\mathrm{m}}$ are lower than in the KBHsSH case. This is not a surprise, as these quantities are related to the value of $|\Delta W|$. Also, as in the case of KBHsSH, we observe a higher distortion of the shape of the disc in the near-horizon region with increasing sphericity $\mathfrak{s}$ (and spin, in this particular case). This is particularly noticeable when plotting the disk morphology in terms of the perimeteral coordinates (cf.~Fig.~\ref{models_Kerr_peri}). For the $a=0.9999$ model the disk is extremely skewed and attached to the BH horizon, particularly in the highly magnetized case in which the values reported in Table~\ref{KBH_disk_parameters} for $R_{\rm in}$ and $R_{\rm c}$ are very close to each other. The appearance of the solution is more disk-like when displayed in terms of the $r$ coordinate, as shown in Fig.~\ref{models_Kerr}, as this radial coordinate expands the near-horizon region. While this coordinate is well suited to do the computations, this is not the case for visualization, where the perimeteral coordinate is preferred since it allows to directly compare the different models as the scale is the same.

To provide additional information for the comparison we show in the bottom panels of Fig.~\ref{radial_profiles} three disk models around Kerr BHs  with the same ADM mass and ADM angular momentum as the KBHsSH cases shown in the upper panels of the same figure. The model on the left plot corresponds to a near-extremal Kerr BH ($a=0.9987$) and the other two have a similar value for the  spin parameter ($a=0.8489$ and $a=0.8941$, for the middle and right plots, respectively). The comparison reveals interesting differences between these models regarding their compactness. The size of the disk in the Kerr case plotted on the left is considerably smaller than its hairy counterpart, KBHsSH model I. In this case, the presence of the scalar field has little effect on the morphology of the disk (as its gravitational field is small) but its effect is nonetheless noted in a reduction of the value of the sphericity (see Table~\ref{density_table}), effectively reducing the effect of the BH spin in the disk (i.e.~increasing its shape). As the mass and angular momentum stored in the scalar field increase, the gravitational field of the scalar field affects the radial morphology of the disk, altering its shape and reducing its extent. Note that both KBHsSH models IV and VII have lesser radial extent than its Kerr BH counterparts with the same ADM mass and angular momentum, even though model VII attains a lower value of the sphericity. These conclusions hold irrespective of the value of the magnetization parameter.

\subsection{Magnetization profiles}

The dependence of the maximum specific enthalpy $h_{\mathrm{max}}$ and the maximum rest-mass density $\rho_{\mathrm{max}}$ with the magnetization parameter is shown in Fig.~\ref{comparison_HBH_Kerr_dens_enth}. The upper panels correspond to the KBHsSH models (I-VII) and the lower ones to our sequence of Kerr BHs with increasing spin parameter. For both cases, an increase in $|\Delta W|$ implies monotonically  higher values for $h_{\mathrm{max}}$ (low magnetization) and also higher values for $\rho_{\mathrm{max}}$ (high magnetization). However, there are quantitative differences between the two cases. For the enthalpy, the values of $h_{\mathrm{max}}$ reached for disks around KBHsSH are much higher than those of the Kerr BH case. This implies that, while the $w = \rho h \simeq \rho$ approximation (employed in~\cite{Komissarov:2006,Gimeno-Soler:2017}) is valid for magnetized disks ($\beta_{\mathrm{m_c}} \sim 1$) around Kerr BHs for values of the spin parameter as high as $a \sim 0.99$, that is not the case for disks around KBHsSH. We note that for the most extreme spin value we can build, $|a-1|=10^{-7}$, the maximum enthalpy for the purely hydrodynamical case is $h_{\rm max}=1.692$. For this case, the maximum density in the extremely magnetized limit reaches a value of $\rho_{\mathrm{max}} = 97$, significantly larger than the value displayed in the left  panel of Fig.~\ref{comparison_HBH_Kerr_dens_enth} for the $a=0.9999$ model.

Figure~\ref{comparison_HBH_Kerr_r_m_max} shows the relative variation of the quotient of the perimeteral radius of the magnetic pressure maximum and the perimeteral radius of the disk centre, $(R_{\mathrm{m, max}}-R_{\mathrm{c}})/R_{\mathrm{c}}$, with the decimal logarithm of the magnetization parameter at the centre of the disk, $\log_{10} \beta_{\mathrm{m_c}}$. The curves plotted correspond to the same KBHsSH and Kerr BH cases as those in figure~\ref{comparison_HBH_Kerr_dens_enth}. For all cases, the radial location of the magnetic pressure maximum decreases with decreasing $\beta_{\mathrm{m_c}}$. In~\cite{Gimeno-Soler:2017} we proved that for $h=1$ disk models in stationary and axisymmetric BH spacetimes, the location of the maximum of the magnetic pressure is identical for all models when $\beta_{\mathrm{m_c}}\equiv 1 / \Gamma - 1=3$. This condition is almost fulfilled for the Kerr BH case  even when $h \neq 1$, with a very slight deviation for cases with very high spin parameter. This cannot be seen clearly in Fig.~\ref{comparison_HBH_Kerr_r_m_max} (even in the inset) but, as an example, for $a=0.9999$, the relative difference of $(R_{\mathrm{m, max}}-R_{\mathrm{c}})/R_{\mathrm{c}}$ with the $h=1$ case is about 0.1\%. (We note that in the radial coordinate of the metric ansatz, the disks are not so skewed and attached to the horizon and the differences would be more visible.) On the other hand, the condition $R_{\mathrm{m, max}}=R_{\mathrm{c}}$ when $\beta_{\mathrm{m_c}}= 3$ is clearly not fulfilled (when $h\neq 1$) for disks built around KBHsSH (see inset in the left panel). At this point, it is relevant to remember that some of the KBHsSH models violate the Kerr bound in terms of the potential. As we mentioned previously, we need a small value of $\Delta W$ for the $h \simeq 1$ approximation to be valid in the non-magnetized regime. Now we can see that, in the KBHsSH case, this approximation is not valid even for mildly-magnetized disks.

\subsection{Torus mass}

In an attempt to gauge the astrophysical relevance of our models, in this section we drop the $\rho_{\mathrm{c}} = 1$ choice we have thus far considered to build the tori and compute their masses and, instead, we assume that the mass of the tori is $M_{\mathrm{T}} = 0.1 M_{\mathrm{ADM}}$ and ask ourselves what are the corresponding values of the central density of each model. The value selected for $M_{\mathrm{T}}$ is, broadly speaking, compatible with the torus masses found through numerical relativity simulations of binary neutron star mergers (see, e.g.~\cite{Rezzolla:2010, Rezzolla:2017} and references therein). Moreover, to avoid complications due to the infinite size of our models, we choose the total potential well as the $90\%$ of its maximum possible value.

Therefore, we compute the mass of the tori around KBHsSH and, for comparison, the corresponding mass for seven disk models around Kerr BHs, each one of them with the same ADM quantities as their KBHsSH counterparts. The resulting values are reported in 
Table~\ref{density_table}. The variables corresponding to the Kerr case are indicated with a `K' superindex in this table. The third  and fifth columns of Table~\ref{density_table} indicate the resulting central densities for the KBHsSH and Kerr BH cases, respectively, in geometrized units. In order to compare these values with those from the end-products of binary neutron star mergers, we need to convert our results to cgs units. To this end, we first need to choose a mass for the scalar field $\mu$, as the maximum ADM mass of KBHsSH depends on $\mu$. In particular, we compute the maximum ADM mass with the following equation (see~\cite{Herdeiro:2015a} and references therein)
\begin{equation}\label{eq:max_mass_HBH}
M^{\rm max}_{\rm ADM} \simeq \alpha_{\mathrm{BS}} 10^{-19} M_{\odot} \left(\frac{\mathrm{GeV}}{\mu}\right)\,,
\end{equation}
for a value of $\alpha_{\mathrm{BS}} = 1.315$ (corresponding to a value of the azimuthal harmonic index $m = 1$). The constant $\alpha_{\mathrm{BS}}$ is computed numerically for rotating boson stars. Note that Eq.~\eqref{eq:max_mass_HBH} corresponds to the maximum mass of a boson star but, as  mentioned in~\cite{Herdeiro:2015a}, this is also the maximum mass for the corresponding hairy BH. Using a value of the mass of the scalar field of $\mu = 2.087 \times 10^{-11}$ eV yields values for the ADM mass of our models that increase from 2.043 $M_{\odot}$ to 4.799 $M_{\odot}$, from model I to VII. This value of $\mu$ is within the mass range suggested by the \emph{axiverse} of string theory (see~\cite{Arvanitaki:2010}) portraying a large number of scalar fields in a mass range from $10^{-33}$ eV to $10^{-10}$ eV. In addition, as mentioned in~\cite{Ramazanoglu:2016}, the value of $\mu$ we choose is compatible with the scalar-field mass range allowed by the observational tests in scalar-tensor theories of gravity. Although in this paper we are working within general relativity, for the low values of the trace of the energy-momentum tensor of our models (in comparison with the values reached for neutron stars) both theories should be indistinguishable. We should note as well that Eq.~(\ref{eq:max_mass_HBH}) is valid for non self-interacting scalar hair. Adding self-interaction terms would produce astrophysically relevant solutions for less extreme values of the scalar-field mass $\mu$~\cite{Delgado:2018}.

Once we compute the new values of the central density in geometrized units, we use the following equation~\cite{RezzollaBook}
\begin{equation}
\rho_{\mathrm{cgs}} = 6.17714 \times 10^{17} \left(\frac{G}{c^2}\right)\left(\frac{M_{\odot}}{M}\right)^2 \rho_{\mathrm{geo}}\,,
\end{equation}
to obtain the value of the central density in cgs units for the different models. These values are reported in columns four and six of Table~\ref{density_table}. The range of values is fairly broad, spanning from $\sim10^{11}$ g cm$^{-3}$ to $\sim10^{14}$ g cm$^{-3}$. This is due to the significant differences in size of the different disks, especially between the Kerr and KBHsSH cases. Comparing these values with those reported in the literature (see~\cite{Rezzolla:2010,Rezzolla:2017}) we conclude that, despite our assumptions, they are in the same ballpark than the central densities found in disks consistently formed through ab-initio simulations of binary neutron star mergers. In particular, changing the distribution of the specific angular momentum from our simplistic constant prescription to a more realistic power-law distribution, may help improve the accuracy of our results. 

\section{Conclusions}
\label{conclusions}

Astrophysical BHs are commonly surrounded by accretion disks, either at stellar-mass scales or at supermassive scales. In the former case, stellar-mass BHs surrounded by thick disks (or tori) are broadly accepted as natural end results of catastrophic events involving the coalescence and merger of compact objects, namely binary neutron stars and BH-neutron star systems (see e.g.~\cite{FaberRasio2012,Paschalidis:2016agf,Rezzolla:2017} and references therein). These systems are traditionally described using the paradigmatic BHs of general relativity, where the spacetime metric is given by the Kerr metric, solely characterized by the BH mass and spin. Upcoming observational campaigns may, however, provide data to discriminate those canonical BH solutions from exotic alternatives as, e.g.~those in which the BHs are endowed with scalar or vector (Proca) hair, recently obtained by~\cite{Herdeiro:2014a,Herdeiro:2016}. It is conceivable that testing the no-hair hypothesis of BHs will become increasingly more precise in the next few years as new observational data is collected in both the gravitational-wave channel and in the electromagnetic channel.

In this paper we have considered numerically generated spacetimes of Kerr BHs with synchronised scalar hair and have built stationary models of magnetized tori around them. Those disks are assumed to be non-self-gravitating, to obey a polytropic equation of state, and to be marginally stable, i.e.~the disks completely fill their Roche lobe. In addition, and for the sake of simplicity, the distribution of the specific angular momentum in the disks has been assumed to be constant. The models have been constructed building on existing approaches presented in~\cite{Komissarov:2006} and~\cite{Gimeno-Soler:2017} which dealt with (hairless) Kerr BHs. An important generalization of the present work compared to the methodology presented in previous works has had to do with the fluid model: while the matter EOS we use is still rather simplistic (a polytropic EOS) the models are allowed to be thermodinamically relativist, as the specific enthalpy of the fluid can adopt values significantly larger than unity. That has led to interesting differences with respect to the findings reported in~\cite{Gimeno-Soler:2017} for the purely Kerr BH case.

We have studied the dependence of the morphology and properties of the accretion tori on the type of BH system considered, from purely Kerr BHs with varying degrees of spin parameter (namely from a Schwarzschild BH to a nearly extremal Kerr case)  to KBHsSH with different ADM mass and horizon angular velocity. Comparisons between the disk properties for both types of BHs have been presented. The sequences of magnetized, equilibrium disks models discussed in this study can be used as initial data for numerical relativity codes to investigate their dynamical (non-linear) stability and can be used in tandem with ray-tracing codes to obtain synthetic images of black holes (i.e.~shadows) in astrophysically relevant situations where the light source is provided by an emitting accretion disk (first attempted by~\cite{Vincent:2016}). In a companion paper we will present the non-constant (power-law) case, whose sequences have already been computed. The dynamical (non-linear) stability of these solutions as well as the analysis of the corresponding shadows will be discussed elsewhere.

\section*{Acknowledgements}

This work has been supported by the Spanish MINECO (grant AYA2015-66899-C2-1-P), by the FCT (Portugal) IF programme, by the FCT grant PTDC/FIS-OUT/28407/2017, by  CIDMA (FCT) strategic project UID/MAT/04106/2013, by CENTRA (FCT) strategic project UID/FIS/00099/2013 and by  the  European  Union's  Horizon  2020  research  and  innovation  (RISE) programmes H2020-MSCA-RISE-2015 Grant No.~StronGrHEP-690904 and H2020-MSCA-RISE-2017 Grant No.~FunFiCO-777740. E.R. gratefully acknowledges the support of DIAS. The authors would like to acknowledge networking support by the COST Action CA16104.  

\begin{appendix}

\section{Finding $l_{\mathrm{mb}}$ and $r_{\mathrm{mb}}$}
\label{ang_mom_appendix}

We start by considering the Lagrangian of a stationary and axisymmetric spacetime
\begin{equation}
\mathcal{L} = \frac{1}{2}\left[g_{tt} (\dot{t})^2 + 2 g_{t\phi} \dot{t}\dot{\phi} + g_{rr} (\dot{r})^2 + g_{\theta\theta} (\dot{\theta})^2 + g_{\phi\phi} (\dot{\phi})^2\right]\ ,
\end{equation}
where $\dot{x^{\alpha}} = \mathrm{d}x^{\alpha}/\mathrm{d}\lambda$ denotes the partial derivative of the coordinates with respect to an affine parameter $\lambda$. We can note that we have two cyclic coordinates ($t$ and $\phi$). Then, the canonically conjugate momentum of each coordinate is conserved, namely
\begin{eqnarray}\label{eq:momenta}
p_t = \frac{\partial L}{\partial \dot{t}} = -E \ ,
\\
p_{\phi} = \frac{\partial L}{\partial \dot{\phi}} = L\ ,
\end{eqnarray}
where we identify the constants of motion as the energy and angular momentum of a test particle.

If we assume motion in the equatorial plane (i.e.~$\theta = \pi/2$, $\dot{\theta} = 0$) we can write the relativistic four-momentum (of a massive particle) normalisation as
\begin{equation}\label{eq:momenta_normalisation}
p_{t}p^t + p_{r}p^r + p_{\phi}p^{\phi} = -m^2\ ,
\end{equation}
where $m$ is the mass of a test particle. Using the defintions of the energy and angular momentum of the particle and taking into account that $p^{\alpha} = \dot{x^{\alpha}}$, we can rewrite the above equation as
\begin{equation}
-E \dot{t} + L \dot{\phi} + g_{rr} \dot{r}^2 = -m^2.
\end{equation}
Now, we can find the expressions for the contravariant momenta $p^{t}$ and $p^{\phi}$ from $p_{\alpha} = g_{\alpha \beta}p^{\beta}$
\begin{eqnarray}
p^{t} = \frac{g_{\phi\phi}E + g_{t\phi}}{g_{t\phi}^2-g_{tt}g_{\phi\phi}} \ ,
\\
p^{\phi} = -\frac{g_{tt}L+g_{t\phi}}{g_{t\phi}^2-g_{tt}g_{\phi\phi}}\ ,
\end{eqnarray}
replace these expressions into Eq.~\eqref{eq:momenta_normalisation} and write the expression for the radial velocity $\dot{r}$
\begin{equation}\label{eq:radial_velocity}
\dot{r} = \left(-m^2+\frac{g_{\phi\phi}E^2+2 g_{t\phi}LE+g_{tt}L^2}{g_{t\phi}^2-g_{tt}g_{\phi\phi}}\right)^{\frac{1}{2}}.
\end{equation}
We want to consider circular orbits, so the radial velocity must be $\dot{r} = 0$. Then, we arrive at
\begin{equation}
g_{t\phi}^2-g_{tt}g_{\phi\phi} = g_{\phi\phi}e^2+2 g_{t\phi}le+g_{tt}l^2\ ,
\end{equation}
where we have introduced the specific energy per unit mass ($e = E/m$) and the specific angular momentum per unit mass ($l = L/m$). Additionally, we are interested in bound orbits. Specifically, we want marginally bound orbits ($e=1$). Taking this into account, we get the following expression for the specific angular momentum
\begin{equation}\label{eq:bound_ang_mom}
l^{\pm}_{\mathrm{b}} = \frac{g_{t\phi} \pm \sqrt{ (g_{t\phi}^2-g_{tt}g_{\phi\phi})  (1+g_{tt}) } }{-g_{tt}}\,
\end{equation}
which corresponds to Eq.~\eqref{eq:mb_ang_mom}. It is well-known that in BH spacetimes there is an innermost circular marginally bound orbit for test particles. Naturally, a marginally bound particle at the innermost circular orbit has to have the smallest possible value of the specific angular momentum (i.e.~a minimum of Eq.~\eqref{eq:bound_ang_mom}). The radial location of said minimum is, obviously, the innermost circular marginally bound radius $r_{\mathrm{mb}}$.
\end{appendix}

\bibliographystyle{apsrev4-1}
\bibliography{references}

\end{document}